\documentclass[fleqn,usenatbib]{mnras}

\usepackage{newtxtext,newtxmath}
\usepackage{multirow}
\usepackage{graphicx}

\usepackage[T1]{fontenc}
\usepackage{booktabs}

\DeclareRobustCommand{\VAN}[3]{#2}
\let\VANthebibliography\thebibliography
\def\thebibliography{\DeclareRobustCommand{\VAN}[3]{##3}\VANthebibliography}

\usepackage{graphicx}	
\usepackage{amsmath}	

\title[Effects of uncertainties in stable mass transfer and stellar winds]{Importance of stable mass transfer and stellar winds for the formation of gravitational wave sources}

\author[A. Dorozsmai, S. Toonen]{
Andris Dorozsmai,$^{1}$\thanks{E-mail: andris@star.sr.bham.ac.uk}
Silvia Toonen,$^{2,1}$
\\
$^{1}$Institute of Gravitational Wave Astronomy and School of Physics and Astronomy, University of Birmingham,\\ Edgbaston, Birmingham B15 2TT, United Kingdom\\
$^{2}$ Astronomical Institute Anton Pannekoek, University of Amsterdam, Science Park 904, 1098 XH Amsterdam, The Netherlands
}

\date{Accepted XXX. Received YYY; in original form ZZZ}

\pubyear{2015}

\begin{document}
\label{firstpage}
\pagerange{\pageref{firstpage}--\pageref{lastpage}}
\maketitle

\begin{abstract}
The large number of gravitational wave (GW) detections have revealed the properties of the merging black hole binary population, but how such systems are formed is still heavily debated. Understanding the imprint of stellar physics on the observable GW population will shed light on how we can use the gravitational wave data, along with other observations, to constrain the poorly understood evolution of massive binaries. We perform a parameter study on the classical isolated binary formation channel with the population synthesis code \textsc{SeBa} to investigate how sensitive the properties of the coalescing binary black hole population are on the uncertainties related to first phase of mass transfer and stellar winds. We vary five assumptions: 1 and 2) the mass transfer efficiency and the angular momentum loss during the first mass transfer phase, 3) the mass transfer stability criteria for giant donors with radiative envelopes, 4) the effective temperature at which an evolved star develops a deep convective envelope, and 5) the mass loss rates of stellar winds. We find that current uncertainties related to first phase of 
mass transfer have a huge impact on the relative importance of different dominant channels, while the observable demographics of  GW sources
are not significantly affected. Our varied parameters have a complex, interrelated effect on the population properties of GW sources. Therefore, inference of massive binary physics from GW data alone remains extremely challenging, given the large
uncertainties in our current models. 
\end{abstract}




\section{Introduction}
Massive stars play an essential role in astrophysics.
They are responsible for the chemical enrichment of the universe via stellar winds and supernovae. They are also progenitors of various interesting astrophysical phenomena, e.g. neutron stars, black holes, gamma ray bursts (e.g. \citealt{Langer_2012}).
However, our understanding of these rare and short-lived objects is still incomplete. The population of merging compact binaries, observed via gravitational waves (GW) offer a unique but indirect way to study the evolution of these objects.
Since the first detection of GWs, about a hundred merging binary black holes have been observed, which makes the inference of the population statistics of black hole-black hole binaries (BH-BH binaries) possible \citep{Abbott_2021, Abbott2023_GWTC3_POP}. 

Numerous formation channels of merging stellar mass binary black holes  have been proposed in the last decades. These include formation scenarios involving isolated, interacting massive binaries (i.e. the classical isolated binary channel, see e.g. \citealt{Paczynski1976IAUS...73...75P}; \citealt{Heuvel76};    \citealt{Tutukov1993MNRAS.260..675T}; \citealt{Dominik_2012}; 
\citealt{Mennekens_2014}; \citealt{Belczynski_2016}; 
\citealt{Eldridge_2016};
\citealt{Mennekens2016};
\citealt{Stevenson_2017};
\citealt{Giaccobbo2018MNRAS.480.2011G};
\citealt{Kruckow_2018}; 
\citealt{KlenckiNelemans2018}; \citealt{Neijssel_2019}; 
\citealt{Petra2020};
\citealt{marchant2021role}; \citealt{Bavera2021};
\citealt{Briel2022};
\citealt{Riley_2022};
\citealt{Briel2023}), massive binaries comprising chemically homogeneously evolving stars  (\citealt{de_Mink_2016}; \citealt{Mandel_2016}; \citealt{Marchant_2016}; 
\citealt{Eldridge_2016}), or  scenarios in which dynamical interactions play a key role in forming GW transients, e.g. in dense environments, such as globular clusters (e.g. \citealt{Kulkarni1993Natur.364..421K}; \citealt{Sigurdsson1993Natur.364..423S}; \citealt{PortegiesMcMillian2000}; \citealt{Rodriguez_2016}; \citealt{DiCarlo2020}), nuclear clusters (e.g. \citealt{AntoniniPerets2012ApJ...757...27A}), AGN discs   (e.g. \citealt{McKerna2020}; \citealt{Bird2017MNRAS.464..946S}; \citealt{Bartos2017ApJ...835..165B}) or scenarios involving hierarchical, field triples (e.g. 
\citealt{Silsbee2017ApJ...836...39S};\citealt{Antonini2017ApJ...841...77A}; 
\citealt{Martinesz_https://doi.org/10.48550/arxiv.2105.01671};\citealt{Alejandro2021}; \citealt{Stegemann2022arXiv220316544S}). The possibility of merging binary black holes originating from population III stars (\citealt{Belc2004};\citealt{Kinugawa10.1093/mnras/stu1022}; \citealt{Inayoshi10.1093/mnras/stx757}) or from primordial black holes (\citealt{BirdPhysRevLett.116.201301}, \citealt{Sasaki2018}) has also been proposed and studied.

The classical isolated binary channel is perhaps the most studied formation path.
In these scenario, typically two main sub-channels are identified. The first one is the CEE channel, in which the key step in the formation of close BH-BH binaries is the so-called common envelope evolution (CEE, e.g. \citealt{Ivanova_2013}). 
Several, earlier population synthesis studies predicted a merger rate for this channel that is broadly consistent with the currently inferred LIGO rate \citep[see e.g.][]{ Mandel_2022}, although these predictions are sensitively dependent on the highly uncertain common envelope efficiency and the binding energy of the envelope of the donor star.
However, recent detailed stellar evolutionary models of \citet{Klencki_2021} and \citet{marchant2021role} showed that the binding energy of evolved stars with radiative envelopes could be underestimated by prescriptions commonly used by rapid population synthesis codes. Furthermore, a deep convective envelope could potentially be developed at a significantly cooler effective temperature (\citealt{Klencki_2020}) than previously assumed. These two developments would imply an appreciably lower predicted merger rate for this channel, possibly orders of magnitude lower than the currently inferred rate.

The second dominant channel (stable channel) involves two subsequent stable mass transfer episodes. The orbit of a binary experiencing a stable phase of mass transfer episode with black hole accretor can shrink significantly, if the mass ratio of the system is sufficiently high. This can lead to the formation of BH-BH binaries that merge due to GWs within the age of the universe \citep{vandenheuvel17}. Earlier studies did not predict this formation path to be significant \citep[see e.g.][]{Dominik_2012, Belczynski_2016}{}. However, the detailed simulations of  \citet{Pavlovskii2017}
 showed that a stable mass transfer episode in a binary comprising an evolved donor star with radiative envelope and a BH accretor is more readily achieved than previously assumed. 
Subsequent studies, with assumptions that are in agreement with the findings of \citet{Pavlovskii2017}, have shown that this formation path can be the dominant channel within the isolated binary scenario (\citealt{KlenckiNelemans2018}, \citealt{olejak2021impact};
 \citealt{Lieke2022ApJ...931...17V};
\citealt{Gallegos-Garcia2021};
 \citealt{Bavera2021}; \citealt{marchant2021role}; \citealt{Andrews_2021}; \citealt{Briel2023}).

The properties of GW sources from the two aforementioned channels are sensitively dependent on various, highly uncertain binary evolutionary phases (e.g. mass transfer episodes).
This, in principle means that observations of merging binary black holes  \citep[e.g.][]{ligovirgo, Abbott_GWTC3} could be used to constrain massive binary physics. Unfortunately, there are currently too many uncertainties in massive stellar evolution to draw any meaningful conclusion \citep[see e.g.][]{Startrack2022ApJ...925...69B}. Nevertheless, it is still essential to understand how uncertainties of binary and stellar physics affect the observable properties of the merging binary black hole population in order to correctly interpret the observed GW data in the future.

In this paper, we perform a parameter study on the classical isolated binary channel, using a rapid population synthesis code, \textsc{SeBa} (\citealt{Portegies96}, \citealt{Toonen_2012}).
In the first part of this paper, we study the uncertainties related to the first phase of mass transfer (i.e. mass transfer episodes between two hydrogen rich stars). For this, we test different assumptions regarding the angular momentum loss mode and the fraction of mass ejected during the mass transfer phase with a non-compact accretor.
We also vary the mass transfer stability criteria of evolved stars with radiative envelopes and make different assumptions about the evolutionary stage at which giant stars develop convective envelopes  to investigate the implications of studies such as \citet{Ge2015}; \citet{Pavlovskii2017};
\citet{Ge_2020}; \citet{Klencki_2021}.
We make model variations using all possible combinations of parameter variations. This allows us to explore the interrelated effects of uncertainties.

In the second part of the paper, we investigate the effects of uncertainties in mass-loss rates of line-driven stellar winds. Both theoretical (\citealt{Krti_ka_2017}; 
\citealt{Sundqvist_2019}; 
\citealt{bjrklund2020new};  \citealt{BJ22https://doi.org/10.48550/arxiv.2203.08218}) and observational (e.g. \citealt{Fullerton2006}) studies suggest that mass loss rates of O/B stars could be overestimated by a factor of 2-3 by the prescription of \cite{Vink_2001}, which is commonly used in stellar evolutionary codes. As we will see, the impact of lowered mass loss rates on the demographics of GW sources sensitively depends on our assumptions regarding other, seemingly unrelated binary physics. Therefore, the second part of the paper shows an example of the importance of interrelated effects of uncertain parameters and it highlights the dangers of devising strategies to infer stellar physics directly from GW data without performing a full parameter study.

The impact of the uncertainties in binary physics on the isolated binary channel has been extensively studied with population synthesis approach in the recent years. However, most of such parameter studies typically concentrated on the episodes following the first phase of mass transfer. For example, the  importance of common envelope evolution was investigated by varying parameters related to the common envelope efficiency \citep[e.g.][]{VignaGomez2018, Bavera2021,Broekgaarden2022BHBHNSNS}, the binding energy of the donor star \citep[e.g.][]{O_Shaughnessy_2008, VignaGomez2018, Dominik_2012, Stevenson2015}, and by making different assumptions on whether Hertzsprung gap donors can survive the common envelope phase \citep[e.g.][]{Dominik_2012, Stevenson2015, Chruslinska2018, VignaGomez2018, Broekgaarden2021_BHNS, Broekgaarden2022BHBHNSNS}. The impact of core collapse was investigated by varying the magnitude of natal kicks received by the stellar remnant \citep[e.g.][]{O_Shaughnessy_2008, Stevenson2015, VignaGomez2018, Broekgaarden2021_BHNS,Broekgaarden2022BHBHNSNS, Ghodla2022, Richards2023}, by applying different supernova mechanisms \citep[e.g][]{Dominik_2012, Stevenson2015, VignaGomez2018, Broekgaarden2021_BHNS,Broekgaarden2022BHBHNSNS,RomanGarza2021,vanSon2022NoPeaksWithoutValleys}, and by varying the maximum neutron star mass \citep[e.g.][]{Dominik_2012, Stevenson2015,Broekgaarden2021_BHNS,Broekgaarden2022BHBHNSNS}. Uncertainties regarding the second (stable) phase of mass transfer were studied by exploring the implications of super-Eddington accretion for BH accretors \citep[e.g.][]{Belczynski_2020, Bavera2021,Briel2023}.

A few studies also considered some of those parameters, which are investigated in this paper. For example, the impact of mass transfer stability of donor stars crossing the Hertzsprung gap was studied in \citet{VignaGomez2018, olejak2021impact,vanSon2022NoPeaksWithoutValleys}. Furthermore, \citet{O_Shaughnessy_2008, Dominik_2012, Broekgaarden2021_BHNS,Broekgaarden2022BHBHNSNS, Startrack2022ApJ...925...69B,vanSon2022NoPeaksWithoutValleys} investigated the impact of different accretion efficiencies for mass transfer episodes with non-compact accretors, while \citet{Chruslinska2018, VignaGomez2018, Startrack2022ApJ...925...69B} tested different assumptions on the angular momentum loss mode during non-conservative mass transfer episodes. The importance of stellar winds on the properties of merging compact objects was also previously explored by \citet{O_Shaughnessy_2008, Dominik_2012, Stevenson_2017, Renzo2017, Belczynski_2020,  Belczynski2020_70_solar_mass, Startrack2022ApJ...925...69B, Broekgaarden2021_BHNS}. Clearly, many previous studies investigated the role of those parameters, which we also consider in this paper. However, this was done typically in a different astrophysical context (e.g. to study the formation of merging double neutron star binaries, see \citealt{VignaGomez2018,Chruslinska2018}) or to investigate the formation of specific systems see e.g. \citealt{Belczynski2020_70_solar_mass,Startrack2022ApJ...925...69B}). More importantly, all of the above mentioned studies typically vary only one parameter with respect to their fiducial models and they never study systematically the importance of the first phase of mass transfer and how the related uncertainties can impact the population of GW sources. 



The paper is organised as following. In section \ref{section:isolated_brief}, we briefly review the classical isolated binary formation channel to introduce the terminology used in this paper. In section \ref{section:methodology}, we describe the code used in this study. In section \ref{sub:popprops}, we show our results that highlight the importance of uncertainties in the first mass transfer phase. In section \ref{sub:stellarwinds}, we show how decreasing the line-driven winds for O/B stars and/or WR stars by a factor of 3 can change the properties of the merging binary black hole population. Finally, in section \ref{sub:conclusion}, we summarise our main findings.

\subsection{The classical isolated binary formation channel}
\label{section:isolated_brief}
In this subsection, we provide a brief overview of the classical isolated formation channel for merging binary black holes. Our purpose is to introduce the terminology used in the rest of the paper. For a detailed overview on this subject, see e.g. \citet[][]{Postnov_2014, m2018merging, Mapelli2020}.

Black hole binaries with circularised orbits merge within the Hubble time (which we define here as 13.5 Gyrs), if their orbital separation is not larger than a few tens of solar radii (\citealt{Peters}; \citealt{m2018merging}). However, the radii of massive stars reaches orders of magnitude larger values than that during their evolution. Therefore, GW sources of the classical isolated binary channel must originate from interacting binaries.

\begin{figure}
\hspace*{-0.5cm}
  \includegraphics[width=\columnwidth]{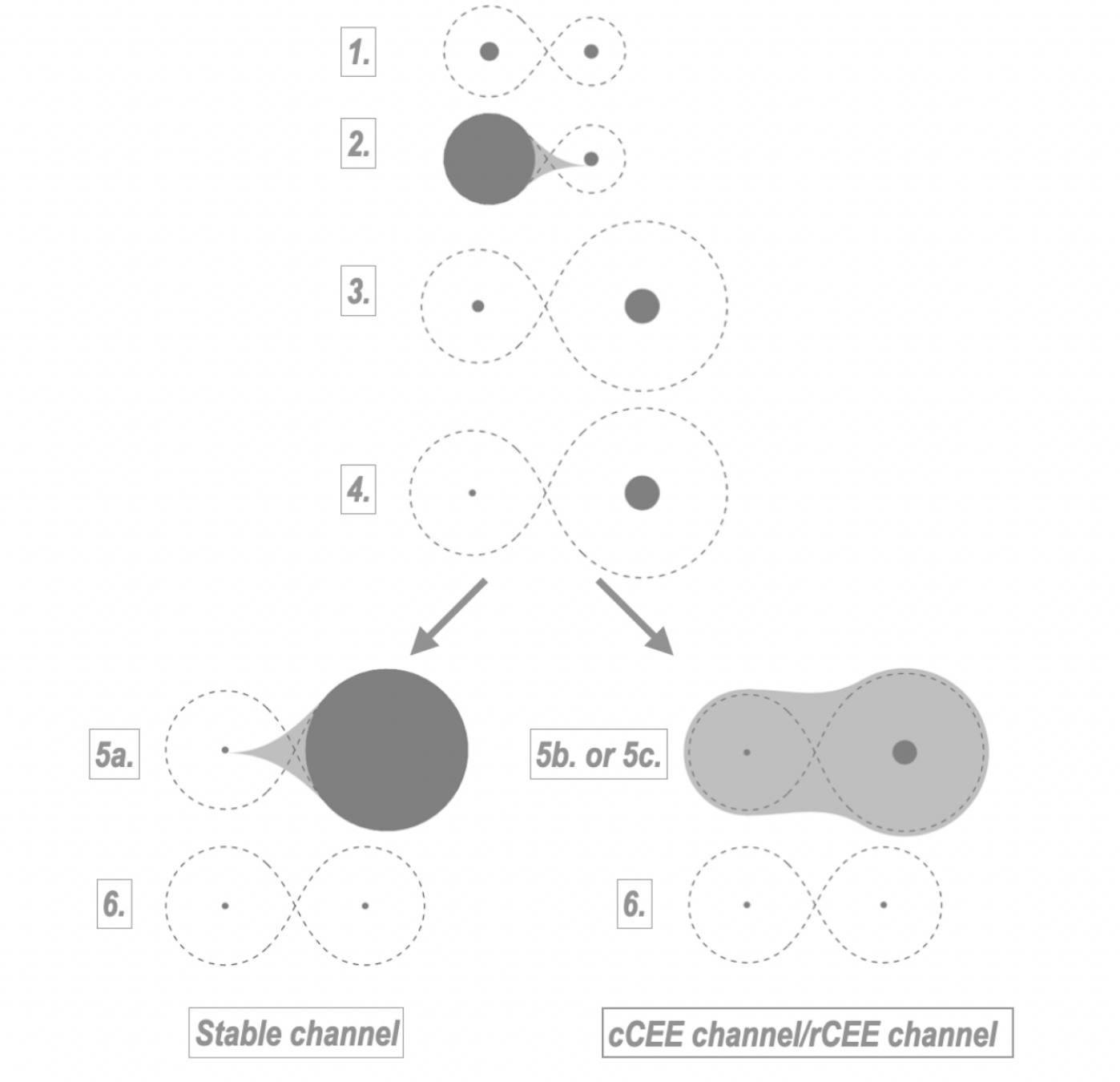}
  \caption{The most common formation channels of gravitational wave sources from isolated binaries as predicted in this paper. cCEE means that the dynamically unstable mass transfer is initiated by a giant donor, which has a deep convective envelope, whereas for rCEE the donor still has mostly a radiative envelope.}
  \label{fig:isolatedchannelsl}
\end{figure}

We show a schematic drawing of the most common formation paths of GW sources according to our simulations in Fig. \ref{fig:isolatedchannelsl}. At zero-age main sequence, the binaries are not interacting and their orbital separation widens due to stellar winds (stage 1). The initial primary star eventually fills its Roche-lobe and consequently the first phase of mass transfer is initiated (stage 2). In the dominant channels considered here, this mass transfer phase always occurs in a dynamically stable manner. 
This phase ends with the initial primary star losing its hydrogen envelope. At this stage, the donor star is a stripped helium star, and depending on the metallicity, it could launch intense stellar winds and therefore may be observed as a Wolf-Rayet star (stage 3). The mass ratio and the orbit of the binary system at this stage depend on how much matter was accreted by the secondary and how much angular momentum was lost by the binary during the first phase of mass transfer.  The stripped helium star eventually forms a black hole (stage 4). It is currently uncertain whether this occurs via a supernova or a direct collapse.

After the inital primary forms a compact object, the secondary star expands as well and initiates the second mass transfer phase (stage 5). Based on the mass ratio of the system and the envelope structure of the donor, this episode can occur in a stable (stage 5a) or unstable fashion (stage 5b or 5c). In case of the latter, a common envelope phase is initiated  (\citealt{Ivanova_2013}). As the common envelope ensuing the binary exerts friction on the system, the period is expected to dramatically decrease. If orbital energy is not used efficiently to unbind the envelope, the change in the orbital separation could be of order $\sim 1000 \,R_{\odot}$. 
This process can lead to an efficient formation of merging binary black holes (i.e. CEE channel). We distinguish two subtypes of the common envelope channel based on the evolutionary phase of the donor star during the second phase of the mass transfer. In the first type (5b), the donor star has a radiative envelope (rCEE), while in the second (5c), the donor star has a deep convective envelope (cCEE).
The mass transfer stability criteria is sensitively dependent on whether the envelope of the donor is mostly radiative or convective. We also note that the binding energy of the envelope could be significantly different for these two types of evolved stars. 

If the second mass transfer phase occurs in a stable manner, the orbit can shrink sufficiently, and thus lead to the formation of a GW source, if the mass ratios of the binary at the onset of the mass transfer phase are relatively high (stage 5a). For example, the orbit of a binary with a mass ratio $q = M_{\rm donor}/M_{\rm accretor} \sim 3$ and $M_{\rm accretor} \approx 30\,M_{\odot}$ at the onset of the mass transfer phase,  shrinks roughly by $\sim 100\,R_{\odot}$, assuming the accretion rate of the black hole is Eddington limited. This implies that binaries with initial orbital separations of a few  $\sim 100\,R_{\odot}$ can form GW sources efficiently via the stable channel. 

In principle, GW sources could also be formed from systems, in which the first phase of mass transfer is dynamically unstable. However, we do not discuss this formation scenario  in this paper. This is because the merger rates associated with this formation scenario are negligible in our models.

By the time the second phase of mass transfer occurs, the initial primary star is typically already a black hole. We note that this can occur for a very large fraction of the parameter space, especially if rejuvenation of the accretor is taken into account after the first phase of mass transfer (see e.g. \citealt{Tout1997MNRAS.291..732T}). For example, systems with an initial primary mass of $M_{\rm ZAMS,1} = 100\,M_{\odot}$ can evolve in such a way, even, if their initial mass ratio are as close to unity as $q_{\rm ZAMS} = M_{\rm ZAMS,2}/M_{\rm ZAMS,1} = 0.99$. We find that only those binaries form gravitational wave sources with non-negligble rates, in which the second phase of mass transfer occurs with a compact accretor. Therefore, in the rest of the paper, when we mention the second phase of mass transfer, we always refer to mass transfer episodes with BH accretors.

\section{\textsc{SeBa} and model variations}
\label{section:methodology}
We use the rapid population synthesis code \textsc{SeBa} for our binary simulations\footnote{ https://github.com/amusecode/seba} (\citealt{Portegies96}, \citealt{Toonen_2012}). An up-to-date description of the code can be found in \cite{Toonen_2012}. In the following sections, we only describe elements which are especially relevant for this study, or which have been changed with respect to \citet{Toonen_2012}. The most relevant parameters related to stellar and binary physics used in all of our model variations are summarised in Table \ref{tab:models}. 

\begin{table*}
\centering
\caption{ Summary of the most important parameters in our rapid population synthesis simulation with \textsc{SeBa}. The top of the table shows the standard model used in this paper, while the bottom of the table shows the model variations. We run simulations on a metallicity grid Z = 0.0001, 0.0003, 0.0005, 0.0007, 0.001, 0.003, 0.005, 0.01, 0.02 with all possible combinations of model variations unless stated otherwise in section \ref{section:methodology}. For the detailed descriptions of stellar wind models see Table \ref{tab:mdotapplied}.
}
\label{tab:models}
\begin{tabular}{@{}lll@{}}
\toprule
\multicolumn{3}{l}{\textbf{Parameters not varied in this paper}} \\ \midrule
\textit{Parameter} &
  \textit{Model/value} &
  \textit{Label} \\
 &
   &
   \\
Single stellar tracks &
  \citet{Hurley_2000} &
  - \\
Tidal interactions &
Orbits are circularised by the time of the\\  mass transfer \citep{Portegies96} &
  - \\
SN prescription &
  Delayed model of \citep{Fryer_2012} &
  - \\
Natal kick velocity distribution &
  \citet{Verbunt2017} &
  - \\
Natal kick velocity scaling for BHs &
  $v_{\rm kick,BH} = (1-f) \cdot v_{\rm kick, NS} \cdot (M_{\rm BH}/M_{\rm NS})^{-1}$ &
  - \\
Common envelope treatment &
  $\alpha$-formalism with $\alpha\lambda = 0.05$ &
  - \\
Angular momentum loss mode with BH or NS accretor &
  $\gamma = M_{d}/M_{a}$ &
  - \\
Accretion efficiency for BH and NS accretors&
  Eddington limited accretion &
  a \\
 \midrule
\multicolumn{3}{l}{\textbf{Model variations}} \\ \midrule
\textit{Parameter} &
  \textit{Model/value} &
  \textit{Label} \\
 &
   &
   
   \\
   Stellar wind model &
1.) $f_{\rm wind}$ = $f_{\rm wind,WR}$ = 1 and $f_{\rm LBV} = 1.5$ &
  Model I \\
 &
2.) $f_{\rm wind}$ = 1/3, $f_{\rm wind,WR}$ = 1 and $f_{\rm LBV} =1.5$ &
  Model II \\
 &
3.) $f_{\rm wind}$  = $f_{\rm wind,WR}$ = 1/3 and $f_{\rm LBV} = 1.5$  &
  Model III \\
 Angular momentum loss mode with non-compact accretors &
 1.) $\gamma = 2.5$ (\citealt{Portegies96}) &
  $\gamma = 2.5$ \\ &
2.) $\gamma = 1$ (\citealt{Podsiadlowski92};  \citealt{Belczynski2008}) & 
  $\gamma = 1$ \\
 Accretion efficiency of non-compact accretors &
 1.) $\beta = 0.3$ &
  $\beta = 0.3$ \\ &
2.) $\beta = 0.7$ &
  $\beta = 0.7$ \\
Mass radius exponent of giants with radiative envelopes \\ &
  1.) $\zeta_{\rm ad,rad} = 4$ &
  $\zeta_{\rm ad,rad} = 4$  \\ &
2.) $\zeta_{\rm ad,rad} = 7.5$ &
$\zeta_{\rm ad,rad} = 7.5$  \\
Boundary of deep convective envelope &
1.) Prescription of \citet{Klencki_2021} &
  $T_{\rm eff,K}$ \\ &
2.) convective above $\rm{log}_{10}T_{\rm eff} = 3.73\,K $ (\citealt{Belczynski2008}) &
  $T_{\rm eff,IT}$ \\ \bottomrule
\end{tabular}
\end{table*}

\subsection{Treatment of binary interactions}
\label{sec:masstransfers}

In this section, we summarise how binary interactions are treated in \textsc{SeBa}. 
We assume tidal interactions circularise the orbit by the onset of the mass transfer (\citealt{Portegies96}). Change in the orbital separation and eccentricity due to gravitational waves emission are calculated according to \citealt{Peters}.


A mass transfer episode occurs, if any of the stars in the binary fill their Roche-lobe. We calculate the Roche-lobe radius according to \citet{EggletonApprox}.
The evolution of the orbital separation during a stable phase of  mass transfer is determined as \citep[e.g.][]{BookCompactXray, 1994vandenheuvel}:
\begin{equation}
\label{eq:dotaa}
    \frac{\dot{a}}{a} = -2 \frac{\dot{M}_{d}}{M_{d}}\left[ 1 - \beta \frac{M_{d}}{M_{a}} - \left(\gamma+\frac{1}{2}\right)\frac{(1 - \beta)M_{d}}{M_{d} + M_{a}} \right],
\end{equation}
where $M_{d}$, $M_{a}$ are the mass of the donor and the accretor star, respectively, $\beta$ is the mass accretion efficiency, ie. the amount of mass that is accreted and $\gamma$ is the ratio of specific angular momentum that leaves the system and the total specific angular momentum of the binary, ie. $dJ/dM_{\rm tot} = \gamma J/M_{\rm tot}$, where J is the angular momentum of the binary and $M_{\rm tot}$ is the total mass of the binary.
Although, $\beta$ and $\gamma$ is expected to depend on the parameters on the binary (see e.g. references in section \ref{subsec:first_phase_of_mass_transfer}), it is commonly assumed that these parameters are constant in rapid population synthesis codes \citep[see e.g. ][]{Riley_2022, Belczynski2008}. 
If $\gamma$ is constant, Equation \ref{eq:dotaa} can be integrated:
\begin{equation}
\label{eq:stable_mt_eq}
\frac{a}{a_i} = \left(\frac{M_{d}}{M_{d,i}}\frac{M_{a}}{M_{a,i}}\right)^{-2}\left(\frac{M_{\rm tot}}{M_{\rm tot,i}}\right)^{2 \gamma +1},
\end{equation}
where the subscript 'i' stands for initial, i.e. at onset of the mass transfer phase.

If the accretor is a black hole, we assume that the accretion is Eddington limited and that the specific angular momentum leaving the system is that of the accretor, which implies $\gamma = M_{d}/M_{a}$ (i.e. so-called isotropical reemission). The change in orbital separation for isotropical reemission in the limit of $\beta \rightarrow0$  \citep[e.g.][]{soberman1997stability, vandenheuvel17}: 
\begin{equation}
    \label{eq:iso}
     \frac{a}{a_{i}} = \frac{M_{\rm tot,i}}{M_{\rm tot}}\left(\frac{M_{\rm d,i}}{M_{d}}\right)^2\exp\left(2\frac{M_{d}- M_{d,i}}{M_{a}}\right).
\end{equation}
We discuss the treatment of dynamically unstable mass transfer episodes in section \ref{subsec:common_envelope}.

When the members of the binary lose mass via stellar winds, we assume that a fraction of it is accreted via Bondi-Hoyle accretion \citep{bondihoyle}, while the rest leaves the system with a specific angular momentum of the donor, which corresponds to $\gamma = M_{a}/M_{d}$.
In this case, the orbit widens as:
\begin{equation}
\label{eq:Jeans}
    \frac{a}{a_{i}} = \frac{M_{\rm tot,i}}{M_{\rm tot}}.
\end{equation}

\subsection{First phase of mass transfer }
\label{subsec:first_phase_of_mass_transfer}

If the accretor is not a remnant, we assume that $\gamma = 2.5$, following \citet{Portegies96}. We also test $\gamma = 1$, following \citet{Podsiadlowski92} and \citet{Belczynski2008}. 
In this study, we test two, constant values for mass transfer efficiency when the accretor is a non-compact object; $\beta = 0.3$ and $\beta =0.7$. When the accretor is a neutron star or black hole, we assume that the accretion is Eddington-limited and $\gamma = M_{d}/M_{a}$.

There are currently numerous uncertainties regarding mass transfer episodes with non-degenerate accretors  \citep[e.g.][]{Langer_2012}. The fraction of the transferred mass that is eventually ejected from the binary and the specific angular momentum that is removed from the system depends on many factors. For example on whether an accretion disk is formed during the mass transfer episode \citep[e.g.][]{LubowShu1975ApJ...198..383L}, on how efficiently the accretor star is spun up due to accretion \citep[][]{Packet1981}, on whether accretion is possible above the critical rotation of the accretor star \citep[][]{PophamNarayan1991ApJ...370..604P} and on the response of the radius of the accretor star on thermal timescale \citep[see e.g.][]{Polsmarinus, Hurley2002, Riley_2022}.


Detailed binary evolution models of massive stars indicate a low mass transfer efficiency ($\beta \approx 0.1$, on average) for mass transfer phases with evolved donors, if it is assumed that accretion is not possible above critical rotation \citep[e.g.][]{Langer_2020}. If the donor star is still on the main sequence, tides can be sufficiently strong to counteract the spinning up, leading to higher mass transfer efficiencies (see e.g. \citealt{KoushikSen2022A&A...659A..98S}). These findings are broadly consistent with a few observational studies (\citealt{Petrovic}; \citealt{Shao_2016} and see also \citealt{2007deMink}). On the other hand, there are theoretical and  observational studies, which conclude near-conservative mass transfers episodes among massive stars \citep[e.g][]{Schootemeijer_2018, Vinciguerra_2020}, therefore a consensus regarding this physical process is still missing.

\subsection{Mass transfer stability criteria and treatment of mass transfer}
\label{subsec:mt_stability}
We determine the stability of mass transfer with the use of the so-called mass-radius exponents (\citealt{soberman1997stability}):
\begin{equation}
    \zeta_L = \frac{d \log R_L}{d \log M}
\qquad
    \zeta_{\rm ad} = \left(\frac{d \log R_{d}}{d \log M}\right)_{\rm ad}
\qquad
    \zeta_{\rm th} = \left(\frac{d \log R_{d}}{d \log M}\right)_{\rm th},
\label{eq:zetas}
\end{equation}
where $R_{d}$ is the radius of the donor star, $\zeta_L$ expresses how the Roche-lobe radius reacts to mass overflow, while $\zeta_{\rm ad}$ and $\zeta_{\rm th}$ expresses how the radius of the donor reacts to mass loss during mass transfer on dynamical, and thermal timescale, respectively.
Three different mass transfer modes can be distinguished: stable mass transfer on nuclear time scale ($\zeta_L \leq \rm{min}(\zeta_{\rm ad},\zeta_{\rm eq})$), stable mass transfer on thermal timescale ($\zeta_{\rm ad}\geq \zeta_L \geq \zeta_{\rm eq}$.) and unstable mass transfer ($\zeta_L > \rm{max}(\zeta_{\rm ad}, \zeta_{\rm eq}$)). In the first two cases, we assume that the mass transfer rate is $\dot{m} = M_{d}/\tau$, where $\tau$ is the nuclear timescale in the first and the thermal timescale in the second case.
If the mass transfer is dynamically unstable, we assume common envelope evolution (see section \ref{subsec:common_envelope}).

As a major simplification, we assume a constant $\zeta_{\rm ad}$ and $\zeta_{\rm th}$ for a given stellar evolutionary phase (these are summarised in Table \ref{tab:zetas}).
Giants with deep convective envelopes tend to have low $\zeta_{\rm ad}$, possibly even negative. Therefore, donor stars of this type are likely to initiate unstable phases of mass transfers. At what stage the deep convective envelope develops in massive stars is still very uncertain. It is common to use effective temperature as a proxy for the evolutionary stage at which such an envelope is developed (we will note this as $T_{\rm eff,boundary})$. We test two assumptions. First, that a deep convective envelope develops at an effective temperature $\rm{log}T_{\rm eff} = 3.73\,K$, following \citet{IvanovaTaam2004} and \citet{Belczynski2008}. In the second model variation, we follow the prescription of \citet{Klencki_2020}. This prescription gives $T_{\rm eff,boundary}$ as a function of luminosity and metallicity. The predicted values of $T_{\rm eff,boundary}$ from \citet{Klencki_2020} are typically considerably cooler than $\rm{log}(T_{\rm eff}) = 3.73\,K$.

\subsection{Common envelope evolution}
\label{subsec:common_envelope}

In this study, we model common envelope evolution by adopting the energy formalism \citep[e.g.][]{Webbink84,Heuvel76}. As the common envelope engulfs and exerts friction on the binary, the orbital separation starts to shrink. It is assumed that a fraction ($\alpha_{CE}$) of the energy liberated from the orbital energy is used to unbind the envelope.
Then the orbital separation by the end of the CEE phase can be given as:
\begin{equation}
    \frac{G M_{d} (M_{d}- M_{\rm d,core})}{\lambda R_{d}} = \alpha_{\rm CE} \left( \frac{GM_{a}M_{\rm d,core}}{2 a_f} - \frac{G M_{d} M_{a}}{2 a_i} \right),
\end{equation}
here, the left hand side term is the binding energy of the envelope. $M_{\rm d,core}$ is the mass of the helium core of the donor, $R_{d}$ is the radius of the donor, $a_i$ and $a_f$ are the initial and final orbital separation, respectively The term $\lambda$ describes the structure of the envelope \citep{Kool90}. Several studies published tabulated or fitted data for $\lambda$ for a range of masses and evolutionary stages \citep[e.g.][]{  dewi,Xu_2010, Loveridge_2011,Claeys_2014,  Kruckow2016,Klencki_2021}. The values of these calculated $\lambda$ parameters can vary over orders of magnitude, depending on the radius and the mass of the donor star and on the metallicity. \citet{Klencki_2021} and \citet{Kruckow2016} predict, however, that for sufficiently massive stars ($M_{\rm ZAMS}\gtrsim30\,M_{\odot}$), $\lambda$ varies only by a factor of a few ($\sim$2-5) as a function of stellar parameters and metallicity, once its radius has expanded sufficiently ($R\gtrsim$ 500-1000 $R_{\odot}$), given that the star still has mostly radiative envelopes (see e.g. Fig. 1 in \citealt{Kruckow2016} and Fig. C.3 in \citealt{Klencki_2021}).

In this study, we assume a constant $\lambda = 0.05$, which is in reasonable agreement with the results of \citet{Kruckow2016} for stars with $M_{\rm ZAMS} \gtrsim30\,M_{\odot}$ and $R\gtrsim500\,R_{\odot}$. We note that we assume that Hertzsprung gap donors cannot survive CEE episodes  \citep[][]{Dominik_2012}. This also implies that donor stars in GW progenitors typically have $R\gtrsim500\,R_{\odot}$ at the onset of the CEE in our simulations.
We assume $\alpha_{CE} = 1$.

\subsection{Mass transfer episode types based on the evolutionary phase of the donor}
\label{sub:typesofmt}

We distinguish the following mass transfer phase types based on the evolutionary stage of the donor star.

\textit{Case A}: the donor is a main sequence star (see e.g. \citealt{KoushikSen2022A&A...659A..98S}).
If the period of the system is sufficiently short, the secondary might also fill its Roche-lobe, leading to the formation of contact systems (\citealt{Pols1994A&A...290..119P};
\citealt{Wellstein2001A&A...369..939W};   \citealt{Menon2021MNRAS.507.5013M}).
 The outcome of Case A mass transfer phase is expected to be very different from those which start with an evolved giant donor (i.e Case B and Case C). As opposed to giants, main sequence stars do not have fully developed helium cores. During Case A mass transfer, fusion in the developing helium core is expected to halt because of the rapidly dropping central temperatures. Consequently, the mass of the naked helium star that is left after the end of Case A mass transfer phase is lower than for binaries experiencing mass transfer episodes with evolved donor stars (e.g. see \citealt{Langer_2020}). 
In populaton synthesis codes like \textsc{SeBa}, it is challenging to model Case A mass transfer episodes accurately. There are two major reasons for this. Firstly,  the stellar tracks of \cite{Hurley_2000} do not track the mass of the developing helium core on the main sequence. The core mass is only determined at the start of the Hertzsprung gap phase. Secondly, a constant mass-radius exponent is assumed for a given stellar evolutionary phase (see subsection \ref{subsec:mt_stability}). This means that the radius response of the donor during a Case A mass transfer phase is assumed to be the same, regardless of how much mass had already left the star (i.e. even, if in principle most of the hydrogen rich mass had already left the star).
The consequence of these two points is that, the amount of mass that is transferred to the accretor can be significantly overestimated, and the mass of the black hole that the donor eventually forms can be severely underestimated by codes like \textsc{SeBa}.
For a different approach in a  binary population synthesis code, see e.g. \citet{Agrawal2023} or codes that use detailed interacting binary stellar models, such as \textsc{BPASS} \citep{Eldridge2017, StanwayEldridge2018}, \textsc{Brussels code} \citep[see e.g.][]{Vanbeveren1998:TheWRandO-typestarpopulationpredictedbymassivestarevolutionarysynthesis, DeDonderVanBeveren2004:BrusselsCode}, \textsc{ComBinE} \citep[][]{Kruckow_2018} or \textsc{POSYDON} \citep{Fragos2023}. In such codes a more physically motivated modelling of Case A mass transfer episode is possible.
To conclude, the outcome of Case A mass transfer episodes predicted by codes based on the stellar tracks of \cite{Hurley_2000} should be treated with caution. Nevertheless, we still show such systems in this work for completeness. 

\textit{Case B}: the donor star is crossing the Hertzsprung gap. During this evolutionary phase of the donor star, a large and rapid increase in stellar radius occurs. 
The mass transfer phase ends with the donor star losing its hydrogen envelope, leaving a naked helium star behind \citep[but see][]{Laplace2020A&A...637A...6L}. Since the Hertzsprung gap phase lasts only for a few $\sim 10^4$ years, the helium core does not have sufficient time to significantly grow and therefore the mass of the helium star (and therefore the black hole that is eventually formed) is not strongly dependent on the exact initial separation of the binary. Furthermore, these systems are not significantly affected by LBV winds (assuming steady mass loss rates). As noted in section \ref{subsec:common_envelope}, we follow \citet{Dominik_2012} and assume that binaries with donor stars crossing the Hertzsprung gap cannot survive CEE.

\textit{Case C}: the donor star is in its core helium burning phase.
We distinguish two sub-categories: (i) Case Cr: core helium burning donor star with radiative envelope, (ii) Case Cc: core helium burning donor star with deep convective envelope. We assume that the mass transfer stability criteria is the same for Case B and for Case Cr mass transfer phases (see Table \ref{tab:zetas}). Yet, there are important differences in the predicted outcome of these two episodes, as the core helium burning phase lasts orders of magnitude longer with a slower expansion in radius. This means that for Case Cr mass transfers, the mass of the remnant that the donor star eventually forms is sensitively dependent on the initial separation, since the mass of the helium core can grow significantly during the core-helium burning phase. Moreover, the effects of LBV winds are no longer negligible. 
Donor stars with deep convective envelopes have very different envelope structures and therefore different mass transfer stability criteria. As already mentioned, unstable mass transfer phases are more readily realised for these systems (see Table \ref{tab:zetas} and \ref{tab:zeta2q}).

\begin{table*}
\caption{A summary of the mass transfer stability criteria used in this study for each stellar evolutionary phase. For HG and CHeB stars with radiative envelopes we show two values (4 and 7.5) as we test both values in our study. HW87 stands for \citet{Hjellming}}
\begin{tabular}{@{}lcccccc@{}}
\toprule
 &
  \textbf{Main sequence} &
  \textbf{Hertzsprung gap} &
  \textbf{Core helium burning phase} &
  \textbf{AGB phase} &
  \textbf{Helium star} &
  \textbf{Helium giant} \\ \midrule
$\zeta_{\rm ad}$ &
  \textit{4} &
  \textit{4 or 7.5} &
  \textit{\begin{tabular}[c]{@{}c@{}}For radiative envelope: 4 or 7.5\\ For fully convective envelope: HW87\end{tabular}} &
  \textit{HW87} &
  \textit{15} &
  \textit{HW87} \\
$\zeta_{th}$ &
  \textit{0.55} &
  \textit{-2} &
  \textit{\begin{tabular}[c]{@{}c@{}}For radiative envelope: -2\\ For fully convective envelope: 0\end{tabular}} &
  \textit{0} &
  \textit{1} &
  \textit{-2} \\ \bottomrule
\end{tabular}
\label{tab:zetas}
\end{table*}

\subsection{Supernova and natal kicks}
The mass of the remnant after core collapse is computed based on the delayed supernova model from \citet{Fryer_2012}.
This prescription determines the remnant mass as a function of CO core mass, which in \textsc{SeBa} is obtained from the fits of \citet{Hurley_2000}.
The kick velocity for black holes is calculated as :
\begin{equation}
\label{eq:kick}
    v_{\rm BH} = (1 - f_b)\left(\frac{M_{\rm NS}}{M_{\rm BH}}\right)v_{\rm kick}
\end{equation}
Where $f_b$ is the fallback, $M_{NS}$ is the canonical neutron star mass $M_{NS} = 1.4 M_{\odot}$ and $v_{\rm kick}$ is a random velocity kick drawn from the distribution inferred by \citet{Verbunt2017} from proper motion measurements of pulsars. The distribution of \citet{Verbunt2017} is a combination of two Maxwellian functions with velocity dispersions of $\sigma$ = 75$\,\rm{km/s}$, and $\sigma$ = 315$\,\rm{km/s}$, and weights of 0.42 and 0.58, respectively. 

\subsection{Stellar wind prescriptions}
\label{subsection:swp}
Massive stars lose a substantial fraction of their mass via stellar winds. 
We can roughly group stellar wind mechanisms into three groups; line-driven winds (which also includes Wolf-Rayet winds), winds of Luminous blue variables (LBVs), and dust-driven winds. Line-driven winds can be further distinguished based on whether they are optically thin (typically stellar winds of main sequence and evolved stars with hydrogen envelopes) or they are optically thick (line-driven winds of stripped helium stars, i.e. Wolf-Rayet winds). 
In the following, we briefly summarise the stellar wind prescriptions we use in this study, while Table \ref{tab:mdotapplied} shows at what evolutionary stage these prescriptions are applied. 

\begin{table*}
\caption{Summary of stellar wind prescriptions applied to different evolutionary stages of massive stars in \textsc{SeBa}. HD stands for Humphreys-Davidson limit. The scaling factor $f_{\rm wind}$ is 1 for the default model and 1/3 for the 'scaled-down' model. $\dot{M}_{\rm V01}$ stands for \citet{Vink_2001}, $\dot{M}_{\rm NJ}$ for \citet{Jager}, $\dot{M}_R$ for \citet{Reimers}, $\dot{M}_{\rm WR}$ for \citet{Sander_2020}, $\dot{M}_{\rm VW}$ for \citet{1993ApJ...413..641V} and  $\dot{M}_{\rm LBV}$ is assumed to be $1.5\cdot10^{-4}\,M_{\odot}\rm{yr^{-1}}$ following \citet{Belczynski_2010}}
\begin{tabular}{|l|l|ll|}
\hline
\textbf{Main sequence} & \textbf{Hertzsprung gap \& Core helium burning} & \textbf{AGB phase}                                     & \textbf{Helium star \& helium giant} \\ \hline
\multicolumn{1}{|c|}{\textit{if $T_{\rm eff} \leq 50 kK$:}} &
  \textit{if $T_{\rm eff} \geq 8 kK$:} &
  \multicolumn{1}{l|}{\textit{By default:}} &
  \textit{Always:} \\
\multicolumn{1}{|c|}{$f_{\rm wind}\cdot\dot{M}_{\rm V01}$} &
  \multicolumn{1}{c|}{$f_{\rm wind}\cdot\dot{M}_{\rm V01}$} &
  \multicolumn{1}{l|}{$\dot{M} = \rm{max}(\dot{M}_{\rm VW},\dot{M}_{R},\dot{M}_{\rm NJ})$} &
  \multicolumn{1}{c|}{} \\ \cline{1-3}
\textit{otherwise:}    & \textit{if $T_{\rm eff} \leq 8 kK$:}                & \multicolumn{1}{l|}{\textit{if star beyond HD limit:}} & \multicolumn{1}{c|}{}                \\
\multicolumn{1}{|c|}{$f_{\rm wind}\cdot\dot{M}_{NJ}$} &
  \multicolumn{1}{c|}{$\dot{M} = \rm{max}(\dot{M}_{NJ},\dot{M}_{R})$} &
  \multicolumn{1}{c|}{} &
  \multicolumn{1}{c|}{\textit{$f_{\rm wind,WR}\cdot\dot{M}_{\rm WR}$}} \\ \cline{1-2}
                       & \textit{if star beyond  HD limit:}              & \multicolumn{1}{c|}{$\dot{M}_{\rm LBV}$}                   &                                      \\
                       & \multicolumn{1}{c|}{$\dot{M}_{\rm LBV}$}            & \multicolumn{1}{c|}{}                                  &                                      \\ \hline
\end{tabular}
\label{tab:mdotapplied}
\end{table*}

Line driven winds of O/B stars are modelled in \textsc{SeBa} using the mass-loss rates from \citet{Vink_2001}, as long as the star is within the grid defined by \citet{Vink_2001}, otherwise, the empirical formula from \citet{Jager} is used. These mass-loss rates are applied until stars reach $T_{\rm eff} \sim 8000K$ (see Table \ref{tab:mdotapplied}).

 We note that \citet{Vink_2001} estimates the global metallicity dependence to be $\dot{M} \propto Z^{0.69}$ for O stars, by assuming a metallicity dependence of the final wind velocity to be $v_\infty \propto Z^{0.13}$, following \citet{Leitherer92}.
This is consistent with the observations as shown by \citet{Mokiem07}, although these results still depend on the findings of \citet{Leitherer92}. We note that however, \citet{Kriticka06} and \citet{bjrklund2020new} find  weaker metallicity dependence of escape velocity, where in the latter, the exponent can even be negative for high luminosity stars.
In \textsc{SeBa}, we assume $\dot{M} \propto Z^{0.85}$ (i.e. ignoring the metallicity dependence of $v_\infty$), so that our choice of modelling is consistent with other population synthesis studies (e.g. \citealt{Dominik_2012, Mapelli_2016, Giacobbo_2017, Stevenson_2017}, but see e.g. \citealt{Eldridge_2016} in which $\dot{M} \propto Z^{0.69}$ is used). We assumed that solar metallicity is Z = 0.02.

If a main-sequence star is outside of the grid defined by \citet{Vink_2001}, we apply the empirical formula of  \citet{Jager} and assume a metallicity scaling $\dot{M} \propto Z^{0.85}$. We note that this is different from the what was originally suggested by \citet{Kudritzki87}, which was $\sim Z^{0.5}$. We do this so that there is a consistent metallicity dependence for optically thin line-driven winds.

LBV stars, stars beyond the Humphreys-Davidson limit, experience very high mass loss rates, in the order of $ \dot{M}_{\rm LBV}\sim 10^{-5}$-$10^{-3} M_{\odot}\rm{yr^{-1}}$, although this is highly uncertain. Even less is known about their possible eruptions, in which huge amount of mass could be lost in a very short time (\citealt{Humphreys};
\citealt{lbvvink2012ASSL..384..221V};     \citealt{Smith}).
For the mass-loss rates of LBV stars, we follow the assumption of \citet{Belczynski_2010}, i.e. the mass loss rate is constant and has a value of $\dot{M}_{\rm LBV} = 1.5\cdot10^{-3}\,M_{\odot}\rm{yr^{-1}}$.

If the star becomes a cool giant (ie $T_{\rm eff} \leq 8000 K$), we calculate the mass loss rate according to Reimer's empirical formula \citep{Reimers}. We compare it with the mass loss from \citet{Jager} and we take the maximum value.
For the mass loss rates of thermally pulsating AGB stars, we use the prescription of \citet{1993ApJ...413..641V}.
Mass loss rates for helium stars and helium giants (Wolf-Rayet star winds) are calculated according to \citet{Sander_2020}.

In section \ref{sub:stellarwinds}, we investigate the impact of uncertainties of these mass loss rates on the demographics of interacting massive binaries and GW sources. We test three stellar wind models, which are the following: 'Model I' with our standard stellar wind model ($f_{\rm wind} = f_{\rm wind,WR} = 1$), 'Model II' with the optically thin line driven winds scaled down by a factor 3 ($f_{\rm wind} = 1/3, f_{\rm wind,WR} = 1$) and 'Model III' where besides the optically thin line driven winds, Wolf-Rayet-like winds are also scaled down ($f_{\rm wind} = f_{\rm wind,WR} = 1/3$). For the exact description of our applied stellar winds prescriptions and scaling factors, see Table \ref{tab:mdotapplied}.
With Model II, we aim to study the implications of
\citet{Krti_ka_2017}; 
\citet{Sundqvist_2019}; \citet{bjrklund2020new} in a simplified way.  These studies found that the prescription of \citet{Vink_2001} systematically overpredicts the mass loss by a factor $\sim 2$-$3$. With Model III. we  investigate the general uncertainties in the mass loss rates of stripped helium stars (see e.g. \citealt{Sander_2020}). The latter becomes especially significant for interacting binaries, as envelope loss due to a mass transfer episode can significantly increase the time spent as a Wolf-Rayet star.

\subsection{Initial conditions}
\label{subs:icamv}
Observations suggest that about half of the stars are in binaries (or in higher order, hierarchical systems) and this multiplicity fraction increases with increasing mass \citep{stellarmult1}.
In particular, \citet{Sana2012} showed
that the binary fraction reaches $f_b \simeq 0.7$ for stars in the mass range $M \simeq 15$-$60\, M_{\odot}$. 
The same observations showed that the orbital period distribution of these young, massive binaries favour short period systems:
\begin{equation}
     f_p (\log p)  \propto ( \log p )^\pi, \quad \quad \text{for $ \log p\in
[0.15, 5.5]$},
\end{equation}
here the period, $p$ is in days and  $\pi = -0.55$. Although, other studies suggested somewhat flatter distributions (e.g.  \citealt{Kobulnicky_2014} or \citealt{Dunstall} based on observations of B stars in the Tarantula Nebula in the Large Magellanic Cloud). 

According to \citet{Sana2012}, the mass ratio distribution of massive stars is nearly flat, $f_q \propto q^{-0.1}$, while the distribution of eccentricities follows $f_e \propto e^{-0.42}$. A similar mass ratio distribution is inferred by \citet{Kobulnicky_2014}, while \citet{Dunstall} finds $f_q \propto q^{-2.9}$ for B stars in the Tarantula Nebula, and \citet{DunstallVIII} finds $f_q \propto q^{-1}$ or O stars in the Tarantula Nebula.

The effects of uncertainties in initial conditions were investigated by \citet{mink2015merger}, and they found that only the changes in the initial mass function alters significantly the properties of the merging compact binary object population \citep[but see][]{Klencki_2018}.
This is an important result, as there are observational and theoretical indications that the initial mass function might not be universal  \citep[see][and references therein]{Chru_li_ska_2020}.
In our simulations, we assume these initial distributions are not correlated, although, this might not be a valid assumption \citep[see e.g.][]{Moe_2017, Klencki_2018}.

In this paper, we apply the following initial conditions for our simulations:
\begin{itemize}
    \item Initial mass function:
    we assume a universal initial mass function of the primary star of \citet{Kroupa} in a mass range of $M_{\rm ZAMS} = 20$ - $100\, M_{\odot}$: $N_{\rm IMF}(M_{\rm ZAMS}) \simeq M_{\rm ZAMS}^{-2.3}$.
    \item Initial mass ratio distribution:
    we assume a uniform mass ratio distribution between 0.1 and 1, where the mass ratio is defined as $q = M_{\rm ZAMS,2}/M_{\rm ZAMS,1}$, that is the ratio of the mass of the initially secondary and the mass of the initially primary star. A flat distribution is in a reasonable agreement with observations of \citet{Sana2012}.
    \item Initial separation distribution: we assume a flat distribution in the logarithmic space of binary separation in the interval of $1 R_{\odot}$-$10^4 R_{\odot}$; $N_{a} \simeq \rm{log}(a)$. This is equivalent to Opik's law \citep[][]{OpikLaw1924}, ie. a uniform distribution in $\rm{log} (p)$. We note that since we sample from the distribution of semimajor axis, some of our systems have sub-day periods. However, we discard any systems that fill their Roche-lobe at zero-age main sequence and we do not take them into account for calculating event rates (see section \ref{subs:rates}).
    \item Initial eccentricity distribution: the initial eccentricity assumed to follow a  thermal distribution \citep{Heggie1975}, i.e. $f_e  \propto e$.
\end{itemize}

\subsection{Simulation setup}
\label{subsec:simulations_setup}

In section \ref{sub:popprops}, we
explore the impact of uncertainties of first phase of mass transfer on the merging binary black hole population. We do this by varying four parameters in our simulations, namely:
\begin{enumerate}
    \item $\gamma$, which  expresses the specific angular momentum lost from the binary during a non-conservative mass transfer phase as a fraction of the total specific angular momentum of the binary. We vary $\gamma$ only for non-compact accretors.
    \item $\beta$, which describes the mass transfer accretion efficiency. We vary $\beta$ only for non-compact accretors.
    \item $\zeta_{\rm ad,rad}$, which  is the mass-radius exponent for giant donors with radiative envelopes. It determines the boundary between stable and unstable mass transfer episodes.
    \item $T_{\rm eff,boundary}$, which is the effective temperature at which a deep convective envelope is expected to develop.
\end{enumerate}

 The details of our model variations are summarised in Table \ref{tab:models}. Using only our standard stellar wind model, we run simulations with all possible combinations of $\gamma$, $\beta$ $\zeta_{\rm ad,rad}$ and $T_{\rm eff,boundary}$ (i.e 16 models in total, named M1..M16, see e.g. Table \ref{tab:rates_of_all}). We simulate $10^6$ binaries at each value of our metallicity grid for each model variations. Our metallicity grid is defined as $Z = [0.02, 0.01, 0.005, 0.003, 0.001, 0.0007, 0.0005, 0.0003, 0.0001]$. To make comparisons with other studies easier, we have converted $\zeta$ to the critical mass ratio for each model variation in table $\ref{tab:zeta2q}$. We calculate the merger rate density in the local universe, as summarised in section \ref{subs:rates}.

 In section \ref{sub:stellarwinds}, in which we investigate the impact of uncertainties in mass loss rates due to stellar winds,  we perform binary evolution simulations for $10^6$ systems at metallicities Z = 0.02, 0.01 and 0.005. We test three stellar wind models (e.g. Model I, II, III, see section \ref{subsection:swp}) and vary $\zeta_{\rm ad,rad}$ and $\gamma$, while assuming the convective envelope prescription of \citet{IvanovaTaam2004} and  $\beta = 0.3$.  
 
We note, however, that with the  metallicity specific star formation rate  model that we assume in this study (see equations \ref{eq:fmet} and \ref{eq:sfr}), GW progenitors formed at $Z\gtrsim0.005$ do not contribute to the merger rate density in the local universe significantly.  Consequently, in our models, different assumptions about mass loss rates of line-driven winds (which are only relevant at $Z\gtrsim0.005$) also do not affect the demographics of GW sources at $z\sim0$ significantly.
However, whether merging binary black holes with masses $M_{\rm BH}\sim30\,M_{\odot}$ could be formed in an environment that is typical for the LMC and the SMC or even in the Milky Way (i.e. Z$\sim$0.02-0.005) remains an important and open question \citep[see e.g.][]{Srinivasan2023}. Furthermore, the models for metallicity-specific star formation rates, as well as the metallicity dependence of stellar winds are highly uncertain \citep[see e.g.][]{ Chruslinska2022arXiv220610622C}.


\begin{table*}
\caption{A summary of our predicted rates for all of our model variation with our standard stellar wind models (i.e. Model I)}

\begin{tabular}{@{}llcccc@{}}
\toprule
     Model name
    & Model description                                                            & \multicolumn{1}{l}{$R_{\rm stable}$ {[}$\rm{Gpc^{-3}yr^{-1}}]$} & \multicolumn{1}{l}{$R_{\rm RCEE}$ {[}$\rm{Gpc^{-3}yr^{-1}}]$} & \multicolumn{1}{l}{$R_{\rm CCEE}$ {[}$\rm{Gpc^{-3}yr^{-1}}]$} & \multicolumn{1}{l}{$R_{\rm total}$  {[}$\rm{Gpc^{-3}yr^{-1}}${]}} \\ \midrule
M1  & $\gamma$ = 2.5, $\beta$ = 0.3, $\zeta_{\rm ad,rad}$ = 4,  $T_{\rm eff}$-K & 13.8                                                            & 5.9                                                           & 3.7                                                           & 23.4                                                              \\
M2  & $\gamma$ = 2.5, $\beta$ = 0.7, $\zeta_{\rm ad,rad}$ = 4, $T_{\rm eff}$-K  & 10.2                                                            & 8.7                                                           & 7.5                                                           & 26.4                                                              \\
M3  & $\gamma$ = 2.5, $\beta$ = 0.3, $\zeta_{\rm ad,rad}$ = 4, $T_{\rm eff}$-IT & 13.4                                                            & 0.7                                                           & 5.0                                                           & 19.1                                                              \\
M4  & $\gamma$ = 2.5, $\beta$ = 0.7, $\zeta_{\rm ad,rad}$ = 4, $T_{\rm eff}$-IT  & 9.1                                                             & 0.9                                                           & 11.0                                                          & 21.0                                                              \\
M5  & $\gamma$ = 2.5, $\beta$ = 0.3, $\zeta_{\rm ad,rad}$ = 7.5, $T_{\rm eff}$-K & 60                                                              & 3.3                                                           & 4.0                                                           & 67.3                                                              \\
M6  & $\gamma$ = 2.5, $\beta$ = 0.7, $\zeta_{\rm ad,rad}$ = 7.5, $T_{\rm eff}$-K & 88.9                                                            & 4.2                                                           & 7.8                                                           & 100.9                                                             \\
M7  & $\gamma$ = 2.5, $\beta$ = 0.3, $\zeta_{\rm ad,rad}$ = 7.5, $T_{\rm eff}$-IT  & 53.9                                                            & 0.6                                                           & 5.0                                                           & 59.5                                                              \\
M8  & $\gamma$ = 2.5, $\beta$ = 0.7, $\zeta_{\rm ad,rad}$ = 7.5, $T_{\rm eff}$-IT  & 77.4                                                            & 0.3                                                           & 10.6                                                          & 88.3                                                              \\
M9  & $\gamma$ = 1, $\beta$ = 0.3, $\zeta_{\rm ad,rad}$ = 4, $T_{\rm eff}$-K     & 2.2                                                             & 11.8                                                          & 19.4                                                          & 33.4                                                              \\
M10 & $\gamma$ = 1, $\beta$ = 0.7, $\zeta_{\rm ad,rad}$ = 4, $T_{\rm eff}$-K     & 2.4                                                             & 12.8                                                          & 15.0                                                          & 30.2                                                              \\
M11 & $\gamma$ = 1, $\beta$ = 0.3, $\zeta_{\rm ad,rad}$ = 4, $T_{\rm eff}$-IT    & 2.1                                                             & 3.5                                                           & 42.2                                                          & 47.8                                                              \\
M12 & $\gamma$ = 1, $\beta$ = 0.7, $\zeta_{\rm ad,rad}$ = 4, $T_{\rm eff}$-IT    & 2.4                                                             & 1.4                                                           & 24.2                                                          & 28.0                                                              \\
M13 & $\gamma$ = 1, $\beta$ = 0.3, $\zeta_{\rm ad,rad}$ = 7.5, $T_{\rm eff}$-K   & 43.9                                                            & 7.4                                                           & 19.0                                                          & 70.3                                                              \\
M14 & $\gamma$ = 1, $\beta$ = 0.7, $\zeta_{\rm ad,rad}$ = 7.5, $T_{\rm eff}$-K   & 73.3                                                            & 6.0                                                           & 13.8                                                          & 93.1                                                              \\
M15 & $\gamma$ = 1, $\beta$ = 0.3, $\zeta_{\rm ad,rad}$ = 7.5, $T_{\rm eff}$-IT  & 45.8                                                            & 3.1                                                           & 40.7                                                          & 89.6                                                              \\
M16 & $\gamma$ = 1, $\beta$ = 0.7, $\zeta_{\rm ad,rad}$ = 7.5, $T_{\rm eff}$-IT  & 76.0                                                            & 0.6                                                           & 25.1                                                          & 101.7                                                             \\ \bottomrule
\end{tabular}
\label{tab:rates_of_all}
\end{table*}

\begin{figure*}
  \includegraphics[width=1.0\textwidth]{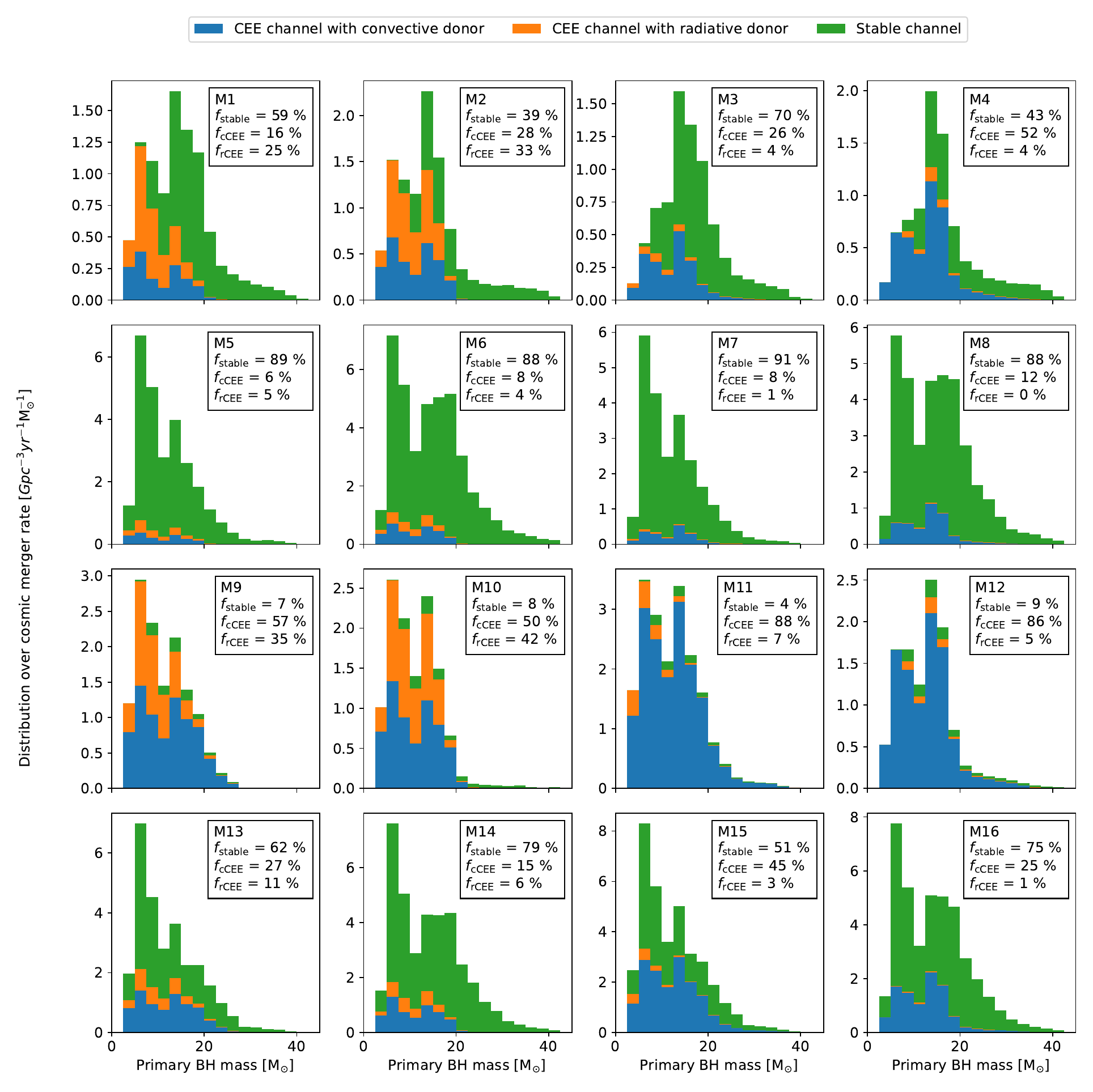}
  \caption{The mass distribution of the primary of the merging binary black hole in the local universe (z = 0) for all of our models with our standard stellar wind models (e.g. $f_{\rm wind} = f_{\rm wind,WR} = 1$, see Table \ref{tab:models}). The distribution is shown by stacked histograms, where each colour indicates a different formation channel. In the legend on the upper right corner in each panel, we show the percentage of each channel. In Fig. \ref{fig:compi}, we provide a comparison between these distributions and the primary mass distribution of BH-BH binaries as inferred from the the third
LIGO–Virgo Gravitational-Wave Transient Catalog (GWTC-3) \citep[][]{Abbott2023_GWTC3_POP}}
  \label{fig:43_hist}
\end{figure*}

\begin{figure*}
  \includegraphics[scale = 0.5]{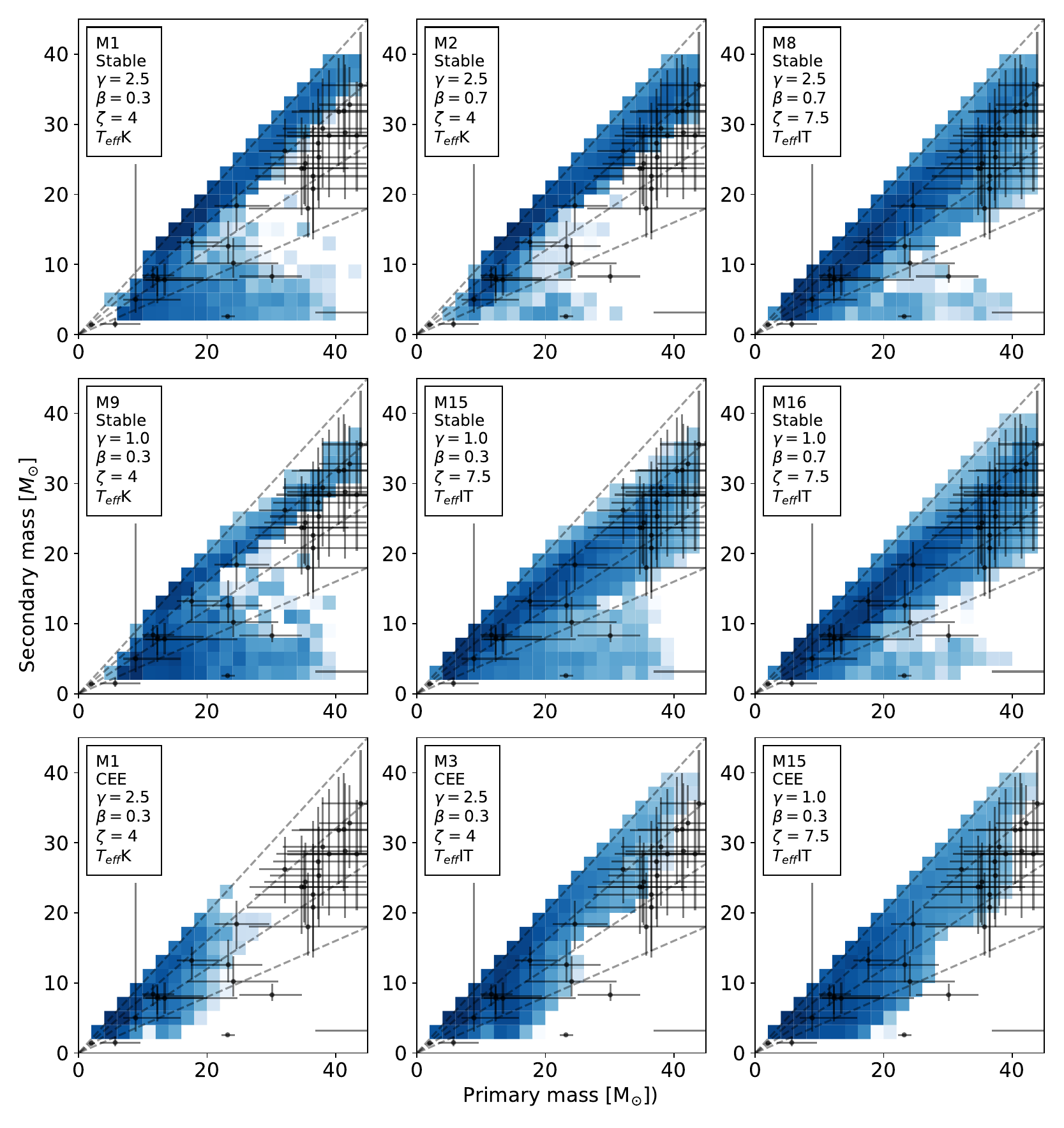}
 \caption{2D Histograms that show the masses of the merging binary black holes for a few selected models at Z = 0.0007. Each histogram has been normalised to the merger efficiency of their own model variation. The dashed lines show constant mass ratios. We also show the observed GW detections from the third
LIGO–Virgo Gravitational-Wave Transient Catalog (GWTC-3) and associated measurement uncertainties \citep[][]{Abbott2023_GWTC3_POP}. We do not show those detections, where any of the inferred merging black hole mass is above $M_{\rm BH} \geq 45\,M_{\odot}$.}
  \label{fig:43_scatter}
\end{figure*}

\begin{figure*}
  \includegraphics[width=\textwidth]{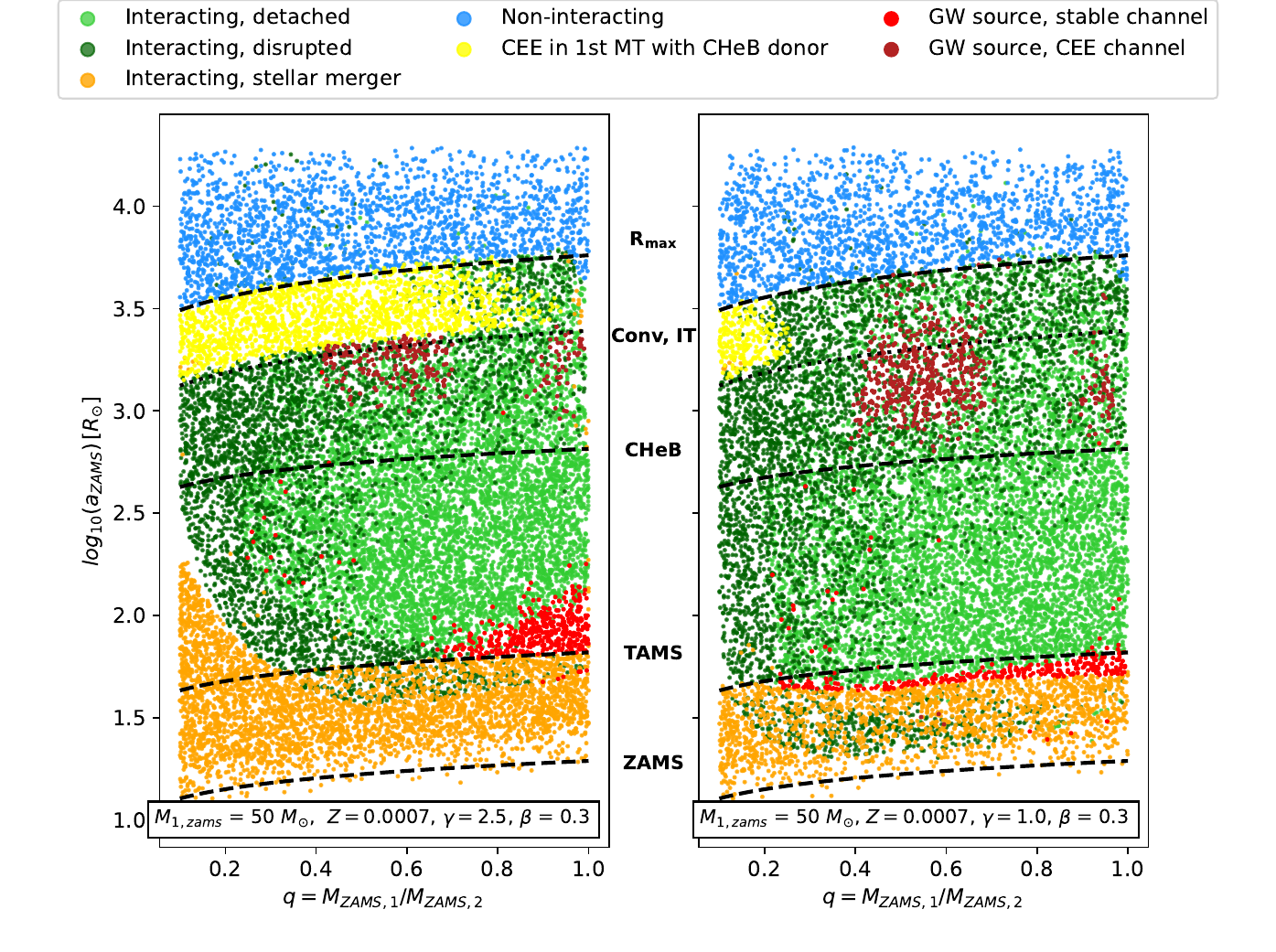}
  \caption{Our simulated binaries in the initial mass ratio - initial separation space. We show the orbital separation after tidal circularisation. The systems are simulated with the following parameters; $\beta$ = 0.3, $\zeta_{\rm ad,rad}$ = 7.5. Furthermore, we have used the assumption of \citet{IvanovaTaam2004} regarding the development of deep convective envelopes. In the left panel we show systems with $\gamma = 2.5$, in the right with $\gamma = 1$.  We show the outcome of each simulated binary with difference colours. We distinguish the following types; (i) interacting, detached: these binaries interact via mass transfer phases at least once and form detached BH-BH binaries which do not merge due to gravitational waves within Hubble time, (ii) interacting, disrupted: these binaries interact at least once and get disrupted due to the supernova kick of one of the binary components, (iii) interacting, stellar merger: these systems interact at least once and merge due to one of their mass transfer phases. Their first mass transfer phase is typically stable or unstable with a MS or HG donor, (iv) non-interacting: none of the stars fill their Roche-lobes, they either stay detached or get disrupted, (v) CEE in 1st MT with CHeB donor: these binaries have a core-helium burning donor at the onset of the first mass transfer phase, and they initiate a dynamically unstable mass transfer (vi) GW source, stable channel: gravitational wave progenitors evolving via two subsequent stable mass transfer phases, (vii) GW source, CEE channel: gravitational wave progenitors evolving via a stable and an unstable mass transfer phase.}
  \label{fig:initial}
\end{figure*}

\section{Results: the impact of uncertainties in stable mass transfer} 
\label{sub:popprops}

In Fig. \ref{fig:43_hist}, we show the primary mass distribution of merging binary black holes in the local universe (z = 0) for the three dominant channels: stable, rCEE, cCEE for all of our model variations. 
In Table \ref{tab:rates_of_all}, we show the corresponding predicted cosmic merger rate densities. 


The total predicted merger rate densities ($R_{\rm total}$) are in a broad agreement with the currently inferred rate from LIGO and Virgo observations \citep[which is $R_{\rm GWTC3} = 28.3^{+13.9}_{-9.1}\rm{Gpc^{-3}yr^{-1}}$][]{Abbott_o3a_2021}. 
$R_{\rm total}$ of model variations with $\zeta_{\rm ad,rad} = 4$ are within a factor of two of this inferred value, while $R_{\rm total}$ of model variations with $\zeta_{\rm ad,rad} = 7.5$ are larger than the observed rate by a factor of 3-4. In Fig. \ref{fig:compi} and section \ref{appendix:additional_figures}, we provide a comparison between the primary mass distribution of our models and inferred distribution from GWTC-3 \citet{Abbott2023_GWTC3_POP}.

Fig. \ref{fig:43_hist} clearly demonstrates that the relative rate of each formation channel varies significantly with different model variations. Parameters $\gamma$ and $\zeta_{\rm ad, rad}$ have the largest impact. The stable channel dominates in 11 out of 16 model variations, and the cCEE channel dominates in the remaining 5. The rCEE channel is non-negligible only in 4 model variations. In 12 model variations, the vast majority of the most massive systems originate from the stable channel \citep[in agreement with][]{Lieke2022ApJ...931...17V, Briel2023}.

While GW sources form most efficiently via the CEE channel in models M1-M3 (see e.g. $\zeta_{\rm ad,rad} = 4$ and $\gamma = 2.5$ models in Fig. \ref{fig:merger_efficiency}), $R_{\rm total}$ is still dominated by the stable channel in these model variations.
This is due to the relatively long formation times associated with the stable channel and  the monotonic increase of the cosmic star formation rate up to $z\sim2$ (see e.g. \citealt{Madau_2014}). Most of the sources of the stable channel are formed at higher redshifts than the sources of the CEE channel, and at these higher redshifts the star formation rate is also higher, which leads to an increased merger rate. \citep[see similar conclusions in ][]{Neijssel_2019, Lieke2022ApJ...931...17V, Briel2022}.



Fig. \ref{fig:43_scatter} shows the predicted mass ratios distribution of GW sources and its dependence on primary BH mass. 
For relatively massive systems ($M_{\rm BH} \gtrsim 20\,M_{\odot}$), this distribution is dependent on the assumptions on the uncertain binary physics. The formation channels can yield a population of merging binary black holes with a relatively narrow range in mass ratios ($0.8 \lesssim q \lesssim 1.0$) or a moderately wide mass ratio distribution  ($0.5 \lesssim q \lesssim 1.0$).  The mass ratio distributions of the most massive GW sources of the stable channel is most sensitive to the assumed mass transfer stability parameter. In models with $\zeta_{\rm ad,rad} = 4$, the typical mass ratios are between $0.8 \lesssim q \lesssim 1.0$, while in models with $\zeta_{\rm ad,rad} = 7.5$, the mass ratio distribution becomes much broader, i.e. $0.5 \lesssim q \lesssim 1.0$. We notice smaller variations in the mass ratio distribution for the CEE channel with different assumptions in binary physics, though model variations with $\gamma = 1$ produce somewhat broader mass ratio distributions than models with $\gamma = 2.5$. The less massive systems generally have much wider mass ratio distributions in all model variations.

Overall, we do not notice significant variations in the merger rate density, BH mass range, or the shape of the mass distributions across different model variations.
Our results, therefore, show that while the main GW observables do not depend sensitively on the uncertainties studied here, the relative importance of dominant channels do.
In other words, the formation paths of the majority of GW progenitors can be entirely different depending on the assumptions on how the first phase of mass transfer proceeds and yet the predicted demographics of merging binary black holes are very 
similar \citep[though we neglect spins in this study, see e.g. ][]{Bavera2021}. This highlights why it is extremely challenging to infer physics of massive binary evolution solely from GW observations, given the huge uncertainties in the current models.

Next, we briefly summarise the most important ways how uncertainties related to the first phase of mass transfer can affect the relative importance of the dominant GW formation channels, while in sections \ref{subsubsec:effect_of_gamma_stable} - \ref{subsubsec:effect_of_Teff}, we discuss these effects in detail.
\begin{enumerate}
    \item The assumed angular momentum loss mode has a strong impact on which formation channel dominates. Typically, the stable channel dominates in models with $\gamma = 2.5$ (see e.g. first row in Fig. \ref{fig:43_hist}), while the cCEE channel  dominates, if $\gamma = 1$, however the latter depends on the assumed $\zeta_{\rm ad, rad}$ as well (compare e.g. third and fourth row in Fig. \ref{fig:43_hist}). 
    \item An increased $\zeta_{\rm ad,rad}$ leads to a higher merger rate of the stable channel. However, this increase is only significant for relatively lower mass black holes ($M_{\rm BH,1} \lesssim 20\,M_{\odot}$). With lower mass transfer efficiencies (e.g. $\beta = 0.3$), we find that the merger rate of binary black holes with $M_{\rm BH,1} \gtrsim 20\,M_{\odot}$ is not affected at all, unless $\gamma$ = 1.0.
    \item The merger rate of the rCEE channel is only non-negligible with the convective envelope prescription of \citet{Klencki_2020}.
    Furthermore, the GW sources of this channel comprise of relatively lower mass BHs (i.e $M_{\rm BH,1} \lesssim 20\,M_{\odot}$) in all of our model variations (see section \ref{subsubsec:effect_of_Teff}).
\end{enumerate}

\subsection{The impact of angular momentum loss mode on the stable channel}
\label{subsubsec:effect_of_gamma_stable}
As evident from Fig. \ref{fig:43_hist}, the merger rate density of the stable channel ($R_{\rm stable}$) depends sensitively on the assumed angular momentum loss mode during the first phase of mass transfer (e.g. compare first with third row).
In our $\gamma = 2.5$ models, this formation channel is particularly efficient and is typcially the dominant out of the three main formation paths considered here. On the other hand, in the $\gamma = 1$ models, the merger rate of this channel is negligible, unless the mass-loss exponent is increased, i.e. $\zeta_{\rm ad, rad} = 7.5$. 

In order to understand the reason for the relation between $\gamma$ and $R_{\rm stable}$, we need to consider the following three important features of this formation channel:
\begin{enumerate}
    \item \textit{The progenitors of the stable channel have relatively short initial orbital separations.} For these systems, the largest values of $a_{\rm ZAMS}$ is a few hundred solar radii (shown in Fig. \ref{fig:initial}). If $a_{\rm ZAMS}$ is much larger than that, the orbit does not shrink sufficiently by two phases of stable mass transfer, such that a merging binary black hole would be formed.
    \item \textit{There is a minimum initial orbital separation for the binaries of the stable channel.}  We identify a minimum $a_{\rm ZAMS}$ associated with this channel,
    below which binaries typically do not form GW sources.
    This minimum $a_{\rm ZAMS}$ roughly coincides with the separation at which the initial primary would fill its Roche-lobe just when it evolved off MS (see Fig. \ref{fig:initial}). If $a_{\rm ZAMS}$ is below this minimum value, the first phase of mass transfer  (Case A) leads to a stellar merger eventually in the vast majority of cases (shown in Fig. \ref{fig:initial}).
    As explained in detail in section \ref{subsec:CaseA}, even if the initial orbit is wide enough for the system to survive the Case A mass transfer phase, the binary still likely merges in the subsequent mass transfer event.
     In this case, the first phase of Case A mass transfer leads to a less massive black hole  from the initial primary (because the formation of the helium core is halted)
     and to a more massive secondary (because more mass is transferred to the accretor) 
     with respect to systems experiencing a mass transfer episodes at later stages.
   This results in a relatively high mass ratio at the onset of the second phase of mass transfer ($q_{\rm MT,2}$) and such systems merge as a result of a dynamically unstable second phase of mass transfer (i.e. typically $q_{\rm MT,2} \gtrsim q_{\rm crit}$ with $\zeta_{\rm ad, rad} =  4$). To conclude, systems with a first phase of Case A mass transfer typically merge before forming BH-BH binaries \citep[see also][]{Gallegos-Garcia2022} and therefore, there exists a minimum initial separation associated with the stable channel. .
   \item \textit{The lower the angular momentum loss is, the wider the orbit becomes during the mass transfer phase.} This means that the orbital separation of binaries  with $\gamma = 1$ typically widens more during the mass transfer phase than with $\gamma = 2.5$. The degree by which the orbit changes due to a first phase of (stable) mass transfer is primarily determined by the initial mass ratio ($q_{\rm ZAMS}$), $\gamma$ and $\beta$ (see right panel of Fig. \ref{fig:qsandas}). Systems that  form GW sources via the stable channel typically have initially near equal masses (0.7$\lesssim q_{\rm ZAMS}\lesssim$1.0, but see e.g. \citealt{Gallegos-Garcia2021}). This is because such binaries develop sufficiently large mass ratios by the onset of the second phase of mass transfer (see left panel of Fig. \ref{fig:qsandas}), and therefore experience efficient orbital shrinking during a stable phase of mass transfer with a black hole accretor (see equation \ref{eq:iso}). For these initial mass ratio ranges, the orbit of the binary significantly widens in the $\gamma = 1$ models, while the net change in the orbital separation is very small in the $\gamma = 2.5$ models (see right panel of of Fig. \ref{fig:qsandas}). We note that, if $q_{\rm ZAMS}$ is very small, the orbit shrinks due to the first phase of mass transfer, even with $\gamma = 1$, but in that case $q_{\rm MT,2}$ will be small and therefore the orbit will widen due to the second (stable) phase of mass transfer and no GW source will be formed (compare left and right panels of Fig. \ref{fig:qsandas}). 
\end{enumerate}

Considering these three points, it is possible to understand why the stable channel is inefficient with $\gamma = 1$. The orbit widens too much due to the first phase of mass transfer, even for those binaries that start out with the minimum $a_{\rm ZAMS}$ associated with this channel. Therefore, the binary black holes that eventually form this way are too wide to merge due to GWs within the Hubble time.

  As shown in Fig. \ref{fig:43_hist}, stable channel can be efficient with $\gamma = 1$, if $\zeta_{\rm ad,rad} = 7.5$. In this case, the significant orbital widening due to the first mass transfer phase can be counteracted by the effect of the second phase of the mass transfer for sources with relatively large mass ratios at the onset of the second mass transfer ($3.2 \lesssim q_{\rm MT,2} \lesssim 5.5$, see equation \ref{eq:iso}). Such sources can form, if the first mass transfer phase is Case B and the mass transfer efficiency is relatively large (e.g. $\beta = 0.7$) or if the first mass transfer is (very late) Case A, which typically leads to large values of $q_{\rm MT,2}$. 

The relationship between $R_{\rm stable}$ and $\gamma$ is therefore dependent on the predicted outcome of Case A mass transfer episodes. We emphasise again our point in section \ref{sec:masstransfers}, that the treatement of Case A mass transfer in stellar evolutionary codes based on \citet[][]{Hurley_2000}, is extremely simplified and its predictions should be treated with caution. While we should expect that the prediction of large values $q_{\rm MT,2}$ following a Case A mass transfer phase is qualitatively true (and theferore a minimum $a_{\rm ZAMS}$ could indeed exist for the stable channel), whether indeed the vast majority of them would be above $q_{\rm crit}$ should be further investigated.

\subsection{The impact of angular momentum loss mode on the CEE channel}
\label{subsubsec:effect_of_gamma_cee}


Fig. \ref{fig:43_hist} shows that the efficiency of the cCEE channel is also sensitively dependent on the assumed $\gamma$, although in an opposite way as for the stable channel. 
$R_{\rm cCEE}$ is about a factor of 8 larger with $\gamma = 1$ than with $\gamma = 2.5$ in our low mass transfer efficiency models. With $\beta = 0.7$, this difference is about a factor of two.

Below we explain the reason for this relationship.
The binaries of the cCEE channel have $a_{\rm ZAMS}$ of a few thousand solar radii. Only in this case, the binaries are sufficiently wide by the onset of the second phase of mass transfer, such that the donor star (i.e. the initial secondary star) fills its Roche-lobe with a deep convective envelope. 

In general, binaries have wider orbital separations at the onset of the second phase of mass transfer ($a_{\rm MT,2}$) with $\gamma = 1$ than with $\gamma = 2.5$. 
Consequently, there are significantly more systems in the latter case for which the second phase of mass transfer is 
 Case Cc (compare the top panels of Fig. \ref{fig:mps}). This also leads to a higher $R_{\rm cCEE}$, since in this channel the second phase of mass transfer is by definition Case Cc (see section \ref{section:isolated_brief}). There are two reasons for this:
\begin{enumerate}
    \item \textit{The rate of unstable first phase of mass transfers of the widest interacting binaries sensitively depends on angular momentum loss.} 
    The binaries with the longest periods that still exchange mass engage in Case Cc first phase of mass transfer (see Fig. \ref{fig:initial}).
    As shown in  Fig. \ref{fig:initial}, in our low mass transfer efficiency model, the majority ($\sim$ 80 per cent) of Case Cc episodes occur in an dynamically unstable way, if $\gamma = 2.5$. For these binaries, the orbital separation drastically decreases due to the first phase of unstable mass transfer and these binaries typically do not form GW sources. On the other hand, the majority of the same mass transfer episodes are stable with $\gamma = 1$, and as a result, the orbit typically widens for these systems. This leads to larger values of $a_{\rm MT,2}$ and consequently a higher rate of Case Cc second phase of mass transfers compared to the models with $\gamma = 2.5$. 
    Therefore, the critical mass ratio associated with a first phase of Case Cc mass transfer is sensitively dependent on the assumed $\gamma$ and $\beta$, (see Fig. \ref{fig:initial} and Table \ref{tab:zeta2q}). 
    This can be understood by considering that with larger angular momentum loss, the orbit shrinks at a faster rate at the beginning of the mass transfer phase and therefore a dynamically unstable mass transfer phase is more easily instigated. 
    \item \textit{The periods of binaries engaging in a first phase of Case Cr mass transfer tend to increase more with lower angular momentum loss.} 
    As a result, the number of binaries that have large $a_{\rm MT,2}$ is higher in the models with $\gamma = 1$ with respect to models with $\gamma = 2.5$, and consequently, so is the rate of a second phase Case Cc mass transfer, since in wider binaries, the donor star fills its Roche-lobe at a later evolutionary stage and therefore it is more likely that this occurs when the star has already developed a deep convective envelope.
\end{enumerate}

    \subsection{The impact of mass transfer stability parameter}
    \label{subsubsec:effect_of_zeta}
    Fig. \ref{fig:43_hist} shows that $R_{\rm stable}$ increases and $R_{\rm rCEE}$ decreases with increasing $\zeta_{\rm ad,rad}$ (compare row 1 and 2, or row 3 and 4 of Fig. \ref{fig:43_hist}). This effect is not surprising; larger $\zeta_{\rm ad,rad}$ translates to larger $q_{\rm crit}$, which implies a larger parameter space for stable mass transfer episodes, in case the donor star is evolved and has a radiative envelope. At the same time, the degree by which the orbital separation shrinks due to a stable phase of mass transfer significantly increases with increasing $q_{\rm MT,2}$ (see e.g equation \ref{eq:iso}), which results in an efficient formation of GW sources via the stable channel. For example, with $q_{\rm MT,2} = 3.2$ (i.e. $q_{\rm crit}$ at $\zeta_{\rm ad,rad} = 4$), the orbit shrinks typically by $\sim$ 200 $R_{\odot}$ due to a stable phase of mass transfer with a BH accretor, while the same with $q_{\rm MT,2} = 5.5$ (i.e. $q_{\rm crit}$ at $\zeta_{\rm ad,rad} = 7.5$) is $\sim$ 1000 $R_{\odot}$. This means that the orbital shrinking due to a stable mass transfer with mass ratios $q_{\rm MT,2}\gtrsim5$ can be as efficient as due to common envelope evolution in our models.

    However, Fig. \ref{fig:43_hist} also shows that only the merger rate of relatively lower mass merging BH-BH binaries (i.e. $M_{\rm BH,1}\lesssim 20\,M_{\odot}$) is significantly affected by an increased $\zeta_{\rm ad, rad}$ (and $\gamma = 2.5$). In particular, when we increase $\zeta_{\rm ad,rad}$ from 4 to 7.5, the merger rates of systems with $M_{\rm BH, 1} \gtrsim20 \, M_{\odot}$ remains practically unchanged in our low mass transfer efficiency model, and increases only by a factor of 3 in our high mass transfer efficiency models (while the $R_{\rm stable}$ for the entire mass range increases almost by a factor of 9).
    The reason for this can be understood by inspecting Fig. \ref{fig:qmax}. At low metallicities, where we expect the vast majority of the GW progenitors to originate from, the maximum $q_{\rm MT,2}$ that binaries engaging in a first phase of Case B mass transfer develop significantly decreases with increasing $M_{\rm ZAMS, 1}$. For example, at Z = 0.0007 and for binaries with $M_{\rm ZAMS,1}\gtrsim60\,M_{\odot}$, this maximum of $q_{\rm MT,2}$ is about 3 and 3.5 for $\beta = 0.3$ and $\beta = 0.7$, respectively. Consequently, increasing $\zeta_{\rm ad,rad}$ from 4 to 7.5 (corresponding to increasing $q_{\rm crit}$ from 3.2 to 5) does not have a significant effect on $R_{\rm stable}$ for the most massive GW progenitors. An exception to this can bee seen in the model variations with $\gamma = 1$. In these models, the most massive GW sources of the stable channel have (a very late) Case A first phase of mass transfer, and develop $q_{\rm MT,2}\gtrsim$3.2. In this case, increasing $\zeta_{\rm ad,rad}$ affects the merger rate of the most massive binary black holes too (see last two panels in the 4th row of Fig. \ref{fig:43_hist})


    \subsection{The impact of different convective envelope prescriptions}
    \label{subsubsec:effect_of_Teff}

     In our models with the convective envelope prescription of \citet{Klencki_2020},
    the maximum primary mass of the BH-BH binaries of the cCEE channel is about $M_{\rm BH,1}\approx 20\,M_{\odot}$ (see e.g. first two columns of Fig. \ref{fig:43_hist}).
    This is about a factor of two lower than with the prescription of \citet{IvanovaTaam2004}.
    This is because the models of \citet{Klencki_2020} predict that most massive stars (i.e. $M_{\rm ZAMS}\gtrsim$ 50$\,M_{\odot}$) either never develop deep convective envelopes or they do at a late evolutionary stage, which is not followed by a significant expansion of stellar radius. This implies that the rate of Case Cc mass transfer episodes with donor stars with $M_{\rm ZAMS}\gtrsim 50\,M_{\odot}$ is negligible and therefore so is $R_{\rm cCEE}$ for $M_{\rm BH,1}\gtrsim20\,M_{\odot}$.

    The merger rate of rCEE is negligible with the prescription of \citet{IvanovaTaam2004}. In these models, stars develop a deep convective envelope soon after the onset of core-helium burning. Therefore, the occurrence rate of a second phase of unstable Case Cr mass transfer is very low. This is no longer true for the models with the prescription of \citet{Klencki_2020}, in which the rCEE channel can account for up to 42 per cent of all mergers (see M10 in Fig. \ref{fig:43_hist}).
     However, the primary BH mass is always $M_{\rm BH,1}\lesssim 20\,M_{\odot}$ for this channel. This can be understood again by inspecting Fig. \ref{fig:qmax}.
The maximum value of $q_{\rm MT,2}$ is either below or slightly above $q_{\rm crit}$ for the most massive systems experiencing a Case Cr mass transfer phase (depending on the assumed mass transfer efficiency). Therefore, very few binaries with $M_{\rm 1,ZAMS}\gtrsim 50\,M_{\odot}$ initiate an unstable Case Cr phase.
 We can also see that the values of $q_{\rm MT,2}$ are considerably lower for Case C than for Case B mass transfer episodes. This is due to the strong LBV winds that decrease the mass ratios of the binaries over time. Therefore, metallicity independent LBV winds also contribute to the low $R_{\rm rCEE}$ in our models.
    


\subsection{Comparison to earlier studies}
\label{subsec:comparison_to_recent_studies}


Relatively early rapid population synthesis studies predicted that merging binary black holes overwhelmingly originate from the CEE channel \citep[see e.g.][]{Dominik_2012, Belczynski_2016, Stevenson_2017}, while the contribution from the stable channel is negligible. On the other hand, \cite{Neijssel_2019} found that the stable channel is the dominant source of merging binary black holes. Qualitatively similar results were found by several subsequent studies \citep[i.e.][]{Gallegos-Garcia2021, olejak2021impact, Lieke2022ApJ...931...17V}. A common interpretation for this difference is that in the latter studies, a significantly higher $\zeta_{\rm ad,rad}$ is assumed (or in case of detailed binary models, computed). For example, \citet{Neijssel_2019} assumes  $\zeta_{\rm ad,rad} = 6.5$ following \citet{Ge2015}, which leads to higher rate of stable mass transfer episodes with significant orbital shrinkage when compared to, for example, \citet{Stevenson_2017}.

Our results confirm that the value of $\zeta_{\rm ad,rad}$ indeed plays an important role for $R_{\rm stable}$ (see discussion in section \ref{subsubsec:effect_of_zeta}). 
However, our models also suggest that the significance of the stable channel is affected by the assumed angular momentum loss mode as well, and we expect this to be also the reason why the models of \citet{Belczynski_2016} predict a negligible $R_{\rm stable}$. In particular,  \citet{Belczynski_2016} assumes $\gamma = 1$ and a critical mass ratio of $q_{\rm crit} = 3$ for HG donors (for comparison, our assumed $\zeta_{\rm ad,rad} = 4$ is equivalent to $q_{\rm crit} = 3.2$, if the accretor is a BH). Therefore, the models of \citet{Belczynski_2016} are fairly similar to our M11 and M12 models (see Table \ref{tab:rates_of_all}). Our simulations of M11 and M12 indicate that the stable channel is essentially negligible, in broad agreement with \citet{Belczynski_2016}. 
However, in the models with a higher angular momentum loss (i.e. $\gamma$ = 2.5), but with the same $\zeta_{\rm ad,rad}$, $R_{\rm stable}$ significantly increases and even dominates for  $M_{\rm BH,1}\gtrsim25\,M_{\odot}$. In conclusion,  the stable channel can embody a significant formation channel for merging binary black holes not only for $\zeta_{\rm ad, rad}\gtrsim 6.5$ but also for strong angular momentum loss modes such as $\gamma = 2.5$.

In the majority of our model variations, the most massive merging binary black holes originate from the stable channel (i.e. $M_{\rm BH,1}\gtrsim25\,M_{\odot}$) in broad agreement with \citet{Neijssel_2019}; \citet{Lieke2022ApJ...931...17V} and  \citet{Briel2022}.
In particular, our Fig. \ref{fig:43_hist} can be directly compared to Fig. 5 of \cite{Lieke2022ApJ...931...17V} and Fig. 7 of \cite{Briel2023}. Our models generally agree more with that of \cite{Lieke2022ApJ...931...17V} than with that of \cite{Briel2023}. This is not surpising, as the models of \cite{Lieke2022ApJ...931...17V} were generated by \textsc{COMPAS} \citep{Riley_2022}, which also use the fitting formulae of \cite{Hurley_2000}. Smaller differences between these models most likely can be attributed to different choices in binary physics assumptions of \cite{Lieke2022ApJ...931...17V}, such as $\zeta_{\rm ad,rad} =  6.5$, isotropic angular momentum loss mode, i.e. $\gamma = M_{\rm d}/M_{\rm a}$ and a mass transfer efficiency, which is related to the thermal timescale of the accretor star. The differences between our predictions and the results of \citet{Briel2023} are more significant. While they predict that the stable channel dominates at high masses, the overall contribution of this channel is still small ($\sim$6.4 per cent). Furthermore, they find several other channels to be efficient, including ones in which only one phase of mass transfer occurs. These differences are most likely due to the following: i) their model is produced with \textsc{BPASS} \citep[][]{Eldridge2017, StanwayEldridge2018}, which uses detailed binary models, ii) they assume that if the accretor star accretes 5 per cent of its initial mass, it will evolve chemically homogeneously (which allows channels with only one mass transfer episode to be efficient), iii) they assume a super-Eddington accretion for BH accretors.

Furthermore, we find non-negligible differences in the reported final mass ratio distributions. In particular \citet{Briel2023} finds that most systems from the stable channel have $q\lesssim 4$ (most likely can be attributed to their assumption of super-Eddington accretion rate), while \citet{Lieke2022ApJ...931...17V} finds a relatively narrow range of 0.5$\lesssim q \lesssim 0.8$. On the other hand, the predicted mass ratios of the systems from the stable channel in our models typically peak at $q\approx$1 and drop rapidly beyond $q\approx 0.8$ with $\zeta_{\rm ad,rad} = 4$ and beyond $q\approx 0.5$ with $\zeta_{\rm ad,rad} = 7.5$ (see Fig. \ref{fig:43_scatter} and Fig. \ref{fig:q43_hist}).
\citet{Lieke2022ApJ...931...17V}  finds a mass ratio distribution of the CEE sources in the range of 0.2$\lesssim q \lesssim 1.0$, which peaks around q$\approx0.3$ and gradually decreases from that value with increasing q. While our predicted mass ratios for this channel are typically in the same range, they peak around $q\approx 1$ in all of our model variations.


\section{Results: the impact of stellar winds}
\label{sub:stellarwinds}
In this section, we investigate how different assumptions about mass loss rates of line-driven winds affect the evolution of interacting massive binaries (subsection \ref{subsec:interacting}) and progenitors of GW sources (subsection \ref{subsec:effect_of_winds_on_mbhbh}). We test three different models, these are Model I with  $f_{\rm wind} = f_{\rm wind,WR} = 1$, Model II with $f_{\rm wind} = 1/3, f_{\rm wind,WR} = 1$, and finally Model III $f_{\rm wind} = f_{\rm wind,WR} = 1/3$. The three different stellar wind models are summarised in Table \ref{tab:mdotapplied} (see also subsection \ref{subsection:swp}). Finally, in subsection \ref{subsec:importance_of_lbv}, we discuss the importance of LBV winds and the Humphreys-Davidson limit on GW sources.

\begin{figure}
\includegraphics[trim=0 0 1cm 0, clip,width=\columnwidth]{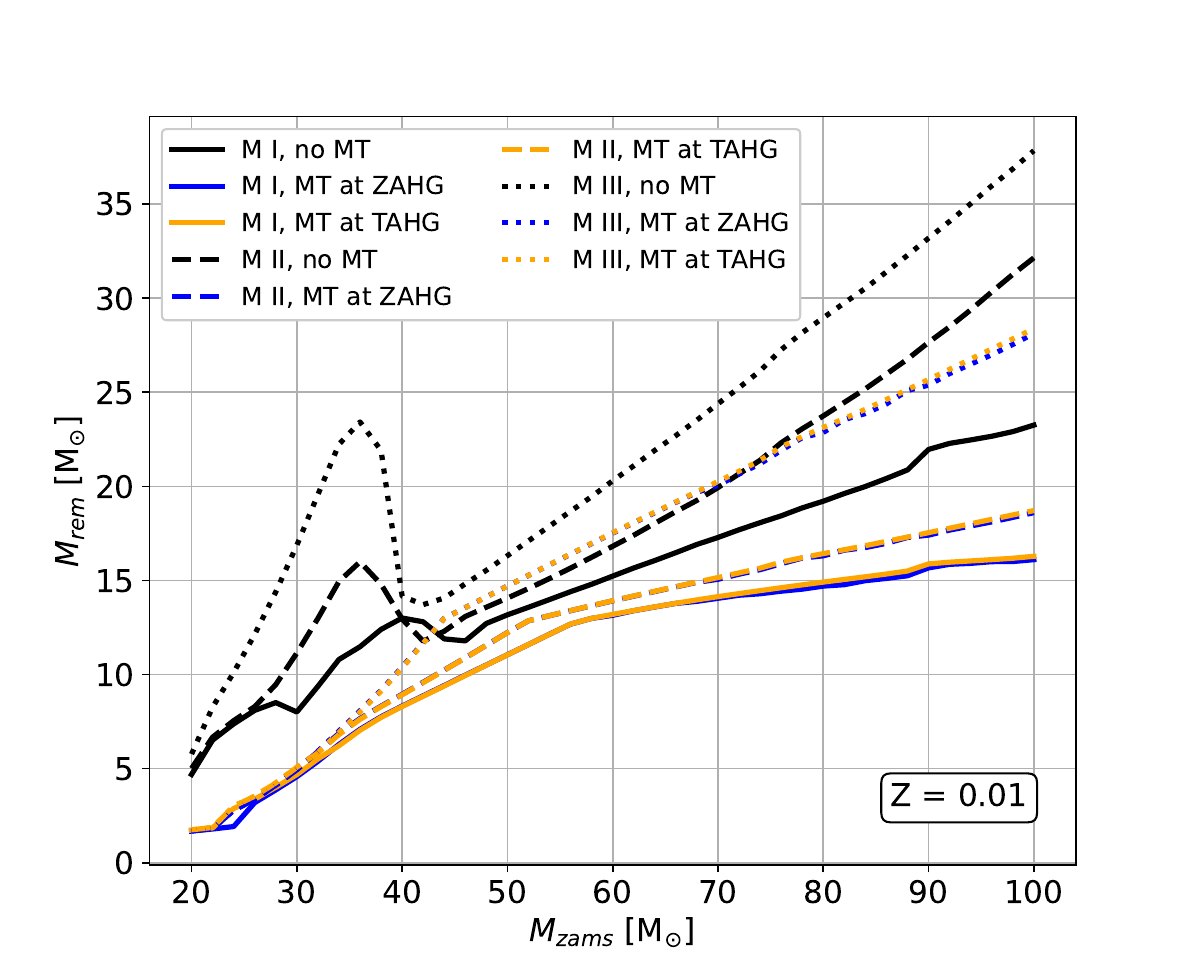}
  \caption{The mass of the remnant as a function of initial mass at Z = 0.01 for stars in interacting binaries and single stars for our three stellar wind models. Model I shown by solid lines, Model II shown by dashed lines and Model III shown by dotted lines. The black lines show the remnant mass for single stars (e.g. their hydrogen envelope is not stripped as a result of a mass transfer episode), the blue lines show the remnant mass for stars, which lose their envelopes just at the moment when they cross the Hertzsprung gap (e.g. at zero-age Hertzsprung gap), the yellow lines show stars, which lose their envelopes at the end of the Hertzsprung gap phase (e.g. terminal-age Hertzsprung gap). 
  Here we ignored the accretion by the BH in an eventual second phase of mass transfer, which is a valid assumption, if the accretion is Eddington-limited}
  \label{fig:Mremnant_interacting}
\end{figure}

\begin{figure}
  \includegraphics[trim=0 0 0.5cm 0, clip,width=\columnwidth]{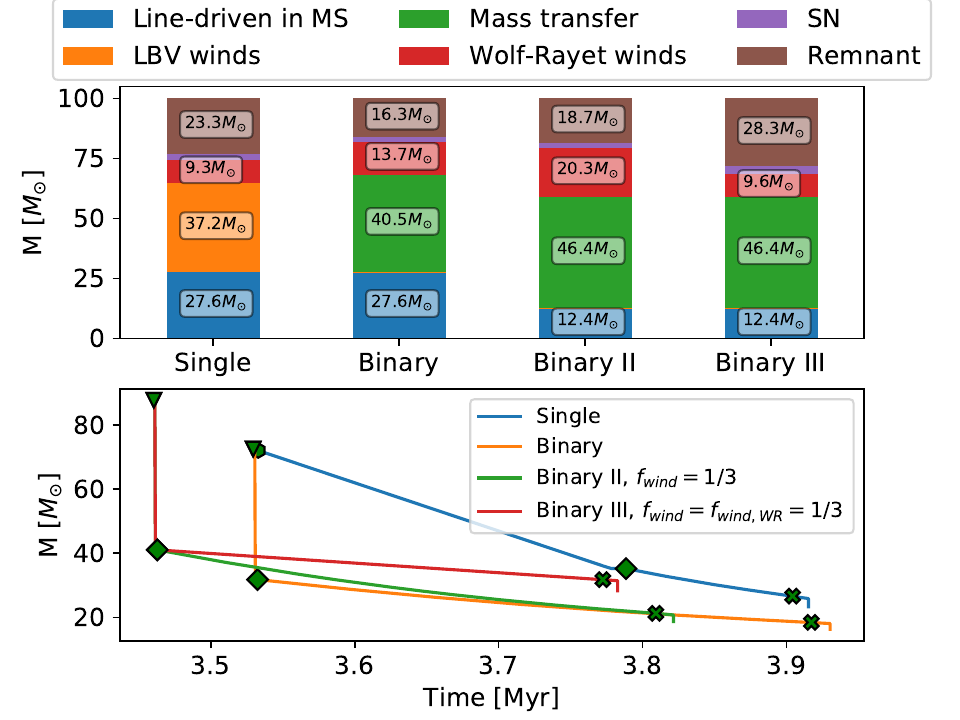}
  \caption{The importance of stellar winds  with and without binary interactions are compared by showing the
  evolution of a star with an initial mass of $M_{\rm ZAMS, 1}$ = 100$M_{\odot}$ at metallicity $Z = 0.01$ as a single star, and as a binary with a companion of $M_{\rm ZAMS, 2} = 90 M_{\odot}$ and with an initial orbital separation of $a_{\rm ZAMS} = 400 R_{\odot}$. 
  We show the binaries for three different stellar wind models. Labels 'Binary', 'Binary II' and 'Binary III' correspond to Model I ($f_{\rm wind} = 1$),  Model II  ($f_{\rm wind} = 1/3$) and Model III $f_{\rm wind} = f_{\rm wind,WR} = 1/3$ ), respectively.
  We assume circular orbits.
  The top panel shows the remnant mass and the mass lost due to different mechanisms. The bottom panel shows the mass of the stars as a function of time. The green symbols indicate the starting points of different stellar evolutionary phases; triangle: hydrogen-shell burning phase, hexagon: core helium burning phase, diamond: helium star, cross: helium giant. For clarity, we also show the curves starting from the hydrogen-shell burning phase.}
  \label{fig:42_single_vs_binary}
\end{figure}

\begin{figure*}
\includegraphics[width=\textwidth]{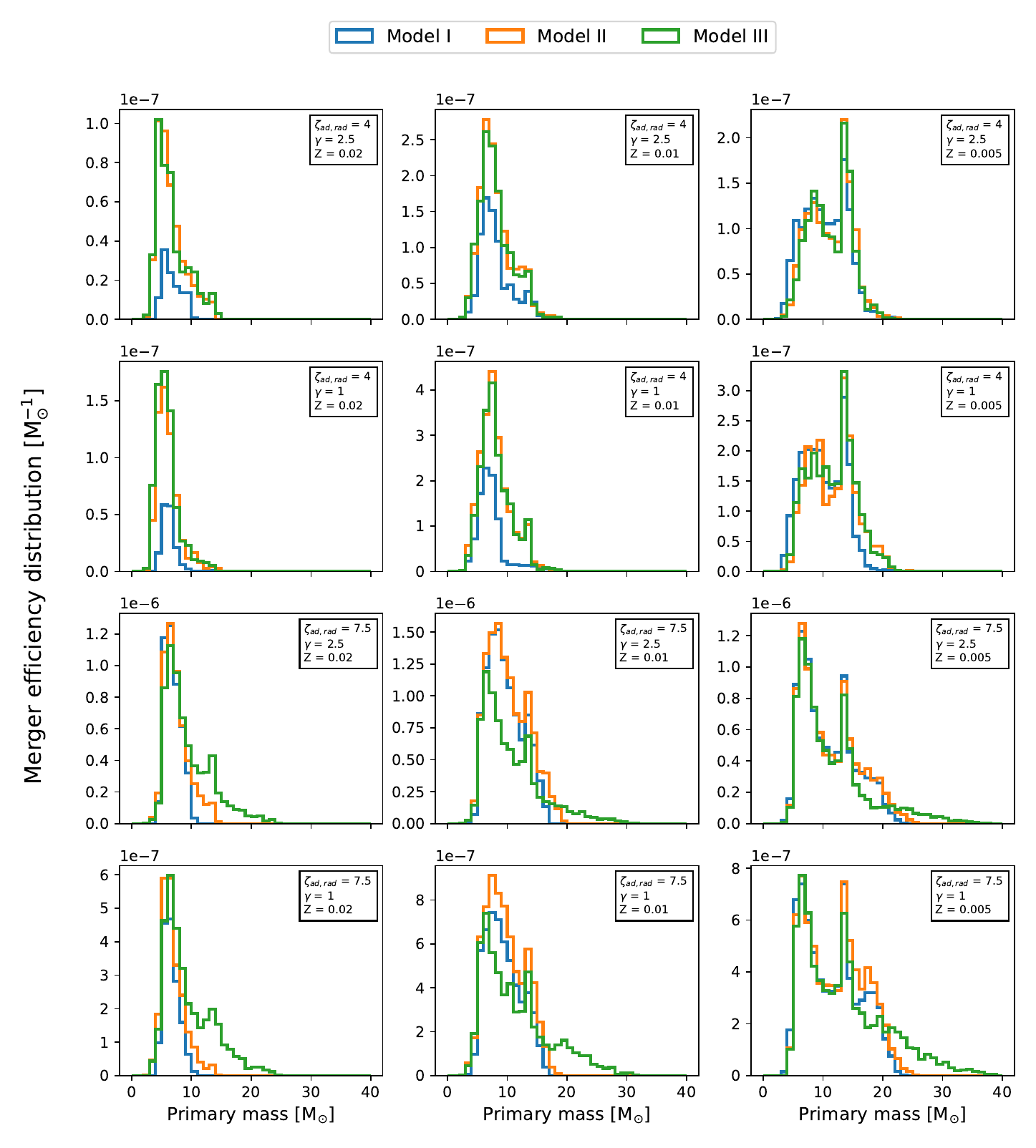} \hfill
  \caption{The primary mass distribution of merging binary black holes for different stellar wind models and different assumptions for mass transfer stability criteria for giants with radiative donors ($\zeta_{\rm ad,rad}$) and angular momentum loss during mass transfer with non-compact accretor ($\gamma$) at metallicities Z = 0.02, 0.01 and 0.005. For all models shown here the convective envelope prescription of \citet{IvanovaTaam2004} and $\beta = 0.3$ is assumed. This histograms are normalised to the merger efficiency.}
  \label{fig:wind_primary_first}
\end{figure*}

\subsection{The effects of stellar winds on binary evolution}
\label{subsec:interacting}

In Figure \ref{fig:Mremnant_interacting}, we show the remnant mass as a function of initial mass for single stars and for stars in interacting binaries at Z = 0.01. We see that stars in interacting binaries produce considerably lower mass black holes compared to their single counterparts \citep[see also for similar conclusions:][]{Woosley2019, Laplace2021:different_to_the_core, VanBeveren1997:The_effect_of_binary_evolution_on_the_theoretically_predicted_distribution_of_etc}.

There are two major reasons for this. Firstly, as the donor star loses its hydrogen envelope as a result of the mass transfer phase, its hydrogen-shell burning is halted and therefore so is the growth of its helium core. As the helium core is expected to grow only slightly during the very short Hertzsprung gap phase, but significantly during the core-helium burning phase (see later Fig. \ref{fig:42_single_vs_binary}), the remnant mass significantly depend on whether the system undergoes Case B or Case C mass transfer.  In particular, at metallicities $Z \gtrsim 0.005$, the radial expansion after the onset of core helium burning of stars with $M_{\rm ZAMS} \gtrsim 40 M_{\odot}$  is negligible  (Fig. \ref{fig:Wind1}). Practically, all  interacting massive binaries with $M_{\rm ZAMS} \gtrsim 40 M_{\odot}$  initiate Case A or Case B mass transfer episodes at these metallicities \citep[see similar discussion in ][]{Vanbeberen2007:The_wolf_rayet_population_predicted_by_massive_single_star_and_mbe}. Consequently, the yellow curves in Fig. \ref{fig:Mremnant_interacting} (corresponding to systems initiating mass transfer at the end of the Hertzsprung gap phase of the donor star) show approximately the maximum remnant mass that stars with $M_{\rm ZAMS} \gtrsim 40 M_{\odot}$ from interacting binaries can have at Z = 0.01. Because of the early envelope stripping, the helium cores of the donor stars cannot grow significantly via hydrogen shell burning. As a result, they also form less massive remnants than their non-interacting counterparts. 

Secondly, the lifetime of the Wolf-Rayet phase of a star in an interacting binary increases compared to that of the single star. The primary star in the interacting binary spends most, if not all of its core helium burning lifetime as a stripped helium star. As a consequence, the star in the binary ends up losing more mass due to Wolf-Rayet winds than its single counterpart. Since the Wolf-Rayet winds directly affect the mass of the helium core, the mass of the black hole sensitively depends on total mass lost during this evolutionary phase.
The single star also loses a significant amount of mass via LBV winds.  However,  this impacts only the hydrogen envelope, and not the mass of the helium core. 

The difference between interacting and non-interactive binary evolution is illustrated in Fig. \ref{fig:42_single_vs_binary}, where we compare the evolution of a single star with an initial mass of $M_{\rm ZAMS} = 100 \,M_{\odot}$, with the evolution of a binary with the same primary mass as the mass of the single star.
The mass of the black hole originating from single stellar evolution is appreciably higher ($M_{\rm BH} = 23.3 \,M_{\odot}$) than the black hole formed through binary evolution ($M_{\rm BH} = 16.2 \,M_{\odot}$, but also compare the black with the yellow lines in Fig. \ref{fig:Mremnant_interacting}). 
In the lower panel of Fig. \ref{fig:42_single_vs_binary}, we  compare the evolution of the masses of these two systems starting from the Hertzsprung gap phase. During this phase, the helium core of the single star grows only slightly, from $M_{\rm core} \sim 31.2 \,M_{\odot}$ to $M_{\rm core} \sim 31.7 \,M_{\odot}$ (also compare the blue with the yellow lines in Fig. \ref{fig:Mremnant_interacting}), while it increases up to $M_{\rm core} \sim 35.2 \,M_{\odot}$ by the end of the core helium burning. As previously mentioned, at these metallicities, the most massive binaries are only expected to undergo Case C mass transfer phase in negligible numbers. This implies that the masses of stripped helium stars ($M_{\rm HE, binary}$), which evolve from an MS star with $M_{\rm ZAMS} = 100\, M_{\odot}$  in an interacting binary is not above $M_{\rm HE, binary} \sim 31.7 \,M_{\odot}$ for the vast majority of the cases. The single star also loses its envelope eventually, mostly due to LBV winds. However, this occurs well after the onset of the core helium burning phase. During this phase, the mass of the helium core can grow uninterrupted. As a result, a more massive  helium star is formed, e.g. in this case  $M_{\rm HE,single} = 35.2 M_{\odot}$.

We also see that while in Model II single stars form appreciably more massive BHs than in Model I, this difference is much smaller for interacting binaries.
As shown in Fig. \ref{fig:42_single_vs_binary}, in Model II, the primary star of the binary system develops a more massive helium core before the envelope loss ($M_{\rm HE,binary} \approx 40.3 \,M_{\odot}$) than in Model I ($M_{\rm HE,binary} \approx 31.7 \,M_{\odot}$). However, this stripped star in the interacting binary loses an enormous amount of mass via Wolf-Rayet winds and ends up with a black hole that is only about 2.4$\,M_{\odot}$ more massive than the black hole evolving from the same system in Model I (compare the solid lines with the dashed lines in Fig. \ref{fig:Mremnant_interacting}).
The decrease in the difference in remnant masses for interacting binary systems compared to single systems is the result of the increased amount of mass lost during the Wolf Rayet phase ($\sim 20.3 \,M_{\odot}$ as opposed to $\sim13.7\, M_{\odot}$).  In Fig. \ref{fig:42_single_vs_binary}, we also show the evolution of the binary with Model III, in which the Wolf-Rayet winds are also scaled down by a factor of three. In this case, the mass of the final black hole is about $M_{\rm BH} \sim 28.3\, M_{\odot}$, which is about $10\,M_{\odot}$ more than than the black hole forming in the binary with Model I  (compare the solid lines with dotted lines in Fig. \ref{fig:Mremnant_interacting}).

\subsection{The effect of stellar winds on merging binary black holes}
\label{subsec:effect_of_winds_on_mbhbh}

\begin{figure*}
\includegraphics[width=1.0\textwidth]{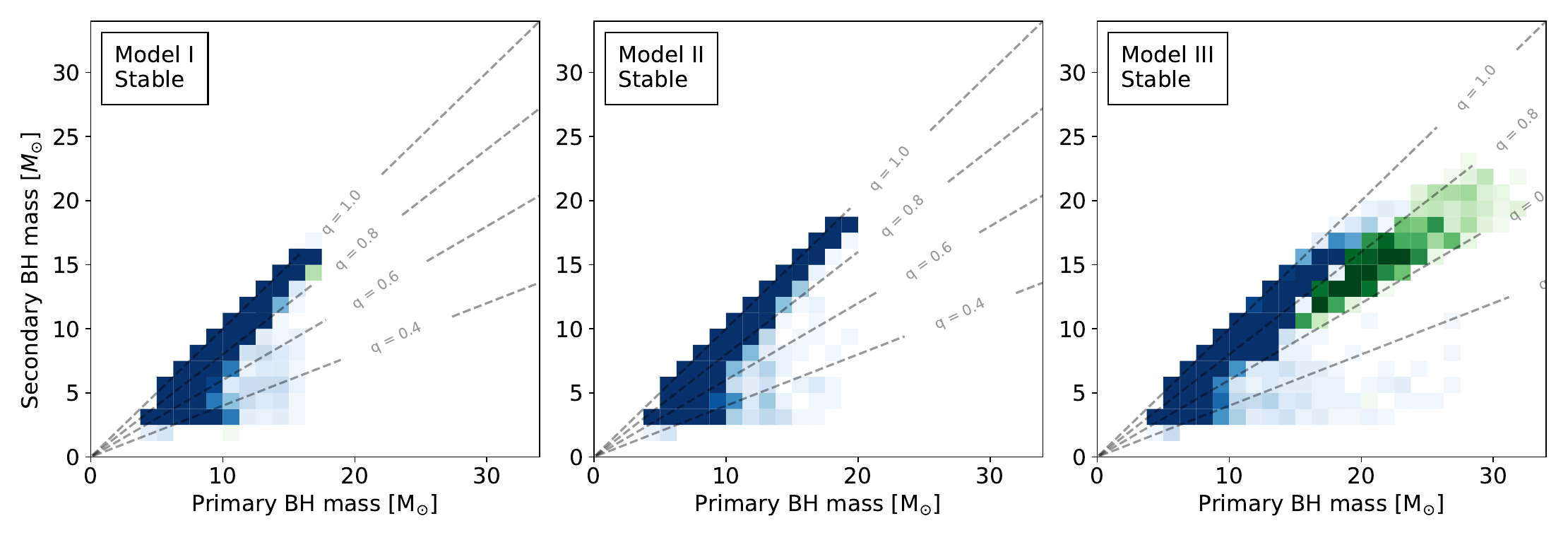} \hfill
\caption{2D histograms of masses of the merging binary black holes for the three stellar wind models at $Z = 0.01$. We have separated sources that have Case A first mass transfer phase (green) and sources that have Case B first mass transfer phase (blue). The histograms have been normalised to the merger efficiency at $Z = 0.01$. }
  \label{fig:windscatter}
\end{figure*}

In this subsection, we discuss how the maximum mass of merging binary black holes are affected with different stellar wind models. In Fig. \ref{fig:wind_primary_first}, we show the primary mass distribution of merging binary black holes for all of our three stellar wind models and with different values of $\zeta_{\rm ad,rad}$ and $\gamma$ at metallicities $Z = 0.02,\, 0.01,\,0.005$. 
The difference in the maximum mass in Model I and Model II is only $\sim\,4\,M_{\odot}$ and therefore not significant. This is not surprising, as we found similar results for interacting binaries in section \ref{subsec:interacting}.

Whether the most massive black holes in Model III form GW sources depends on what we assume about the mass transfer stability criteria for giants with radiative donors. With $\zeta_{\rm ad,rad} = 4$, the most massive systems in Model III form BH-BH binaries that are too wide to merge within Hubble time, or experience stellar merger and never form BH-BH binaries. Therefore, in that case, the maximum masses of the GW sources do not differ significantly between Model II and Model III. 
However, with $\zeta_{\rm ad,rad} = 7.5$, the primary masses of merging binary black holes can reach up to $M_{\rm BH}\sim\,30\,M_{\odot}$ for Model III at $Z = 0.01$, which is significantly larger than that of the most massive GW progenitors in Model II ($M_{\rm BH}\sim\,20\,M_{\odot}$). In order to understand this in more detail, let us consider how the most massive merging black hole binaries are formed at metallicties $Z \gtrsim 0.005$:
\begin{itemize}
    \item \textit{The most massive GW sources form via the stable channel.} This is due to our assumption that envelope ejection during CEE is only possible with core-helium burning donors. At such high metallcities the expansion of radius in such donors is negligible for $M_{\rm ZAMS} \gtrsim 40\, M_{\odot}$ (see Fig. \ref{fig:Wind1}) and therefore so are the rates of Case C mass transfer events for such systems.
    \item \textit{Binaries with the most massive stars only form GW source, if they have $q_{\rm MT,2}\gtrsim4$-$5$.} The orbit only shrinks efficiently due to a stable phase of mass transfer with a BH accretor, if $q_{\rm MT,2}$ is large (see equation \ref{eq:iso}). In relatively high metallicity environments ($Z \gtrsim 0.005$), the orbital separations of binaries with the most massive initial masses are typically so wide at the onset of the second phase of mass transfer  due to stellar winds (see Figure \ref{fig:GetFormationHistory}), that only systems with $q_{\rm MT,2}\gtrsim4$-$5$ form  BH-BH binaries that merge within the Hubble time.
\end{itemize}
Considering these two points, we can understand why binaries with the most massive stars do not form GW sources with $\zeta_{\rm ad, rad} = 4$. In these model variations $q_{\rm crit}$ = 3.2 at the onset of the second mass transfer phase (see Table \ref{tab:zeta2q}). Consequently, a binary with $q_{\rm MT,2}\approx4$-$5$ experiences an unstable phase of mass transfer and the system merges before it could form a BH-BH binary. If $q\lesssim q_{\rm crit}$, then the BH-BH binary that is formed is too wide to merge within the Hubble time.




In Fig. \ref{fig:windscatter}, we show a  2D histograms of the masses of the merging binary black holes at Z = 0.01 for each stellar wind model. We distinguish sources based on the type of the first mass transfer episode (Case A is shown by green and Case B is shown by blue). The most massive systems in Model I and Model II experience a Case B first phase of mass transfer. The most massive black holes from these models have mass ratios $0.8\lesssim q_{\rm final} \lesssim 1.0$. On the other hand, the gravitational wave sources with the most massive primaries are predicted to form in a very different way in Model III. These binaries have their first mass transfers with (late) main sequence donors and the mass ratio distribution of the merging binary black holes are in the range of $0.6 \lesssim q_{\rm final} \lesssim 0.8$.
In Figure \ref{fig:GetFormationHistory}, we show typically formation histories of the most massive merging binary black holes from each mode and we discuss their evolution in detail in section \ref{sec:mostmassivestory}.

\subsection{The effect of LBV winds on the merging binary black hole population}
\label{subsec:importance_of_lbv}
Stars that cross the Humphreys-Davidson limit are predicted to lose a significant amount of mass via LBV stellar winds. However, the underlying mechanism for the mass loss, the predicted mass loss rates and its metallicity dependence are extremely uncertain \citep[see e.g.][]{Smith}.

If the progenitors of merging binary black holes have sufficiently wide initial orbital separations, such that the donor stars cross the Humphreys-Davidson limit, before they initiate a mass transfer episode, then LBV winds will affect the demographics of this merging binary black hole population. This can occur in the following ways.
Firstly, the range  of the mass ratio distribution is decreased to lower values at the onset of the second mass transfer by the LBV mass loss rates. This affects the number of mergers of the rCEE channel (see also e.g. Fig. \ref{fig:qmax} and discussion in section \ref{subsubsec:effect_of_Teff}).
Secondly, intense mass loss rates widen the orbit. This can increase the number of binary black holes which are too wide to merge within Hubble time.
Finally, extremely high LBV mass loss rates can also affect the maximum size that the stars eventually reach. In principle, if the mass loss rates are sufficiently high ($\dot{M}_{\rm LBV}\sim10^{-3}M_{\odot}\rm{yr^{-1}}$), the red-ward evolution of massive stars could be truncated by LBV winds, as they would lose their hydrogen-rich envelopes soon after the onset of core-helium burning. 

\begin{figure*}
    \includegraphics[width=1.0\textwidth]{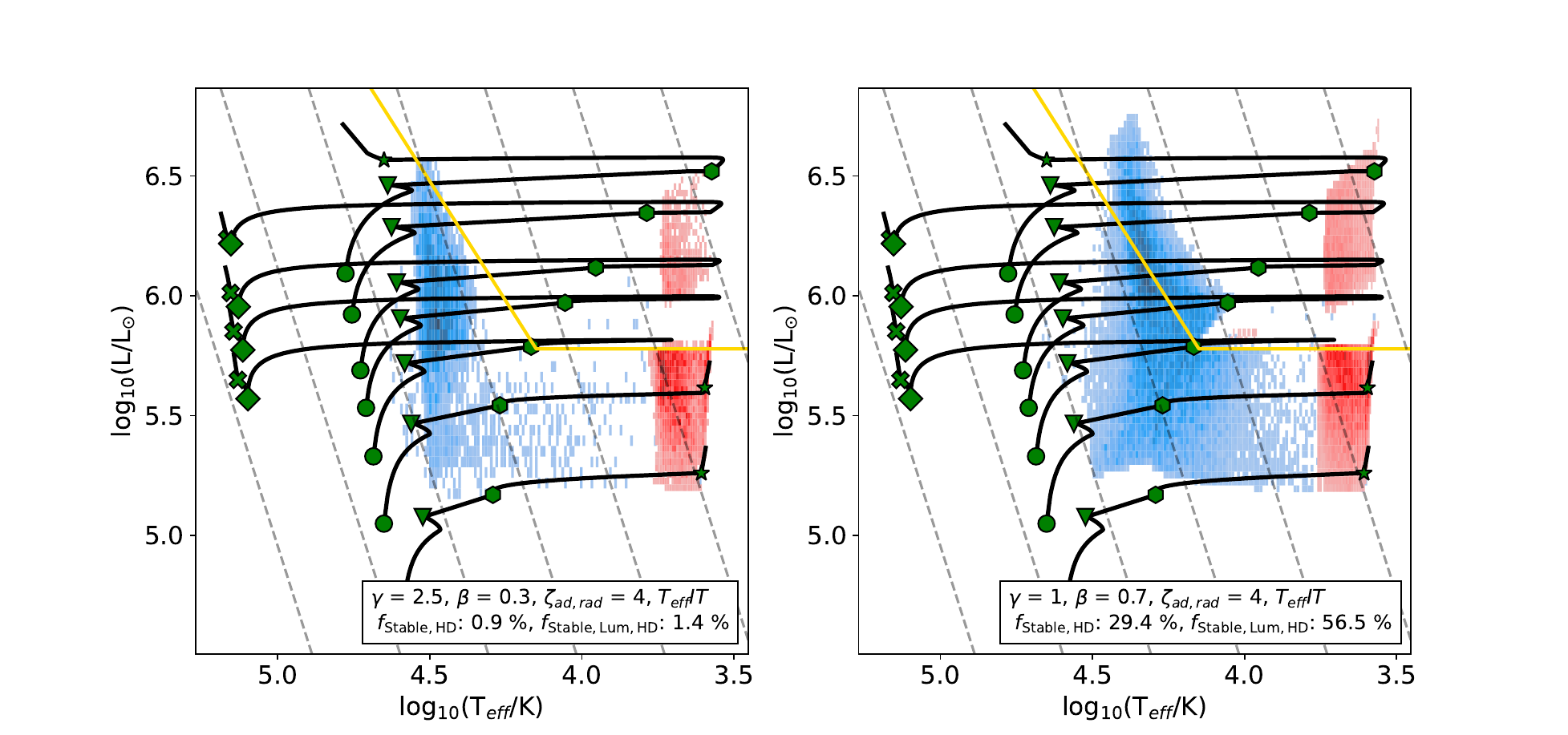} 
    \caption{We show the positions of the donor stars at the onset of the second mass transfer phase of GW progenitors in the HD diagram at Z = 0.0007 with 2D histograms for two different model variations. The blue 2D histogram shows the systems of the stable channel, while the red shows the systems of the CEE channel. The yellow line shows the Humphreys-Davidson
     limit.  We also show a few stellar tracks with the same masses as in Fig. \ref{fig:Wind1}, as well as the their most importnt evolutionary steps with green shapes (see Fig. \ref{fig:Wind1}). In the legend $f_{\rm Stable, HD}$, expresses the number of systems in  the stable channel, in which any of the stars cross the Humphreys-Davidson
 as a fraction of all stable channel GW sources. $f_{\rm Stable, Lum HD}$ is the same but expressed as a fraction of only those stable channel sources, in which any of the stars in the binary would cross the Humphreys-Davidson
 limit, if it evolved without any binary interactions}
    \label{fig:HD_crossing_per_model}
\end{figure*}


The latter is why LBV winds have been associated with the lack of observed red super giants above luminosities of $\rm{log}(L/L_{\odot}\approx5.8$ (see e.g. \citealt{Lamers1988ApJ...324..279L}, but also see \citealt{Higgins2020A&A...635A.175H}; \citealt{Gilkis2021MNRAS.503.1884G}; \citealt{Sabhahit2021MNRAS.506.4473S} for different possible scenarios).  In the context of gravitational wave progenitors, \citealt{Mennekens_2014} found that, if the LBV mass loss rate is in the order of $\dot{M}_{\rm LBV} = 10^{-3}\,M_{\odot}\rm{yr^{-1}}$ or higher, the merger rate of binary black holes drastically decreases. However, we show that, if the BH-BH mergers are dominated by the stable channel, then this is not necessarily true. This is because for such binaries both of the mass transfer phases occur in relatively small orbits, typically when the donor is at the beginning of its hydrogen shell burning phase. Therefore, stars in a large fraction of these systems never cross the Humprehys-Davidson limit.

In Fig. \ref{fig:HD_crossing_per_model}, we show the donor stars at the onset of the second phase of mass transfer in the Hertzsprung-Russel diagram for the stable and the CEE channel at Z =  0.0007 for two model variations. We chose these particular models because they have the lowest and the largest number of GW progenitors of the stable channel, in which any of the stars cross the Humphreys-Davidson limit during their evolution. 
We also show the percentage of the binaries in the stable channel, in which any of the stars evolves beyond the Humphreys-Davidson limit ($f_{\rm Stable, HD}$). These are $f_{\rm Stable, HD} = 0.9$ per cent in the model with $\gamma = 2.5$ and $f_{\rm Stable, HD} = 29.4$ per cent in the model with $\gamma = 1$. This already indicates that in some of our model variations, only a negligible number of GW progenitors of the stable channel is affected by LBV winds. 
We note that this fraction is dependent on the mass of the stars, In particular, in the model with the $\gamma = 1$, all stars with donor mass above 80 $M_{\odot}$ initiate the mass transfer phase after they crossed the Huprheys-Davidson limit.

We also show $f_{\rm Stable HD}$ for all model variations at all metallcities in Fig. \ref{fig:HD_crossing_per_model_extensive}. This clearly demonstrates that especially at lower metallicties (at which most GW sources are predicted to originate from), the majority of the systems in the stable channel never become LBV stars. 

As evident from Fig. \ref{fig:HD_crossing_per_model}, the most massive GW progenitors in the CEE channel initiate the second phase of mass transfer well beyond the Humphreys-Davidson limit, since the donors of these systems typically have a deep convective envelope before they fill their Roche-lobes. 

If the LBV mass loss rates were a magnitude higher than our assumed value, all binaries in the CEE channel with stars that cross the Humphreys-Davidson limit would lose their hydrogen rich envelopes before those would become mostly convective. This would decrease the predicted merger rate and the maximum black hole mass of this formation channel, broadly consistent with \citet{Mennekens_2014}.
On the other hand, this is not the case for the stable channel, as a significant fraction of these systems never evolve beyond the Humphreys-Davidson limit. Even those systems which do, the stars of these binaries typically spend there less than $\sim10$ per cent of their Hertzprung gap lifetime. This means that even, if the mass loss rate is $\dot{M}_{\rm LBV} = 10^{-3}\,M_{\odot}\rm{yr^{-1}}$, the donor would only lose $\sim1\,M_{\odot}$ mass, and therefore the predictions for the stable channel are little affected by a steady, intense LBV mass loss. We show this in Fig \ref{fig:mass_lost_increased_lbv}, where we determined the mass that would be lost by the donor stars of the stable channel due to LBV wind with a mass loss rate of $\dot{M}_{\rm LBV} = 10^{-3}\,M_{\odot}\rm{yr^{-1}}$ for the $\gamma = 2.5$, $\beta = 0.3$ models at each metallicity. We see that even with such an enormous mass loss rate, the vastr majority of the  donor stars would only lose 1-2 $M_{\odot}$. 

\section{Conclusion}
\label{sub:conclusion}
We performed a parameter study on the classical isolated binary formation channel of gravitational sources. Our primary aim was to investigate how sensitively the demographics of merging binary black holes depend on current uncertainties related to the first phase of mass transfer (between two hydrogen rich stars) and on stellar winds. For this, we used the rapid population synthesis code \textsc{SeBa} to simulate the evolution of massive binaries over a metallicity range $Z = 0.0001 - 0.02$ with several model variations each with different assumptions about the binary physics. In the first part of the paper (section \ref{sub:popprops}), we  varied the (i) angular momentum loss mode and (ii) the mass transfer efficiency during first phase of mass transfer, (iii) the mass transfer stability criteria for giants donors with radiative envelopes and (iv) the effective temperature at which evolved stars develop deep convective envelopes. 
In the second part of the paper (section \ref{sub:stellarwinds}), we also varied the mass loss rates of line-driven winds (see Table \ref{tab:models} for an overview for all of our model variations). 

In our model variations, we identify two dominant evolutionary paths, one involves two stable mass transfer phases (stable channel), and one in which a stable mass transfer episode is followed by an unstable mass transfer phase (CEE channel with two variant: cCEE and rCEE, see see Fig. \ref{fig:isolatedchannelsl}).

We find that current uncertainties related to first phase of (stable) mass transfer have a huge impact on the relative importance of different dominant channels, while the observable properties (i.e. merger rate, mass and mass ratio distribution) of merging binary black holes are not significantly affected. 
This implies that models with different binary physics assumptions might yield the same predicted demographics of merging binary black holes, the origin of the majority of GW progenitors could be entirely different. This shows why it is very challenging to infer physics of massive binary evolution solely from GW observations in a meaningful way, given the large uncertainties in our current models (see also \citealt{Startrack2022ApJ...925...69B}).

Our results are in broad agreement with recent studies that have shown that the stability of mass transfer and the structure of the envelope of the donor star plays a significant role in determining the relative importance of the stable and the CEE channel \citep[see e.g.][]{Neijssel_2019, Klencki_2020, Bavera2021, marchant2021role, olejak2021impact, Lieke2022ApJ...931...17V, Briel2023}. Furthermore, we have demonstrated that uncertainties regarding the modelling of the first phase of mass transfer, such as the angular momentum loss mode and mass transfer efficiency adds an additional layer of complexity in predicting the dominant formation channel of GW sources from isolated binaries.

In order to break the degeneracy between the uncertainties in modelling mass transfer and the demographics of the population of merging binary black holes, it is clear that we need more constraints than gravitational wave data is currently supplying us, for example, constraints from electromagnetic observations. 
In particular, regarding the first mass transfer phase, observations of the mass ratios and periods of WR-O/B binary systems could offer invaluable clues about the physics of a mass transfer episode between two hydrogen rich stars. This is because WR-O/B binary systems have not yet experienced core-collapse, and therefore their mass ratios and orbital separations are directly related to the mass transfer efficiency and the angular momentum loss during the first phase of mass transfer. While, only about a few dozens of WR-O/B are known, for which the relevant parameters (such as mass ratio, period, rotational period) are reliable inferred \citep[see e.g.][]{vanderHucht2001, Crowther2007}, related studies have already lead to very important insights about binary physics \citep[see e.g.][]{VanBeveren1997:The_effect_of_binary_evolution_on_the_theoretically_predicted_distribution_of_etc, Vanbeveren1998:MassiveStars, Petrovic,  Eldridge2009, Shara2017:spin_rates_of_o_stars_in_wr+o_binaries, Vanbeveren2018:spin_rates_and_evolution_of_O_components_in_wr+o_binaries,Shao_2016,  Shenar2016}. Our study demonstrates that a larger, unbiased and more complete catalogue of WR-O/B systems with reliably measured parameters could significantly improve our understanding about the origin of merging stellar mass binary black holes.

Below, we summarise our most important results:
\begin{itemize} 
    \item \textit{Impact of angular momentum loss mode on the dominant channels (section \ref{subsubsec:effect_of_gamma_stable} and \ref{subsubsec:effect_of_gamma_cee}):}
   Angular momentum loss during the first phase of  mass transfer determines how the orbital separation changes during mass exchange and has an indirect effect on the rate of stable and unstable mass transfer phases. Consequently, it has also a strong impact on the merger rate of different formation channels.  We find that typically the stable channel dominates in our models with $\gamma  = (dJ/dM_{\rm tot})/(J/M_{\rm tot})$ = 2.5, while, the merger rate of cCEE channel remains low. On the other hand, in our models with $\gamma  = 1$, the merger rate of the stable channel is typically negligible, while the cCEE channel is efficient.
    \item \textit{Impact of mass transfer stability criteria (section \ref{subsubsec:effect_of_zeta}):} The merger rate of the stable channel increases with increasing $\zeta_{\rm ad,rad}$. This is because the degree of orbital shrinkage due to a stable mass transfer phase with a black hole accretor becomes substantially more efficient with increasing mass ratios \citep[see e.g.][]{vandenheuvel17, Pavlovskii2017, Ge_2020}. However, we find that only the merger rate of relatively lower mass BH-BH binaries is significantly affected (i.e. $M_{\rm BH} \lesssim 20\,M_{\odot}$), when $\zeta_{\rm ad,rad}$ is increased from 4 to 7.5 (corresponding to a $q_{\rm crit}$ of 3.2 and 5.5). This is because at lower metallicities (which are the most relevant for GW sources), the maximum mass ratios  of the most massive binaries at the onset of the second mass transfer phase is limited to 3-3.5, depending on the mass transfer efficiency.
    \item \textit{Line driven stellar winds (section  \ref{subsec:interacting} and \ref{subsec:effect_of_winds_on_mbhbh}):} Although, decreasing the mass loss rates of optically thin line driven winds by a factor of three significantly increases the masses of the black holes formed without mass exchange (see Fig. \ref{fig:Wind1}), we do not find an appreciable difference in the masses of merging binary black holes (Fig. \ref{fig:wind_primary_first}). The primary reason for this is that stars in interacting binaries experience a longer Wolf-Rayet phase due to envelope stripping than their single counterparts. Although, with weaker winds on the main sequence, stars develop a more massive helium core, they also experience significantly higher Wolf-Rayet mass loss rates after the envelope stripping, since the more massive helium core will be more luminous after envelope loss, and this leads to considerably higher mass loss rates \citep[see similar results in][]{Woosley2019, Laplace2021:different_to_the_core}.
    However, if the Wolf-Rayet mass loss rates are simultaneously decreased too, we find a significant increase in the masses of gravitational wave sources, but only in the model variations with $\zeta_{\rm ad,rad}=7.5$. 
    \item \textit{LBV winds (section \ref{subsec:importance_of_lbv}):}
    Intense LBV mass loss rates can have a significant effect on the CEE channel. If the mass loss rates are above $\dot{M}_{\rm LBV} \sim 10^{-3}\,M_{\odot}\rm{yr^{-1}}$, the merger rate of the CEE channel becomes negligible \citep[see also][]{Mennekens_2014}. On the other hand, the stable channel is not appreciably affected, as these binaries have so short periods, such that their envelopes are stripped as a result of a mass transfer episode before or very soon after they cross the Humpyhreys-Davidson limit.
\end{itemize}



\section*{Acknowledgements}

 Computational work was performed University of Birmingham's BlueBEAR HPC service. AD acknowledges the anonymous reviewers for their many insightful comments and suggestions.
 ST acknowledges support from the Netherlands Research Council NWO (VENI 639.041.645 and VIDI 203.061 grants

 \section*{Data Availability}

The data underlying this article will be shared on reasonable request to the corresponding author.




\bibliographystyle{mnras}
\bibliography{example} 




\appendix

\section{The effects of stellar winds on single stellar evolution}
\label{sub:single}

In this section, we briefly summarise how single massive stars evolve with our three different stellar wind models (for description of these models, see section \ref{subsection:swp}). Although, the impact of stellar winds on single massive stars has already been studied \citep[see e.g.][]{Belczynski_2010,Renzo2017}, we give this summary so that the results of \ref{subsec:interacting} and \ref{subsec:effect_of_winds_on_mbhbh} can be contrasted to the results from single stellar evolution.  

\begin{figure*}
  \includegraphics[trim=0 0 0.5cm 0, clip,width=\columnwidth]{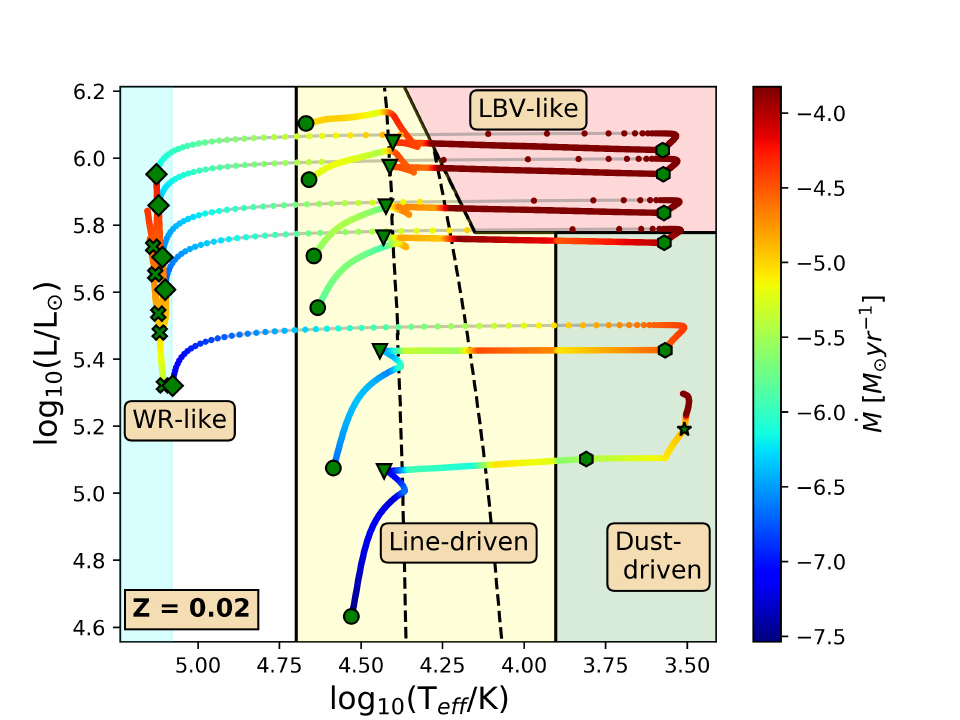} \hfill
  \includegraphics[trim=0 0 0.5cm 0, clip,width=\columnwidth]{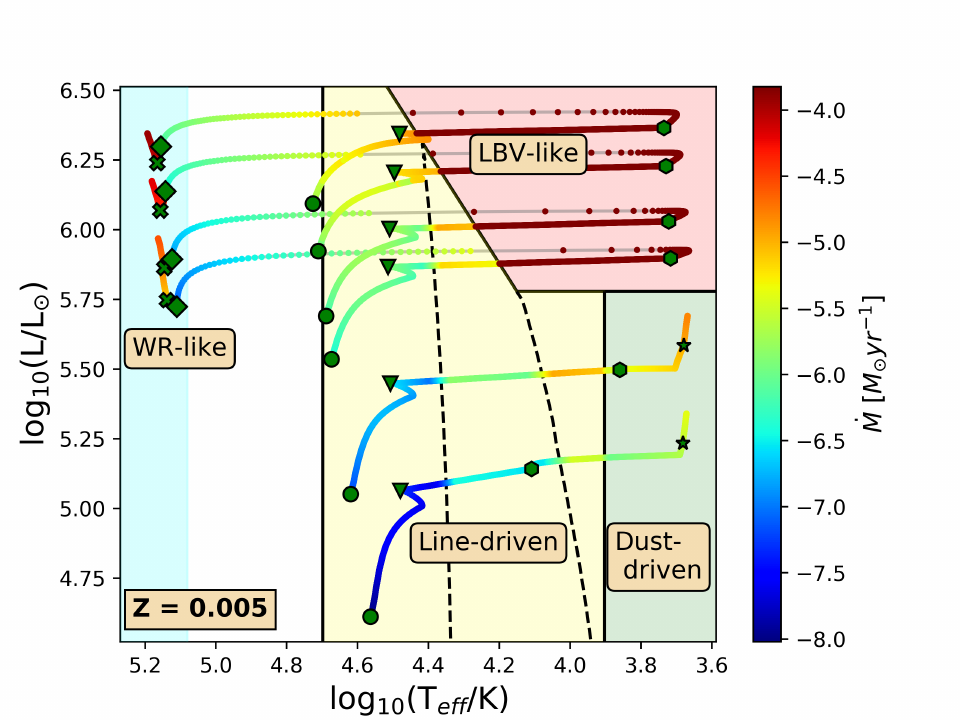}
  \caption{Hertzsprung-Russel diagrams for massive stars with the mass-loss rate colour-coded in the tracks. The stars have $M_{\rm ZAMS}$ = 20, 30, 50, 60, 80, 100 $M_{\odot}$. Different shaded regions denote different stellar wind mechanisms operating on stars. Within the 'Line-Driven' range, the two dashed lines show where the first and the second bi-stability jump occurs. Green figures indicate the starting points of different stellar evolutionary phases; circle: main-sequence, triangle: hydrogen-shell burning phase, hexagon: core helium burning phase, star: AGB, diamond: helium star, cross: helium giant.}
  \label{fig:Wind1}
\end{figure*}

In order to understand the importance of different stellar wind prescriptions, we show where different stellar wind mechanisms dominate in the the Hertzsprung-Russel diagram in Fig. \ref{fig:Wind1}, along with the predicted mass loss rates throughout the evolution of a few selected massive stars.

\begin{figure*}
  \includegraphics[trim=0 0 0.5cm 0, clip,width=\columnwidth]{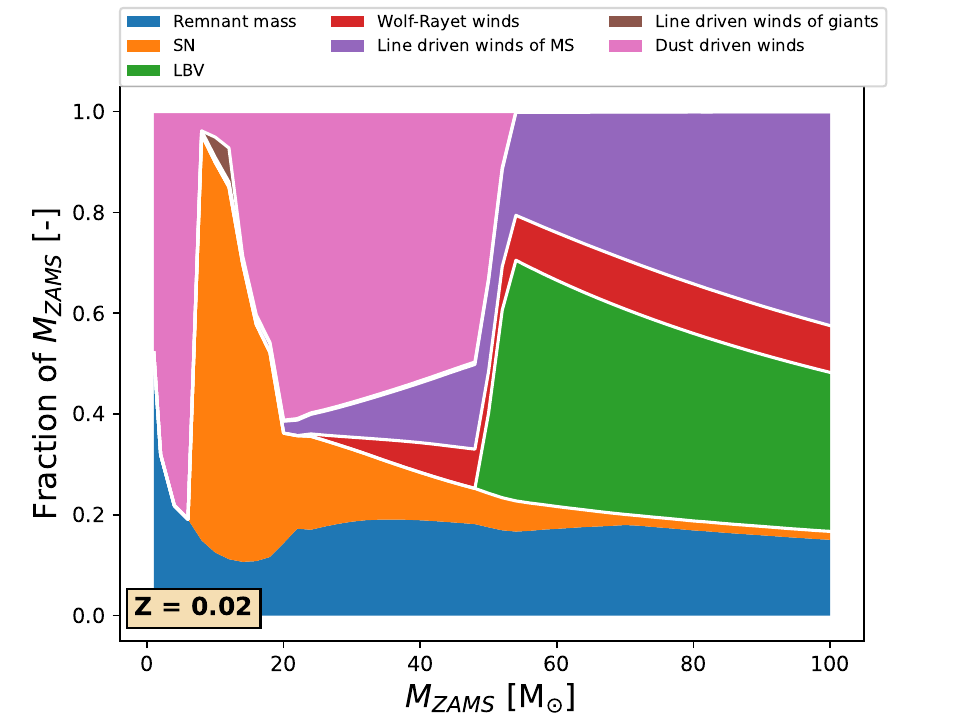} \hfill
  \includegraphics[trim=0 0 0.5cm 0, clip,width=\columnwidth]{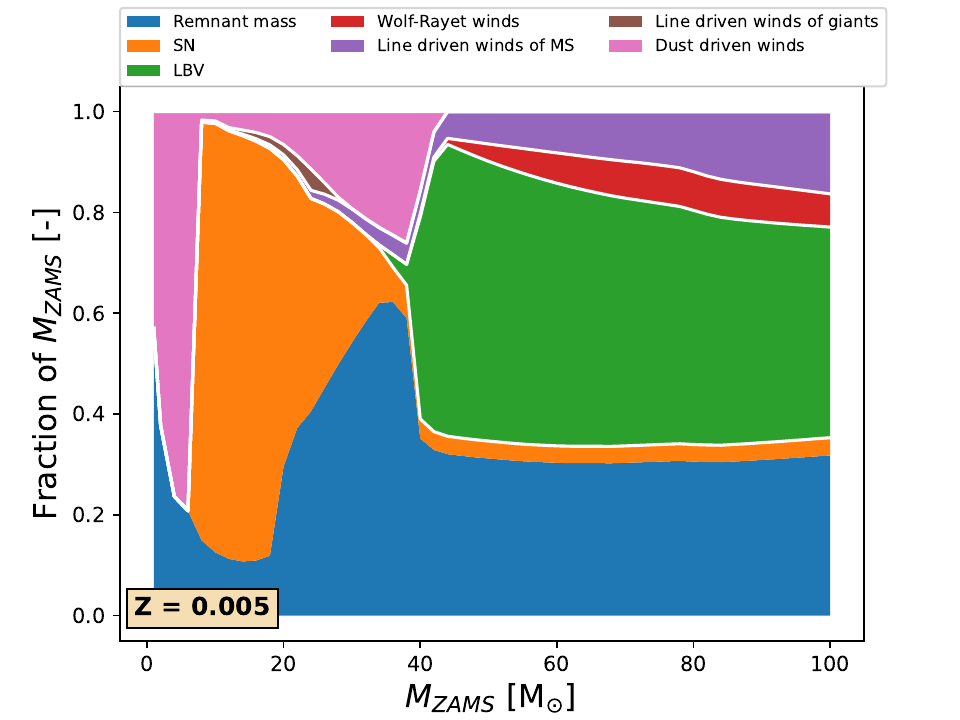}
  \caption{Fraction of mass lost due to different mass loss mechanisms during the entire life of a non-interacting star at Z = 0.02 and Z = 0.005 as a function of zero-age main sequence mass.}
  \label{fig:Wind2}
\end{figure*}

In Fig. \ref{fig:Wind2}, we show the mass lost due to the different wind mechanisms as a fraction of the initial mass for single stars at metallicities $Z = 0.02$ and $Z = 0.005$.
Up to $M_{\rm ZAMS} = 20\, M_{\odot}$, the mass of the remnant is almost completely determined by the mass lost due to supernova and dust-driven winds. 
Above $M_{\rm ZAMS} = 40$-$50\, M_{\odot}$, only a negligible amount of mass is lost during remnant formation as our model \citep[delayed model from][]{Fryer_2012} predicts direct collapse. This implies that no mass is ejected in the process and the only mass loss is due to neutrino losses, which is assumed to be 10 per cent of the proto-NS mass. 

The mass loss due to line-driven winds during the main-sequence phase is substantially more significant for stars with initial masses above $M_{\rm ZAMS} = 40$-$50\, M_{\odot}$.
Furthermore, the stars in this mass range eventually cross the Humphreys-Davidson limit, which leads to severe mass losses during the LBV phase. The masses of black holes which form via direct collapse (i.e. from progenitors with $M_{\rm ZAMS}\gtrsim 40\, M_{\odot}$) end up losing 20-40 per cent of their initial mass via line-driven winds during their main sequence phase. This already indicates that lowering the optically thin line-driven winds by a factor of three can have significant effect on the mass spectrum of the most massive black holes.

\begin{figure*}
  \includegraphics[trim=0 0 0.5cm 0, clip,width=\columnwidth]{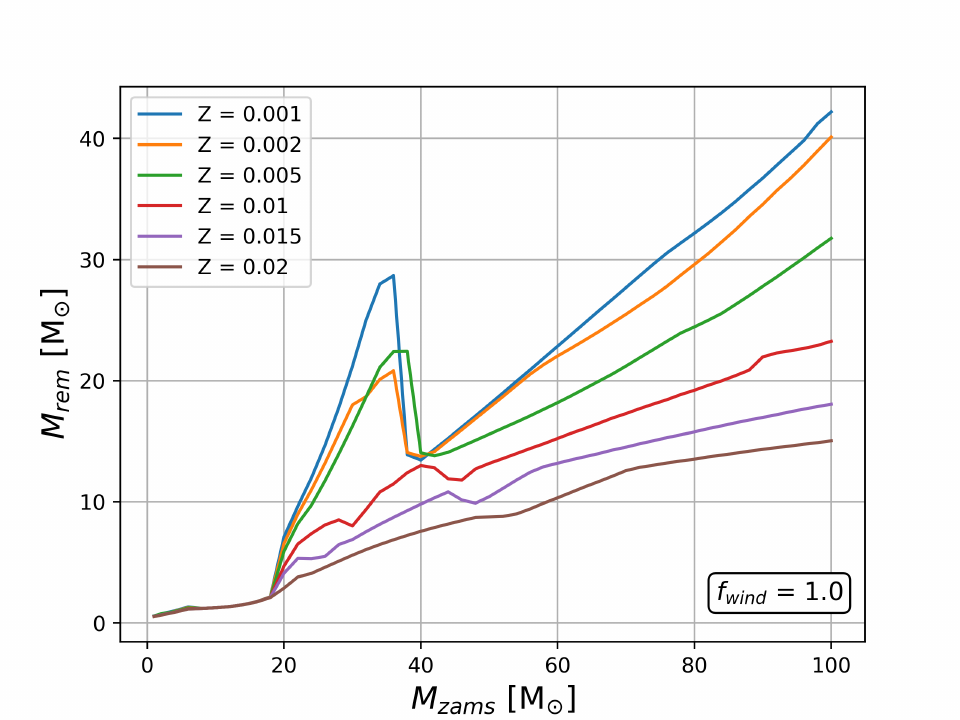} \hfill
  \includegraphics[trim=0 0 0.5cm 0, clip,width=\columnwidth]{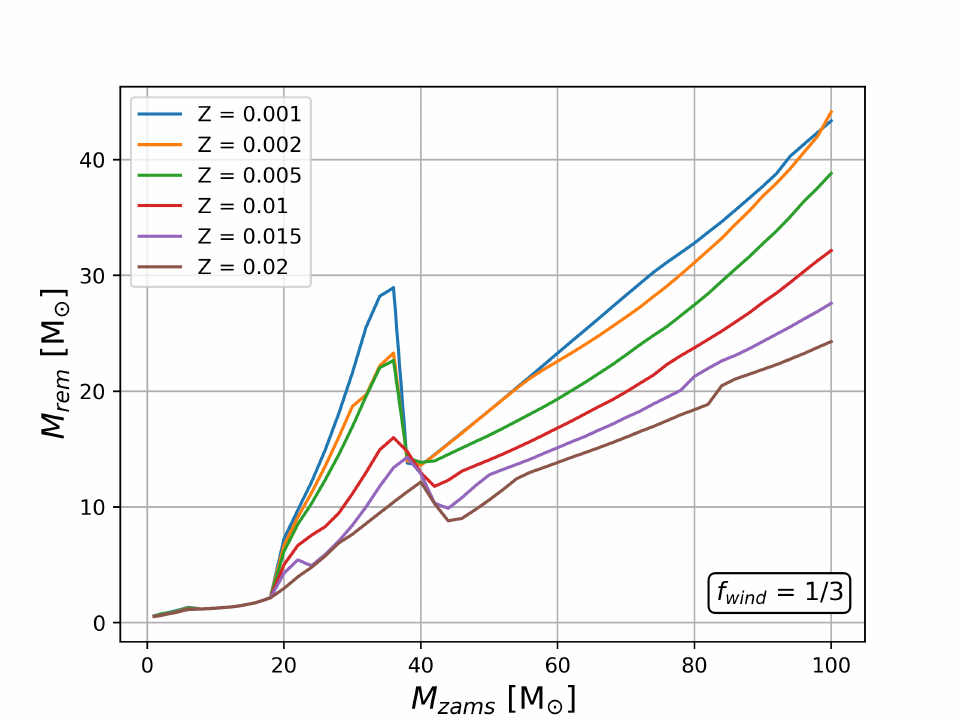}
  \caption{Zero age main sequence mass and remnant mass functions for different metallicities for single stellar evolution. We assume that black hole formation occurs above $M_{\rm ZAMS} = 20 M_{\odot}$.}
  \label{fig:single_bh_spectrum}
\end{figure*} 

In Fig. \ref{fig:single_bh_spectrum}, we show the mass spectrum of black holes formed from single stars for a wide range of metallicities for Model I  and Model II \citep[for comparison see][]{Belczynski_2010, Giacobbo2018_winds}.
Model II produces moderately more massive black holes in the mass range $ 40\, M_{\odot} \lesssim M_{\rm ZAMS} \lesssim 70\, M_{\odot}$ compared to the standard wind model at metallicities $Z \gtrsim 0.005$. However, above $M_{\rm ZAMS} \gtrsim 70\, M_{\odot}$, the differences become increasingly more appreciable. In particular, the most massive star we simulated ($M_{\rm ZAMS} = 100 \,M_{\odot}$) using Model I yields  $M_{\rm BH} \approx 14 M_{\odot}$ at solar metallicity. This increases up to $M_{\rm BH} \approx 24 M_{\odot}$ for the Model II.  At $Z = 0.01$ black holes even as massive as $M_{\rm BH}\gtrsim 30\,M_{\odot}$ are formed in Model II.
Since the initial masses of massive stars follow a distribution of $N \sim M_{\rm ZAMS}^{-2.3}$ (\citealt{Kroupa}), we therefore expect that only the high mass tail of the black hole mass distribution would be affected significantly when scaling down optically thin line driven stellar winds.
Below $Z\approx0.002$, the differences between the two stellar wind models become negligible. 

\section{Merger rate density in the local universe}
\label{subs:rates}

The merger rate density is an important characteristics of a synthetic population of GW sources, as it allows comparison to the catalogue of gravitational wave source detections made by LIGO and Virgo \citep[e.g.][]{Abbott_o3a_2021}.
We calculate the merger rate density in the local universe (i.e. $z\approx0$) for all of our model variations presented section \ref{sub:popprops}.
In order to determine this quantity, we follow a formalism  similar to the one outlined in \citet{Dominik_2015}.

First, we define the merger efficiency as the number of mergers originating from a black hole binary with masses $M_{\rm BH,1}$ and $M_{\rm BH,2}$ at metallicity $Z$ as the fraction of the full parameter space (i.e as the fraction of all the stars born, assuming the initial conditions introduced in subsection \ref{subs:icamv}):
\begin{equation}
\label{eq:mereff}
    \epsilon_{\rm mer} (M_{\rm BH,1}, M_{\rm BH,2}, Z) = f_{\rm pm} \cdot \frac{N_{\rm merger}}{N_{\rm simulated}},
\end{equation}
where $N_{\rm merger}$ is the number of mergers originating from a BH binary with given masses at a given metallicity, $N_{\rm simulated}$ is the number of sampled systems in our simulation and $f_{\rm pm}$ is the simulated parameter space as a fraction of the complete parameter space. We determine the latter as:
\begin{equation}
    f_{\rm pm} =  f_{\rm bin}\cdot\int_{1\,R_{\odot}}^{10^4\,R_{\odot}} f_{\rm a}(a) da\cdot\int_{0.1}^{1} f_{\rm q}(q) dq\cdot\int_{\rm 20\,M_{\odot}}^{100\,M_{\odot}} N_{\rm IMF}(M) dM,
\end{equation}
here $N_{\rm IMF}$ is the normalised initial mass function of \citet{Kroupa}, and has been normalised in an interval of  0.08$\,M_{\odot}$-100$\,M_{\odot}$, $f_q$ is the normalised mass ratio distribution, which has been normalised in the interval of 0-1, and $f_a$ is the normalised semimajor axis distribution, which has been normalised between 10-$10^6\,R_{\odot}$, $f_{\rm bin}$ is the binary fraction, which we assume to be 0.7, following \cite{Sana2012}. 

The merger efficiency can be converted to a merger rate density by assuming a metallicity specific star formation density model $\textrm{SFRd}^{*}(Z,z)$:
\begin{equation}
    \frac{dN}{dVdzdt} = \iiiint\frac{\textrm{SFRd}^{*}(Z,z_{\rm ZAMS})}{\Tilde{M}}\cdot \epsilon_{\rm mer}  dZ dM_{\rm BH,1} dM_{\rm BH,2}dt_{\rm d},
\end{equation}
where $t_{\rm d}$ is the so-called delay time, which is the time between the zero-age main sequence of the stars in the binary and the time of the merger due to GWs, $z_{\rm ZAMS}$ is the redshift at zero-age main sequence and it is a function of the time delay, $\Tilde{M}$ is the average mass of all stellar systems born, such that $\textrm{SFRd}(Z,z)^{*}/\Tilde{M}$ gives the average number of stars born at redshift $z$ and in the metallicity bin centred around $Z$.

Delay time can be directly obtained from the population synthesis simulations. In order to determine $z_{\rm ZAMS}$ for a given $z$ and delay time, we use the relationship for lookback time:
\begin{equation}
    t_{d} = \frac{1}{H_0}\int_{z}^{z_{\rm ZAMS}} \frac{dz'}{(1 + z') E(z')},
\end{equation}
where $E(z) = \sqrt{\Omega_m(1+z)^3 + \Omega_{\lambda}}$, with $\Omega_M = 0.3$, $\Omega_{\lambda} = 0.7$ and $H_{0} = 70 \rm{kms^{-1}Mpc^{-1}}$.

The star formation rate density in a metallicity bin centered at $Z$ with a width of $2\Delta Z$ is $\int_{\Delta Z}^{\Delta Z} \rm{SFRd}(z)\cdot f_{\rm met}(Z,z) dZ$. Here $f_{\rm met}(Z,z)$ gives the distribution of metallicities of binaries at redshift $z$ as:
\begin{equation}
    \label{eq:fmet}
    f_{\rm met}(Z,z) = \frac{1}{\sigma \sqrt{2\pi}}\exp\left(\frac{(\log_{10}(Z) - \mu(z))^2}{2\sigma^2}\right),
\end{equation}
where $\sigma = 0.5$ and the mean metallicity varies with redshift as given by \citet{Madau_2017}:
\begin{equation}
    \label{eq:sfr}
    \mu(z) = \log_{10}(Z_{\odot}\cdot 10 ^{0.153 - 0.074z^{1.34}}) - 0.5ln(10)\sigma^2,
\end{equation}
where we took $Z_{\odot} = 0.02$. Furthermore, the star formation rate density $\textrm{SFRd}(z)$ is given by \citet{Madau_2014}:
\begin{equation}
    \textrm{SFRd}(z) = \frac{0.01\cdot(1 + z)^{2.6}}{1 + ((1+z)/3.2)^{6.2}} \, M_{\odot}\rm{yr^{-1}}Mpc^{-3}.
\end{equation}
We note again that the metallicity specific cosmic star formation rate is very uncertain and this uncertainty has an important impact on the predicted merging binary black hole population \citep[see e.g.][]{Chru_li_ska_2020, Petra2020, Briel2022, Chruslinska2022arXiv220610622C}


\begin{table*}
\caption{Conversion between the mass transfer stability criteria for giants and the critical mass ratio ($q_{\rm crit} = M_{d}/M_{a}$) for different assumptions on the mass transfer efficiency and the angular momentum loss mode during the mass transfer phase. The last column reflects a binary with a black hole accretor whose mass accretion is Eddington-limited and the expelled mass having the specific angular momentum of the accretor. 
Donors with radiative envelopes are characterised by $\zeta_{\rm ad,rad}$, whereas donors with convective envelopes are described by $\zeta_{\rm HW}$. For the latter, we assume a core mass to total mass ratio of 0.45-0.63, which we find is typical for giants at $\rm{log}(T_{\rm eff})\approx 3.73\,K$ with initial masses $M_{\rm ZAMS} = 50 - 100\, M_{\odot}$ with a metallicity independent LBV mass loss rate of $\dot{M}_{\rm LBV} =  1.5\cdot10^{-4} M_{\odot}\rm{yr^{-1}}$.}
\begin{tabular}{|c|c|c|c|c|c|l|}
\hline
{\textbf{Mass transfer stability criteria}} & \multicolumn{6}{c|}{\textbf{Angular momentum loss}}                                                                                   \\ \cline{2-7} 
                                                           & \multicolumn{2}{c|}{$\gamma = 2.5$} & \multicolumn{2}{c|}{$\gamma = 1.0$} & \multicolumn{2}{c|}{$\gamma = M_{d}/M_{a}$}    \\ \cline{2-7} 
                                                           & \multicolumn{6}{c|}{\textbf{Mass transfer efficiency}}                                                                                \\ \cline{2-7} 
                                                           & $\beta = 0.3$    & $\beta = 0.7$    & $\beta = 0.3$    & $\beta = 0.7$    & \multicolumn{2}{c|}{\textit{Eddington limited accretion}} \\ \hline
$q_{\rm crit}$ with $\zeta_{\rm ad,rad} = 4$                         & $3.8$            & $2.9$            & $6.2$            & $3.4$            & \multicolumn{2}{c|}{$3.2$}                                \\ \hline
$q_{\rm crit}$ with $\zeta_{\rm ad,rad} = 7.5$                      & $8.7$            & $5.2$            & $11.6$           & $5.7$            & \multicolumn{2}{c|}{$5.5$}                                \\ \hline
$q_{\rm crit}$ with $\zeta_{\rm ad} = \zeta_{\rm HW}$                  & $0.9 - 0.7$      & $0.9 - 1.1$      & $1.5 - 1.9$      & $1.1 - 1.4$      & \multicolumn{2}{c|}{$1.4 - 1.7$}                          \\ \hline
\end{tabular}
\label{tab:zeta2q}
\end{table*}

\begin{figure*}
  \includegraphics[width=1.0\textwidth]{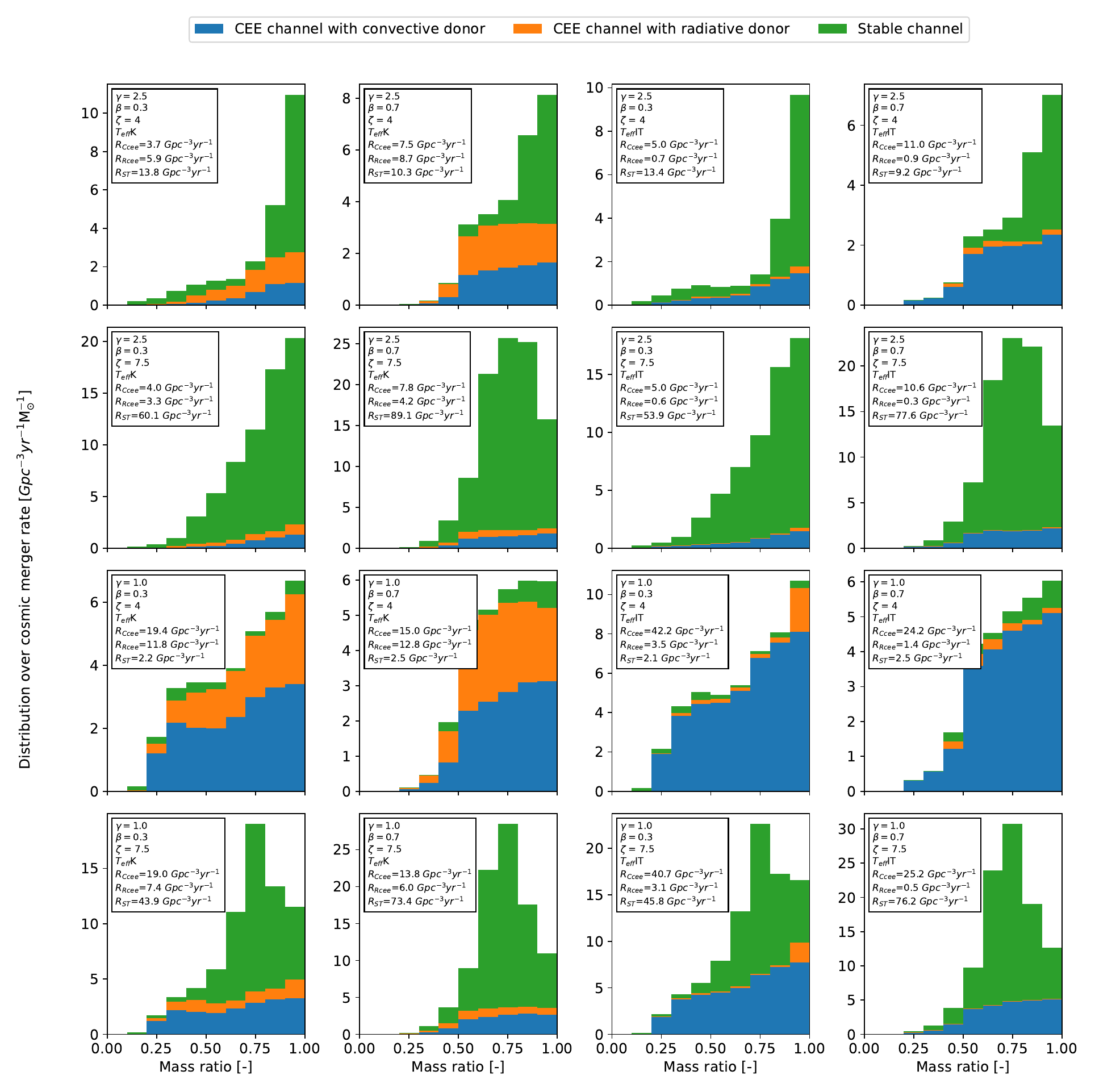}
  \caption{The mass ratio distribution of the primary of the merging binary black hole in the local universe (z = 0) for all of our models with our standard stellar wind models (e.g. $f_{\rm wind} = f_{\rm wind,WR} = 1$, see Table \ref{tab:models}). The distribution is shown by stacked histograms, where each colour indicates a different formation channel.}
  \label{fig:q43_hist}
\end{figure*}


\section{The effect of the first phase of mass transfer} 
\label{sub:effect_of_first_mt}
In this subsection, we discuss in detail how the first phase of mass transfer affects the mass ratio and orbital separation distributions of binaries. In the following, we restrict the discussion to sources with $M_{\rm ZAMS,1} \gtrsim 45 \,M_{\odot}$ at $Z = 0.0007$. We do this so that the  the impact of the first mass transfer phase can be more easily understood, since at this metallicity the stellar winds are negligible and for this mass range the orbit is not modified by any natal kicks. Therefore, the impact of the first mass transfer episode can be more easily understood.. However, we note that results presented here are still qualitatively true for all binaries. 

In the next section, we present examples for Case B and Case C mass transfer episodes. For this, we assume that the donor fills its Roche-lobe when it reaches the midpoint of its Hertzsprung gap lifetime. As noted in subsection \ref{sec:masstransfers}, the outcome of Case B mass transfer epsidoe is not sensitivey dependent on when exactly the Roche-lobe overflow occurs (or equivalently, on the initial orbital separation).

For the examples of Case C mass transfer episodes, we assume that the donor fills its Roche-lobe when it reaches an effective temperature of $\rm{log}(T_{\rm eff}) = 3.73\,K$. The latter criteria does not strictly imply that the donor is in its core helium burning phase. In particular, at Z = 0.0007, binaries with $M_{\rm ZAMS,1} \gtrsim 90\,M_{\odot}$ are still in their Hertzsprung gap phase at this effective temperature. Nevertheless, we still choose this criteria, as we find that it is fairly typical for GW sources of the CEE channel to initiate the first mass transfer phase at this stage. We also note, that outcome of the Case C mass transfer phases is more sensitively dependent the initial separation (see discussion \ref{sec:masstransfers}).

 The distribution of mass ratios and orbital separation of massive binaries at the onset of the second phase of mass transfer are distinct for each formation channel.
Therefore, the distribution of these parameters helps us to understand the importance of each channel. The shape of this distribution is primarily determined by the initial conditions and the first phase of mass transfer. The progenitors of each sub-channel introduced in section \ref{section:isolated_brief} have the following range of mass ratios at the onset of the second phase of mass transfer:
\begin{itemize}
    \item cCEE channel: in this scenario the second phase of mass transfer is Case C, unstable and the donor star has a deep convective envelope, i.e. $q_{\rm MT,2}>q_{\rm crit}$ with $\zeta_{\rm ad} =\zeta_{\rm HW}$
    
    \item rCEE channel: the second phase of mass transfer is Case C, unstable but the donor star has a radiative envelope, i.e. 
    $q_{\rm MT,2}>q_{\rm crit}$ with $\zeta_{\rm ad} =\zeta_{\rm ad,rad}$.
    \item stable channel: 
     The binary has a relatively short period (i.e. Case B first phase of mass transfer). The binary has a relatively large $q_{\rm MT,2}$. The orbit only then can shrink efficiently due to a stable mass transfer (see Equation \ref{eq:iso}). This means there is a minimum value of $q_{\rm MT,2}$ associated with this channel (for a given $M_{\rm ZAMS,1}$). This value depends on $\gamma$ and $\beta$ and on the predicted outcome Case A mass transfer episodes in a complicated way, but it is typically $q_{\rm MT,2, min}\gtrsim2.2$ for  at Z$\lesssim$0.005 (see Fig. \ref{fig:mps}). Therefore: $q_{\rm MT,2,min}<q_{\rm MT,2}<q_{\rm crit}$ with $\zeta_{\rm ad} =\zeta_{\rm ad,rad}$.
    \end{itemize}
We note that a stable phase of mass transfer does not typically change the orbit by orders of magnitude. Therefore, the types of the first and the second phases of mass transfers are the same in most cases. For example, most CEE sources have a Case C first phase of mass transfer.


\begin{figure*}
  \includegraphics[trim=0 0 0.0cm 0, clip,width=\columnwidth]{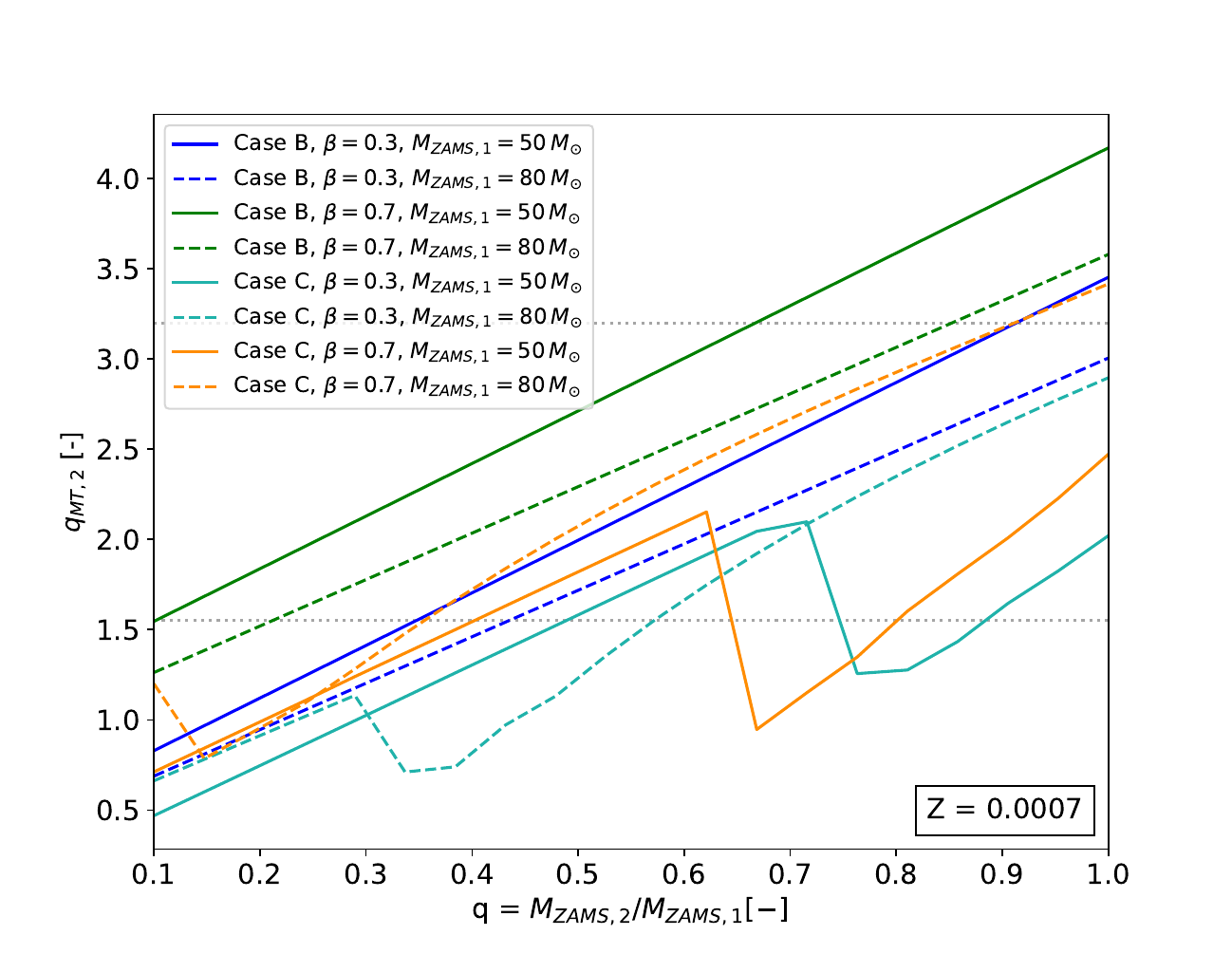}\hfill
  \includegraphics[trim=0 0 0.0cm 0, clip,width=\columnwidth]{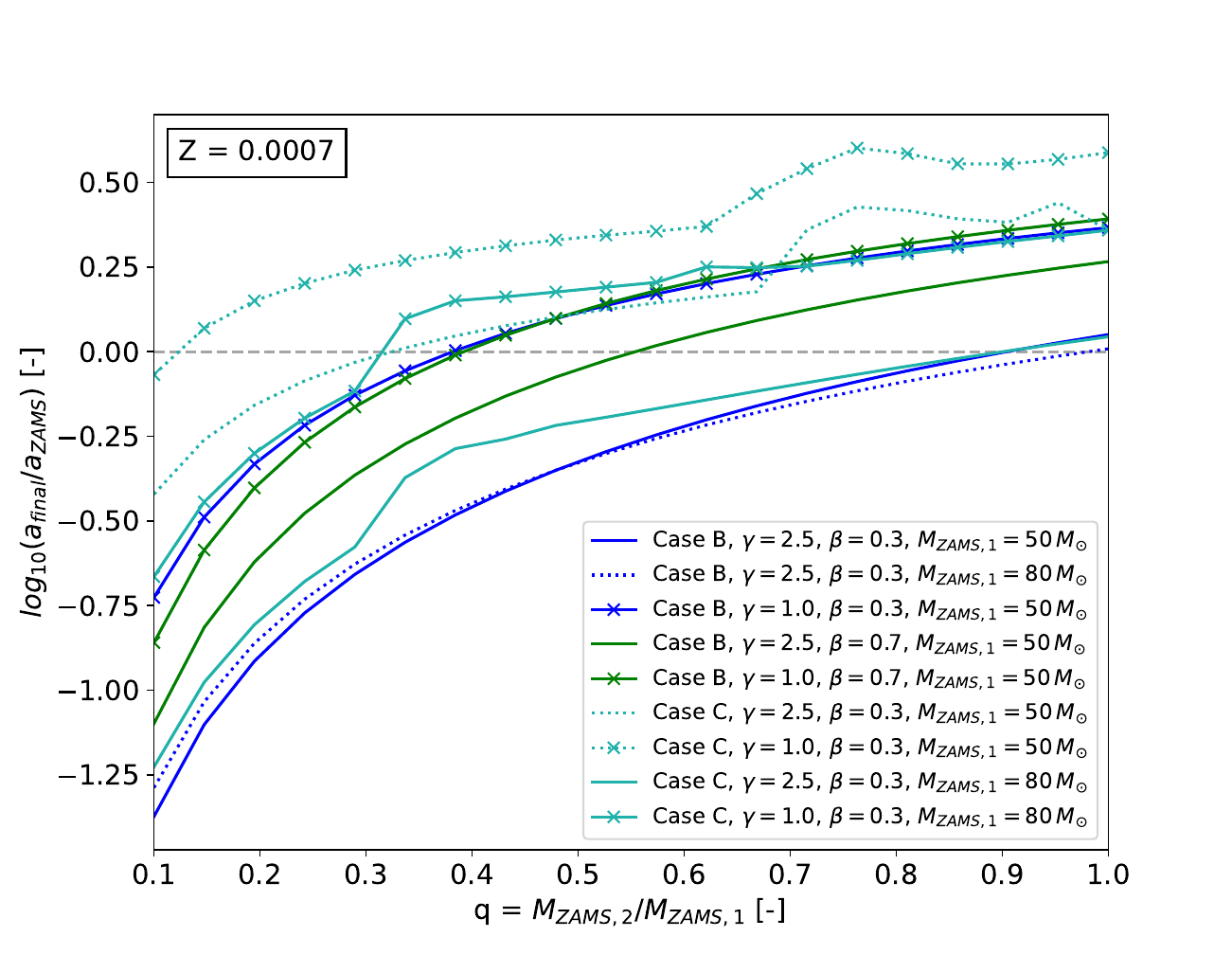}
  \caption{Left panel: the relationship between the initial mass ratio  ($q_{\rm ZAMS} = M_{\rm ZAMS,2}/M_{\rm ZAMS,1}$) and the mass ratio at the onset of the second mass transfer phase ($q_{\rm MT,2} = M_{d}/M_{a}$). We have neglected the effects of supernova but not the stellar winds. We show these relations for different mass transfer types,  mass transfer efficiencies and initial masses. For Case B, we assume that the mass transfer phase starts at the half of the Hertzsprung gap phase lifetime of the donor. In the other case, we assume that the mass transfer phase occurs just before when the donor reaches an effective temperature of $\rm{log}_{10}T_{\rm eff} = 3.73 K$. This coincides with the stage where donors develop deep convective envelopes according to \citet{IvanovaTaam2004}. For the majority of the binaries, this occurs with a core-helium burning donor ($M_{\rm ZAMS,1} \gtrsim 90 M_{\odot}$ at Z = 0.0007). We chose this criteria, because we find that most of the gravitational wave sources from the CEE channel roughly initiate their first mass transfer phase at this stage. Right panel: The relationship between the the initial mass ratio and the shift in the orbital separation in log space for different  different mass transfer types,  mass transfer efficiencies, angular momentum losses and initial masses. Here we have always assume stable mass transfer phase regardless of the assumed $\zeta_{\rm ad,rad}$. We have neglected the effects of supernova but not the stellar winds. The upper horizontal, dotted line shows $q_{\rm crit} = 3.2$, which corresponds to the critical mass ratio for a mass transfer phase with black hole accretor and giant donors with radiative envelopes with $\zeta_{\rm ad,rad} = 4$. The lower bottom line show the critical mass ratio when the donor has a deep convective envelope. See Table \ref{tab:zeta2q}}
\label{fig:qsandas}
\end{figure*}

\begin{figure}
  \includegraphics[width=\columnwidth]{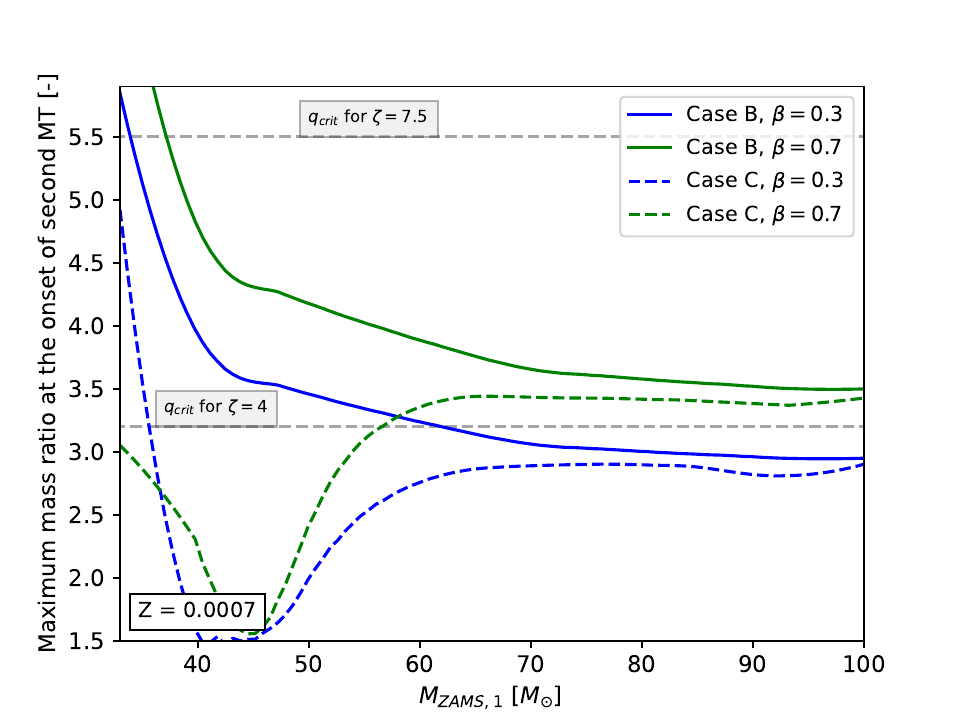}
  \caption{The estimated maximum value of mass ratio at the onset of the second mass transfer phase ($q_{\rm MT,2} = M_{d}/M_{a}$) that a system can reach with a given initial primary mass, mass transfer efficiency and mass transfer phase type. The maximum $q_{\rm MT,2}$ is achieved by binaries that have initially near equal masses. How exactly we defined 'Case B' and 'Case C' mass transfer episode types are explained in the caption of Fig. \ref{fig:qsandas}.}
\label{fig:qmax}
\end{figure}


\subsection{Effect on the mass ratio}
\label{subsec:effectonthemassratio}
In the left panel of Fig. \ref{fig:qsandas}, we show the relation between the initial mass ratio $q_{\rm ZAMS}\equiv M_{\rm ZAMS,2}/M_{\rm ZAMS,1}$ and that at the onset of the second mass transfer phase $q_{\rm MT,2}\equiv  M_{d}/M_{a}$. This figure shows the effect of different mass transfer efficiencies and different initial masses for Case B and Case C mass transfer types.



Fig. \ref{fig:qsandas} tells us what fraction of the binaries can potentially form sources of each channel, and what their initial mass ratio range is. 
For example,  as we will see, GW sources of the stable channel typically require $2.2\lesssim q_{\rm MT,2} \leq q_{\rm crit}$. If $\beta = 0.3$, we expect the most massive sources of the stable channel to evolve from binaries with $q_{\rm ZAMS} \gtrsim 0.8$. On the other hand, the formation via the rCEE channel, which requires $q_{\rm MT,2} > q_{\rm crit}$, is not possible at all with mass transfer efficiencies as low as $\beta = 0.3$. Even if $\beta = 0.7$, the formation becomes only possible if $q_{\rm ZAMS} \gtrsim 0.85$ and  $M_{\rm ZAMS,1} \gtrsim 60\,M_{\odot}$.

We can see that the curves for Case C mass transfer episodes do not have a simple linear relationship, as is the case for Case B. Instead a dip can be observed in the curves at a given range of initial mass ratio. The position of this dip shifts to lower values with increasing mass. This is due to the effect of LBV mass loss. The dip occurs at the lowest initial mass ratio at which the secondary star is still massive enough after the first mass transfer phase, such that it evolves beyond the Humphreys-Davidson limit and eventually loses a significant fraction of its envelope before the onset of the second phase of the mass transfer. 
At the same time, the mass of the accretor black hole at the second mass transfer phase (formed by the initial primary star) is not significantly affected by LBV winds. 
This implies that the range of distribution of $q_{\rm MT,2}$ of binaries with a Case C first phase of mass transfer are narrowed down by the LBV winds.

The maximum $q_{\rm MT,2}$ that can be reached by binaries at Z = 0.0007 is shown in  Fig. \ref{fig:qmax}. The maximum $q_{\rm MT,2}$ is given as a function of $M_{\rm ZAMS,1}$ for a given mass transfer efficiency and mass transfer type. The maximum occurs for near equal mass binaries.
We expect the results to depend only weakly on the metallicity as long as $Z\lesssim 0.002$, because the $M_{\rm ZAMS} - M_{\rm BH}$ relation does not significantly change at this metallicity range (see discussion later in section \ref{sub:stellarwinds}).

We can use Fig. \ref{fig:qmax} to asses the importance of $\zeta_{\rm ad,rad}$. A higher $\zeta_{\rm ad,rad}$ implies a higher critical mass ratio and thereby increases the parameter space for stable mass transfer phase. The degree of orbital shrinkage during stable phase of mass transfer with a black hole accretor increases exponentially with increasing mass ratio (Equation \ref{eq:iso}). 
Hence, we may expect that the merger rate of the stable channel could increase significantly with increasing $\zeta_{\rm ad,rad}$. 
 However, we also have to take into account the limit on the maximum mass ratio at the onset of the second phase of mass transfer. This value is determined by $\beta$ and the $M_{\rm ZAMS} - M_{\rm BH}$ relation. 
The maximum mass ratio of binaries with Case B mass transfer phase is below 3.2, if $M_{\rm ZAMS,1} \gtrsim 60 \, M_{\odot}$ and $\beta = 0.3$. Therefore, in our low mass transfer efficiency models, increasing $\zeta_{\rm ad,rad}$ from 4 to 7.5  only affects the evolution of binaries with $M_{\rm ZAMS,1} \lesssim 60\,M_{\odot}$ .

Fig. \ref{fig:qmax} also shows the importance of the mass transfer efficiency on the rCEE channel. We find that with low accretion efficiencies, the formation of GW sources via the rCEE channel is not possible for almost the entire mass range. This can be seen from Fig. \ref{fig:qmax} as the curve for Case C mass transfer for $\beta \leq 0.3$ and $\zeta_{\rm ad,rad} = 4$ is below the critical mass ratio for $M_{\rm ZAMS,1} \gtrsim 40\, M_{\odot}$.  If $\zeta_{\rm ad,rad}$ is increased up to $7.5$, then the formation of such systems become impossible even with larger mass transfer efficiencies.

\subsection{The effect on the orbital separations}

The relationship between the initial orbital separation and the orbital separation at the onset of the second phase of mass transfer ($a_{\rm MT,2}$) can be approximated as:
\begin{equation}
    \frac{a_{\rm MT,2}}{a_{\rm init}} = f_{\rm wind,post-MT,1}\cdot f_{\rm MT,1}\cdot f_{\rm wind,pre-MT,1},
\end{equation}
if the effects of supernova are negligible.
Here, $f_{\rm wind,post-MT,1}$ and $f_{\rm wind,pre-MT,1}$ describes by what factor the orbit widens due to stellar winds from the birth of the binary until the onset of the first mass transfer phase and from the end of the first mass transfer phase until the onset of the second mass transfer phase, respectively. These changes in orbit are described by Equation \ref{eq:Jeans}. The terms $f_{\rm wind,post-MT,1}$ and $f_{\rm wind,pre-MT,1}$ are negligible for binaries evolving through Case B mass transfer at metallicities $Z \lesssim 0.005$. On the other hand, if the first mass transfer phase is a Case C, then LBV winds contribute to the widening of the orbit appreciably at all metallicities.
The term $f_{\rm MT,1}$ is the change in the orbit during the first mass transfer phase. In principle, these are due to the combined effects of stellar winds and the mass transfer phase itself. However, the effects of the former are typically negligible especially if the mass transfer episode proceeds on thermal timescale. The change in the orbit due to the mass transfer episode is determined by Equation \ref{eq:stable_mt_eq}.
In all cases, $f_{\rm MT,1}$ is the dominant term, i.e. $|f_{\rm MT,1}| \gg \rm{max}(|f_{\rm wind,post-MT,1}|, |f_{\rm wind,pre-MT,1}|)$.

In the right panel of Fig. \ref{fig:qsandas}, we show $\rm{log}(a_{\rm MT,2}/a_{initial})$ as a function of $q_{\rm ZAMS}$ for several model variants and for both Case B and Case C mass transfer types. We see that for Case B mass transfer phase, the change in the orbital separation is mostly determined by the mass ratio of the system,  and it is only very weakly independent on the masses of the binary. The latter no longer holds if the first mass transfer phase is a Case C. Here the mass of the initially primary star does affect the amount of orbital widening. There is an upward shift in the curves of $\rm{log}(a_{\rm 2ndMT}/a_{initial})$ for donor stars with strong LBV winds. This effect is reduced for more massive donor stars, as the amount of mass lost due to LBV winds before the onset of the second mass transfer phase (at $\rm{log} T_{\rm eff} =3.73\, K$) decreases with increasing initial mass.

In general, the orbital separations of systems with low initial mass ratios shrink, whereas the opposite is true for binaries with mass ratios close to one. This is also clear from Equation \ref{eq:stable_mt_eq}, which shows that as long as $M_{d} > M_{a}$ the separation of the system decreases during the mass transfer phase. Where the turnover occurs between widening and shrinking, however, is dependent on the assumed angular momentum loss mode $\gamma$ and accretion efficiency $\beta$. As we increase $\beta$, the effect of the first mass transfer phase starts to converge to the fully conservative case, and therefore the difference between the angular momentum loss becomes less important.

In the context of gravitational wave sources, there is an important point to make regarding the stable channel. In the initial mass ratio range, where we expect the GW sources of stable channel to originate from (i.e. from relatively similar initial masses), the model with $\gamma = 2.5$ and $\beta = 0.3$ shows only an insignificant change in the orbit. The largest widening occurs in models with $\gamma = 1$. This shows why the stable channel is the most efficient in the $\gamma= 2.5$ and $\beta = 0.3$ model, and why it is typically inefficient with $\gamma = 1$.

\subsection{Case A mass transfer episode}
\label{subsec:CaseA}
 The outcome of a Case A mass transfer phase is much more sensitively dependent on the initial separation and the mass ratio than that of the previously discussed mass transfer types (see also  subsection \ref{sub:typesofmt}). This means that it is challenging to show simple relations for these kind of systems as we did in Fig. \ref{fig:qsandas} and \ref{fig:qmax} for Case B and Case C.

To demonstrate typical outcomes of Case A mass transfer phases, we show the evolution of a few selected binaries until the second mass transfer phase as simulated by \textsc{SeBa} in Fig. \ref{tab:CaseA}. These results show two important characteristics of systems evolving through a first phase of Case A mass transfer. Firstly, they develop $q_{\rm MT,2}$ that are significantly larger than that of their Case B counterparts. These mass ratios are often above $q_{\rm crit}$. Secondly, if they survive the Case A mass transfer phase, their orbit widens more compared to systems with Case B mass transfers. These two points imply that the majority of these systems initiate an unstable second phase of mass transfer with an HG donor, which results in stellar merger. All of the systems shown in Fig. \ref{tab:CaseA} would indeed merge with $\zeta_{\rm ad,rad} = 4$, while one of them would merge with $\zeta_{\rm ad,rad} = 7.5$ as a result of the second mass transfer phase. We, however, stress again that such results from rapid population synthesis codes following the stellar tracks of \citet{Hurley_2000} should be treated with caution. These stellar tracks do not track the developing helium core during main sequence and consequently the mass of the stripped donor after the Case A mass transfer might be severely underestimated.

\begin{figure*}
  \hspace*{1cm}
  \includegraphics[width=0.9\textwidth, height=0.8\textheight]{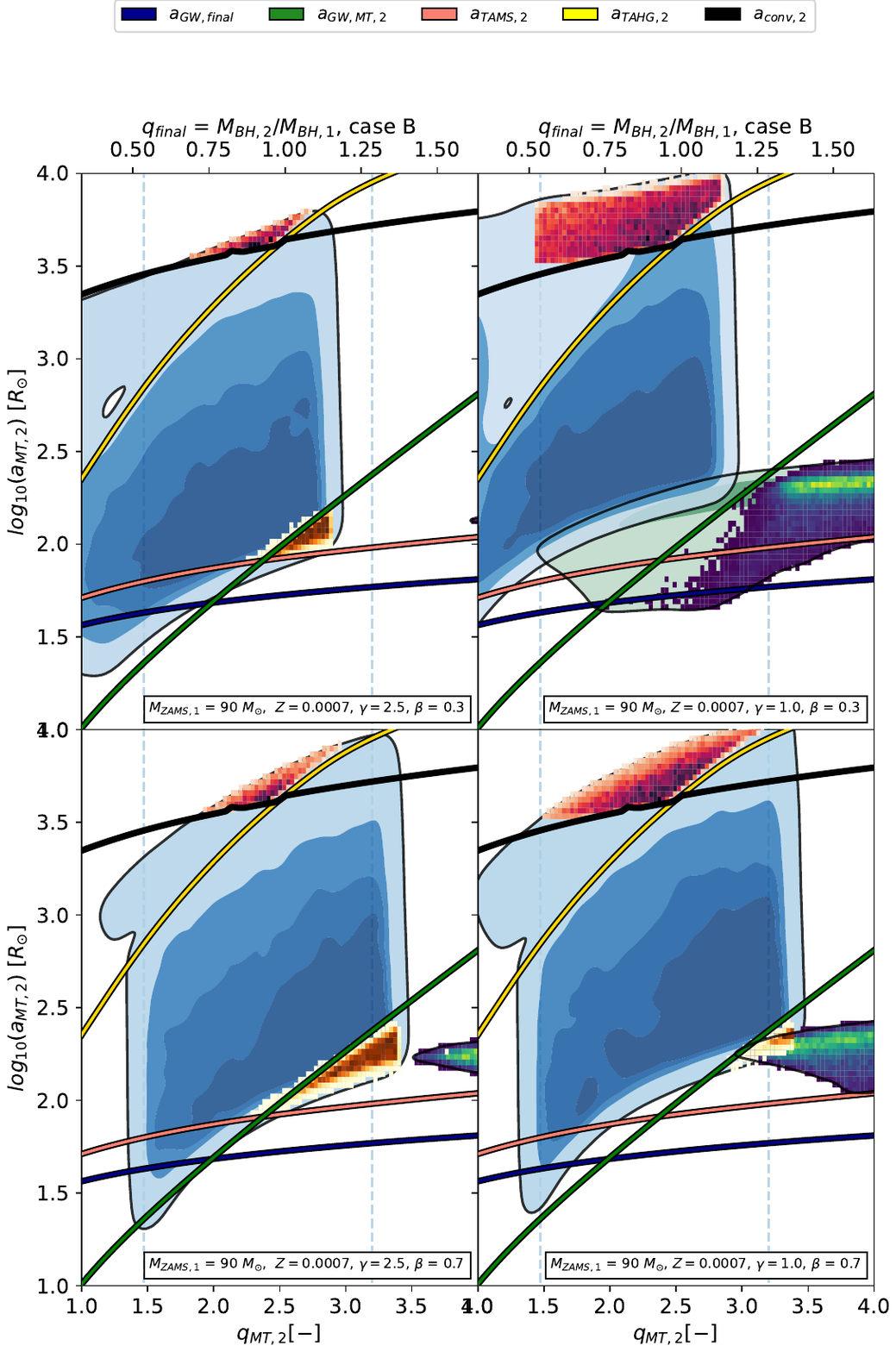}
  \caption{We show the density contours of BH-star binaries, at the onset of the second phase of mass transfer in the mass ratio - orbital separation space for four different model variations. Contours with blue colour denote binaries for which the first phase of mass transfer is either Case B or Case C, while contours with green colour show systems for which this is Case A. In this Fig., we limit our population to systems for which the initial primary star had a mass of $M_{\rm ZAMS,1} = 90\,M_{\odot}$. The metallicity is $Z = 0.0007$ and the convective envelope prescription is of \citet{IvanovaTaam2004}. The $a_{\rm GW,final}$ line shows the separation that binaries have by the time they form BH-BH binaries, if they were to merge due to gravitational waves just within Hubble time. For this, we estimate the remnant mass of the donor assuming that the second phase of mass transfer takes place at the midpoint of the hydrogen shell burning phase. The  $a_{\rm GW,MT,2}$ line indicates the separation that the same binaries have at the onset of the second phase of mass transfer. We assume stable mass transfer (i.e. Equation \ref{eq:iso}) and we neglect the effect of supernova kick on the orbit. The $a_{\rm TAMS,2}$ line indicates the orbital separation at which the second phase of mass transfer occurs when the donor just evolved off the main sequence. The $a_{\rm TAHG,2}$ is the orbital separation at which the second mass transfer phase occurs with a donor, which just evolved off the Hertzsprung gap phase, while $a_{\rm conv,2}$ is the orbital separation at which the donor fills its Roche lobe just before it develops a deep convective envelope. Furthermore we also show 2D histograms of the GW sources; CEE channel (dark red), stable channel with Case B second mass transfer phase (orange), stable channel with Case A second mass transfer phase (blue). These histograms have been normalised to one.}
\label{fig:mps}
\end{figure*}

\label{subsec:impact_cee}

\section{Evolution of massive binaries with different stellar wind models}
\label{sec:mostmassivestory}

Here, we discuss the evolution of the systems, which form merging binary black holes with the most massive black hole binary in each of our stellar wind models.
In order to gain a more detailed understanding of the combined effect of $\zeta_{\rm ad,rad}$ and different stellar wind models, we show the typical formation histories of sources with the most massive primary black holes for each of our stellar wind models in the Appendix in Fig. \ref{fig:GetFormationHistory},. In these scenarios, we assume $\gamma = 2.5$, $\beta = 0.3$ and $\zeta_{\rm ad,rad}$ = 7.5. 
Fig. \ref{fig:GetFormationHistory} shows that the formation path of the most massive sources in Model I and Model II are fairly similar. The evolution starts out with an almost equal mass binary, as this leads to the highest mass ratios at the onset of the second mass transfer phase. On the other hand, the minimum initial separation (so that the first mass transfer is Case B) of such sources is different. Decreasing the optically thin stellar winds by a factor of three increases the radii of stars at the end of the main sequence. In particular, at Z = 0.01, for $M_{\rm ZAMS} = 100\,M_{\odot}$, $R_{\rm TAMS}$ can increase from $83.2\,R_{\odot}$ to $110\,R_{\odot}$ when switching from Model I to Model II. Therefore, binaries of Model II need to start out with a higher initial separation, because of their larger $R_{\rm TAMS}$.
 It is also confirmed, that these sources indeed require large mass ratios of $q\sim4-5$ at the onset of the second mass transfer. The formation of such systems is not possible in the $\zeta_{\rm ad,rad} = 4$ model as they would undergo unstable mass transfer and would merge during CEE. We note that primary masses of $\sim 16-18\, M_{\odot}$ for Model I with $\zeta_{\rm ad,rad} = 4$ can still be seen  in Fig. \ref{fig:wind_primary_first}, however these systems have secondary black hole masses $M_2 \sim 5-7\,M_{\odot}$. The formation of such systems is possible because they start out with a high initial mass ratio and therefore the orbit significantly shrinks after the first mass transfer (even if $\gamma = 1$).

Binaries with initial parameters as shown for Model II in Fig. \ref{fig:GetFormationHistory} do not form gravitational wave sources in Model III. The evolutionary path of such a binary would be exactly the same as for Model II until the formation of the first naked helium star. However, because of the decreased Wolf-Rayet-like mass loss rates, the initial primary star forms a 28.0 $M_{\odot}$ black hole. The intially secondary star becomes a 93.5 $M_{\odot}$ massive HG giant at the start of the second mass transfer. The mass ratio of 3.3 is not sufficiently high enough to shrink the orbital separation ($\sim420\,R_{\odot}$) enough so that the $28.0\,M_{\odot}$ - $30.5\,M_{\odot}$ BH-BH would merge within Hubble time. 

Instead, the sources with the most massive primaries in Model III form with similar initial masses as in Model II but at shorter orbital separations ($a\sim 180-200\,R_{\odot}$). At such a separation the first mass transfer occurs with a MS donor. As explained earlier, this halts the growth of the helium core. It is predicted that the mass transfer episode proceeds while the donor eventually evolves off the main sequence and it finally ends with the HG donor losing its hydrogen envelope, leaving behind a $26.8\,M_{\odot}$ massive naked helium star. With the lowered Wolf-Rayet-like winds, this collapses into a $19.8\,M_{\odot}$ black hole. At the beginning of the second mass transfer, the donor star is a $96.3\,M_{\odot}$ HG giant. The mass ratio $\sim4.9$ is high enough to shrink the orbital separation ($a\sim370.8\,R_{\odot}$) significantly. The same system in Model II would have a larger mass ratio because of the lower mass of the black hole ($M_1\approx 14.9\,M_{\odot}$) and the mass transfer would become unstable, which results in a stellar merger. As systems with such high masses always require mass ratios $q_{\rm MT,2}>4$, it is clear why there is such a huge difference in the maximum masses of merging binary black holes between $\zeta_{\rm ad,rad} = 4$ and $\zeta_{\rm ad,rad} = 7.5$ for Model III.

\begin{figure*}
\includegraphics[width=1.0\textwidth]{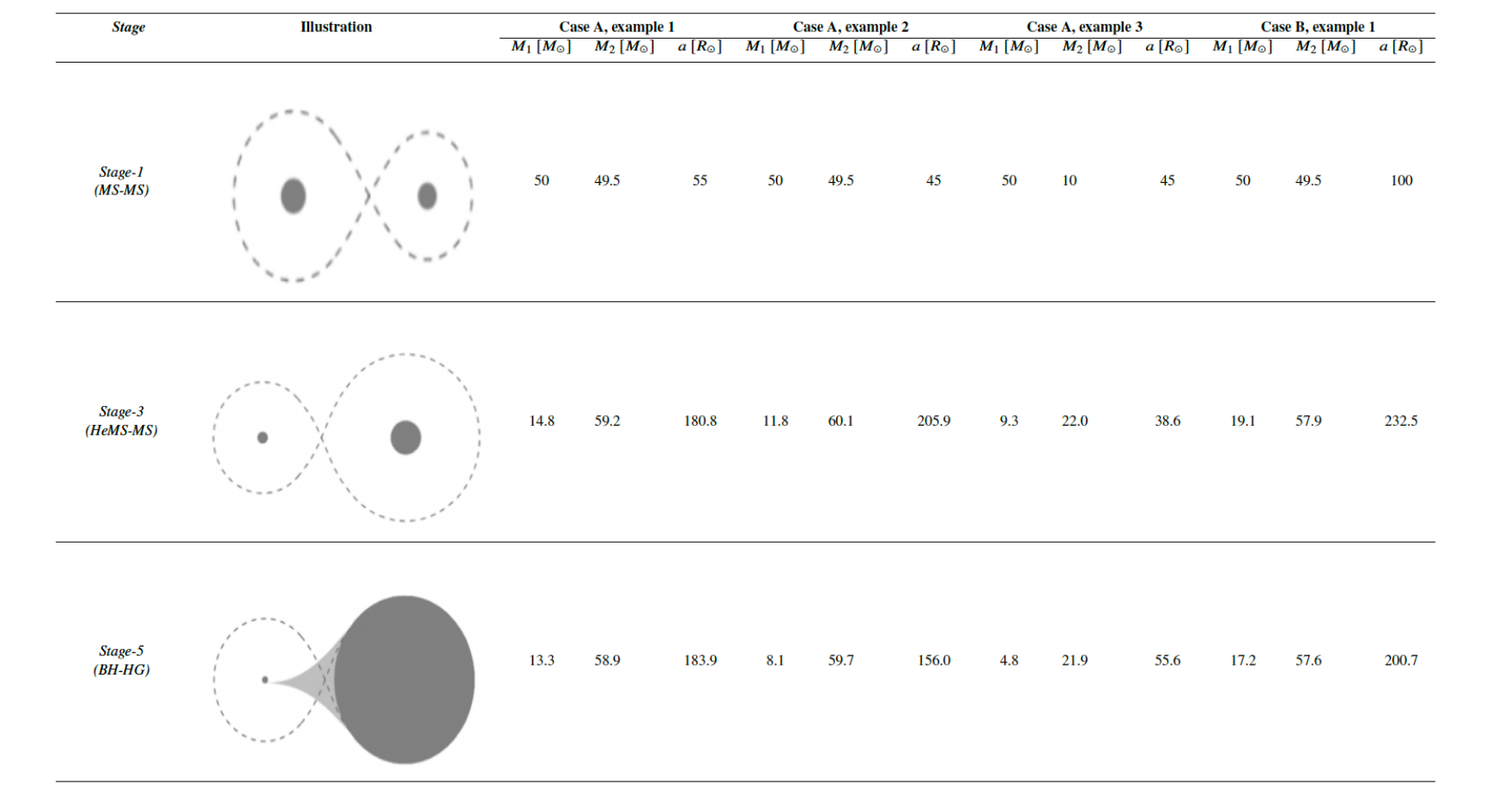} \hfill
  \caption{A summary of the evolution of a few selected binaries going through Case A mass transfer phases compared with a binary evolving through a Case B mass transfer episode. We assume $\gamma = 1$ and $\beta = 0.3$. We only show the evolution until the initial secondary fills  its Roche-lobe The numbers denoting the evolutionary stage are the same as shown in Fig. \ref{fig:isolatedchannelsl}. Therefore Stage 1 refers to the binary at its birth, Stage 2 is at the end of the second mass transfer phase and Stage 3 is just before the second mass transfer episode. We note that for Case A binaries the first mass transfer phase is interrupted at the end of the main sequence of the donor, as its radius starts shrinking. Shortly after, the radius starts to expand rapidly again with the beginning of the Hertzsprung gap phase and therefore refills its Roche-lobe. This second stage ends with the donor losing its hydrogen envelope. For Case A binaries stage 3 refers to stage when the donor becomes a stripped helium star.}
  \label{tab:CaseA}
\end{figure*}

\begin{figure*}
\includegraphics[width=1.0\textwidth]{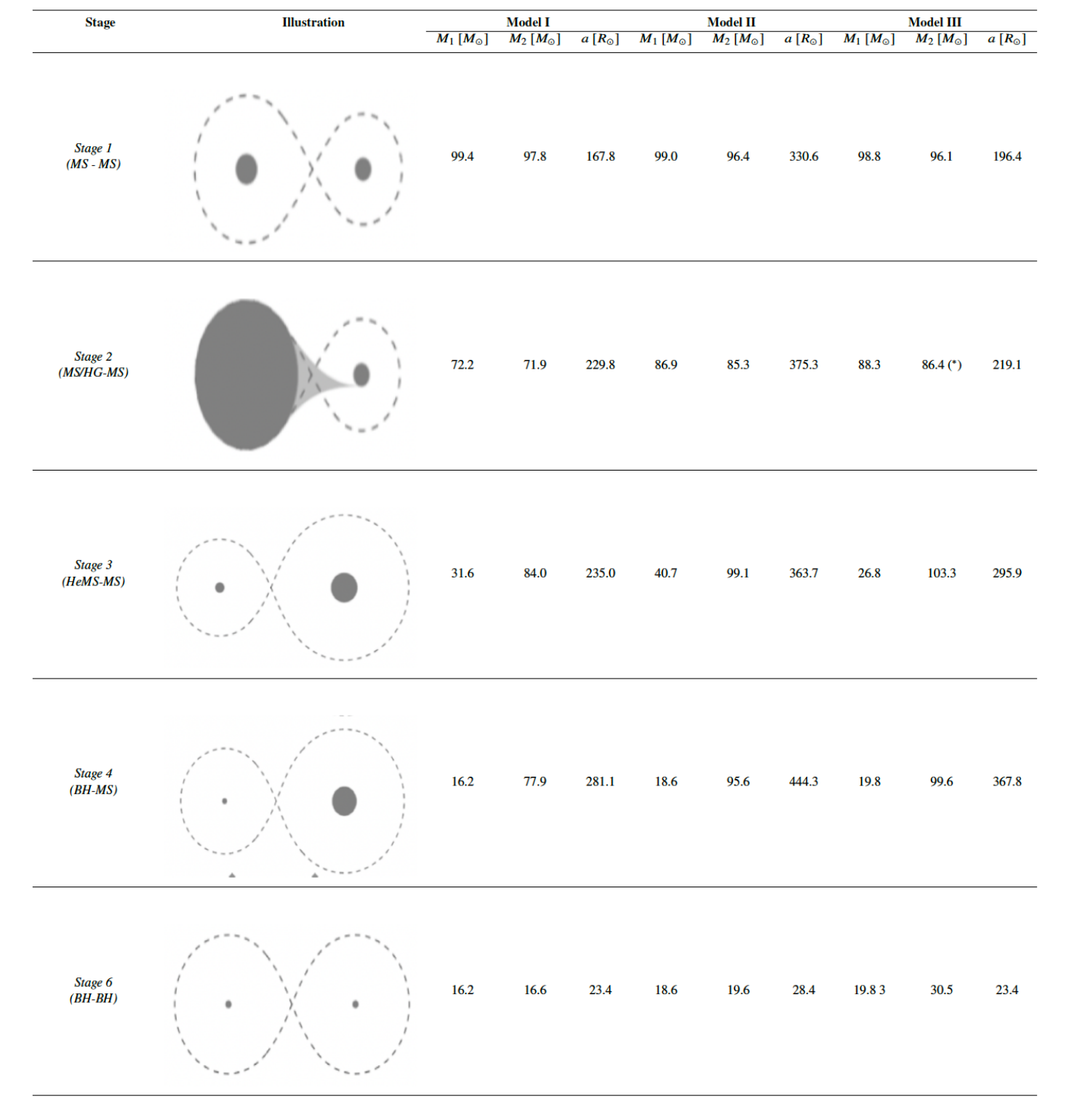} \hfill
  \caption{Typical formation histories of merging binary black holes with the most massive primary black holes according to our three different stellar wind models. For brevity, we only show a few stages of the most important steps in the evolution of these systems. The stage numbers correspond to the evolutionary stages as shown in Fig. \ref{fig:isolatedchannelsl}. For all the sources here, we assume $\gamma = 2.5$, $\beta = 0.3$ and $\zeta_{\rm ad,rad}$ = 7.5. The asterisk for Stage 2 for Model III means that the first mass transfer phase is Case A}
  \label{fig:GetFormationHistory}
\end{figure*}

\section{Additional figures }
\label{appendix:additional_figures}

\begin{figure*}
\includegraphics[width=\textwidth]{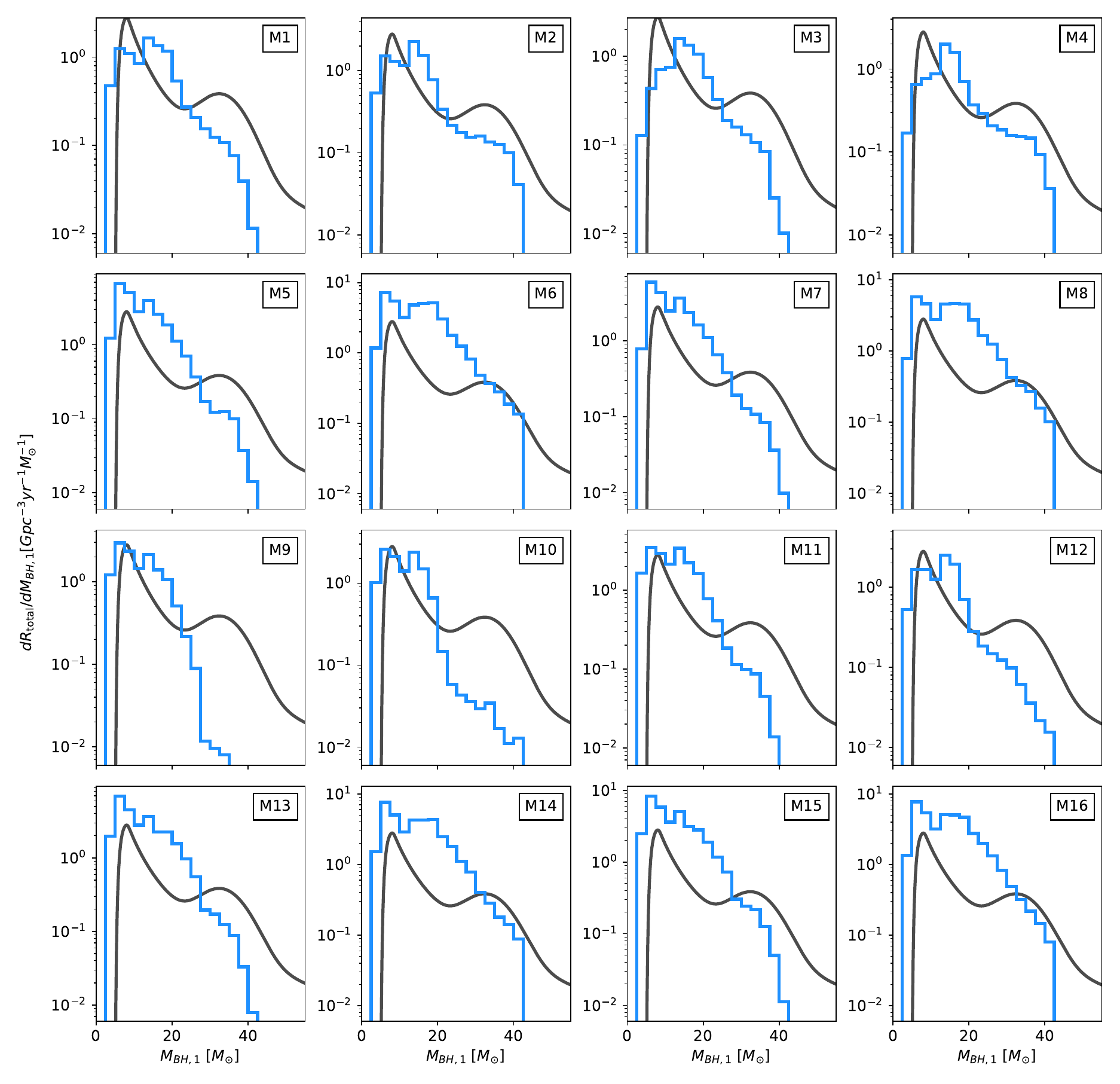} \hfill
  \caption{We compare the distribution of primary BH masses of our 16 different model variations (blue) discussed in section \ref{sub:popprops} (see also Table \ref{tab:rates_of_all})  with  the inferred primary mass distribution of BH-BH binaries of the GWTC-3 catalogue \citep[][]{Abbott2023_GWTC3_POP}, based on the "Power Law + Peak" model (black).}.
  \label{fig:compi}
\end{figure*}

\begin{figure*}
\includegraphics[width=\textwidth]{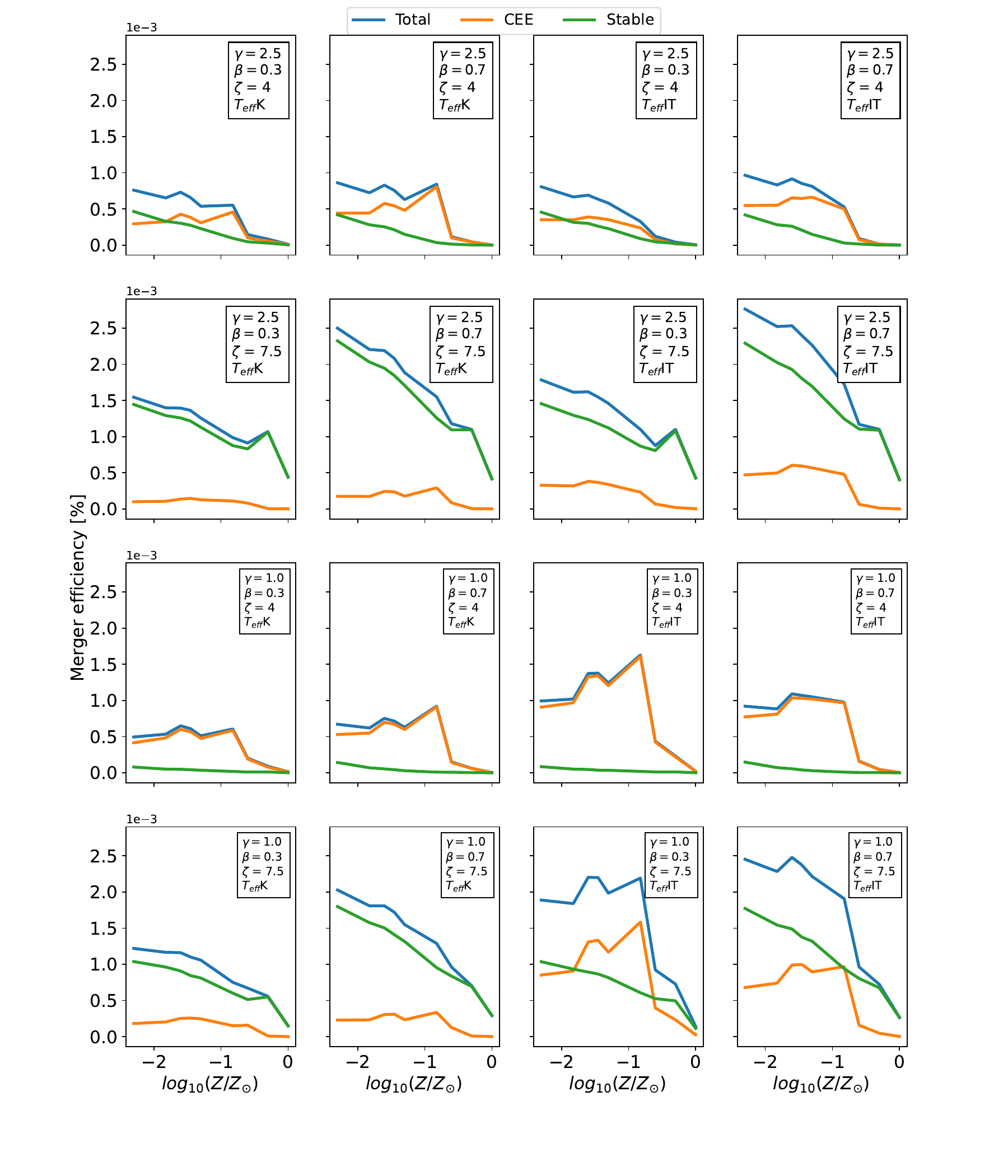} \hfill
  \caption{Merger rate efficiencies of merging binary black holes our different model variations as a function of metallicity. We consider those systems merging binary black holes, for which the time between zero-age main sequence and merger due to GWs is equal or less than the 14 Gyr. Merger rate efficiency is defined in Equation \ref{eq:mereff}.}.
  \label{fig:merger_efficiency}
\end{figure*}

\begin{figure*}
  \includegraphics[width=1.0\textwidth]{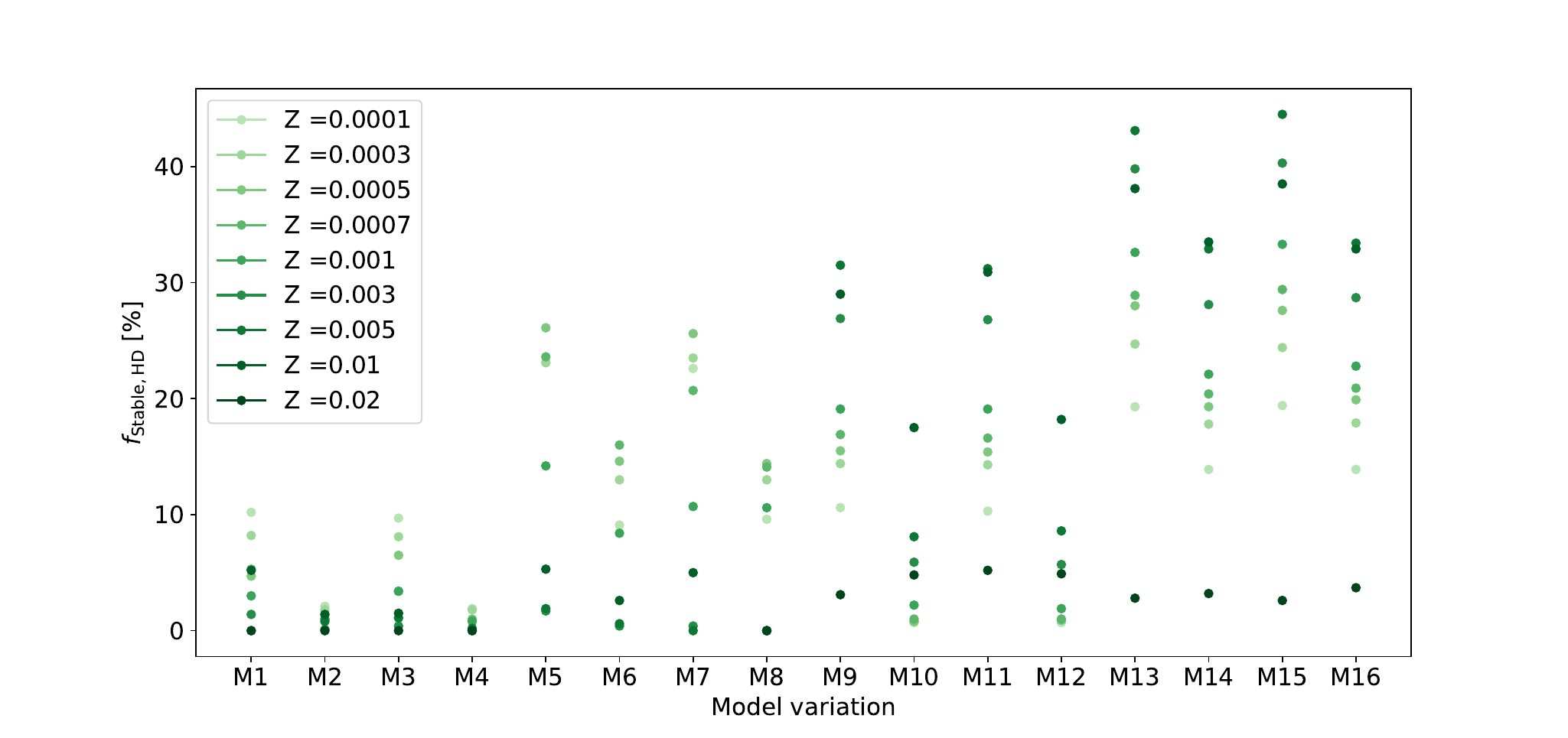}
  \caption{The percentage of the binaries in the stable channel,
in which any of the stars evolves beyond the Humphreys-Davidson
limit for all of our model variations at each metallicity. See Table \ref{tab:rates_of_all} for the meaning of the labels of the model variations.}
\label{fig:HD_crossing_per_model_extensive}
\end{figure*}

\begin{figure}
\includegraphics[trim=0 0 1cm 0, clip,width=\columnwidth]{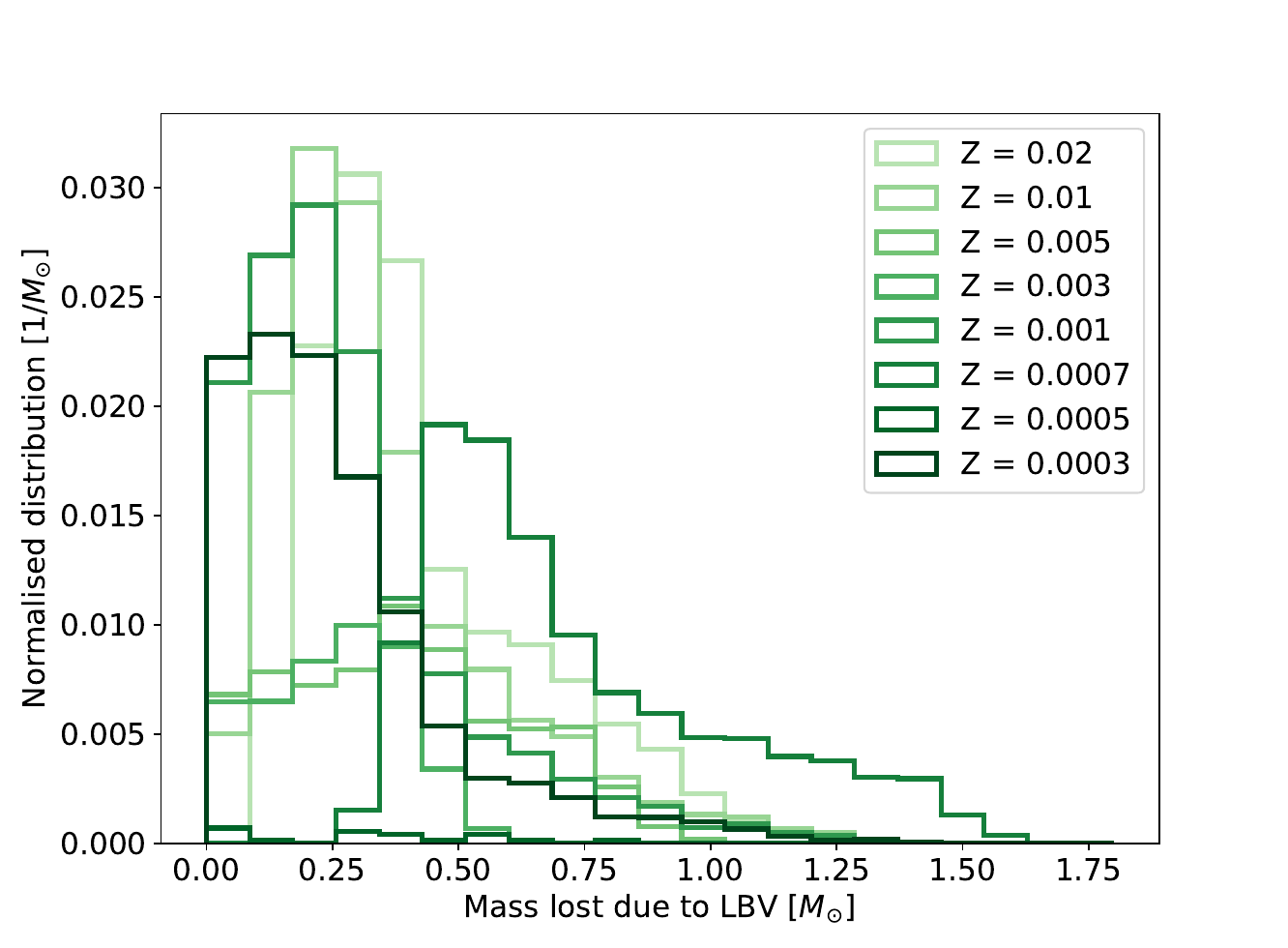}
  \caption{The estimated total mass lost due to LBV winds with $\dot{M_{\rm LBV}}=10^3M_{\odot}\rm{yr^{-1}}$ of those stars in the systems of the stable channel sources, which cross eventually the Humphreys-Davidson limit in the $\gamma = 2.5$, $\beta = 0.3$, $\zeta_{\rm ad,rad} = 4$, $T_{\rm eff}$IT model variation.}
  \label{fig:mass_lost_increased_lbv}
\end{figure}

In this subsection, we present additional figures that are helpful (but not necessary) to understand the main results of this work. In Fig. \ref{fig:compi}, we compare the primary BH mass distributions of our models  (shown in Table \ref{tab:rates_of_all}) with the inferred distribution of the GWTC-3 catalog \citep{Abbott2023_GWTC3_POP}. We represent the inferred distribution with the "Power Law + Peak" model (see e.g. equation B4 in \citealt{Abbott2023_GWTC3_POP}). The values of the model parameters are: $\alpha = -3.5$, $m_{\rm min} = 4.6\,M_{\odot}$, $m_{\rm max} = 86\,M_{\odot}$, $\lambda_{\rm peak} = 0.038$, $\mu_{m} = 34\,M_{\odot}$, $\sigma_m = 5.7\,M_{\odot}$, $\delta_m = 4.82\,M_{\odot}$. For simplicity, we ignored uncertainties. We normalised the distribution to the inferred merger rate density of BH-BH mergers; $R_{\rm GWTC3} = 28.3\,\rm{Gpc^{-3}yr^{-1}}$.
While the upper end of the inferred distribution reaches  $M_{\rm BH,1} = 86\,M_{\odot}$, the maximum BH mass of our model variations is only $M_{\rm BH,1} = 42\,M_{\odot}$, since we do not sample stars more massive than $M_{\rm ZAMS} = 100\,M_{\odot}$. To make the comparison easier, we only show the figures up to $M_{\rm BH,1} = 60,M_{\odot}$ .
None of our models can reproduce the peak at 34$\,M_{\odot}$. If this feature is indeed related to (pulsational) pair instability supernova \citep[PPISN ][]{Farmer2019, Marchant2019}, then this is not surprising . Firstly, we do not model PPISN in this study. Secondly, since we do not consider stars with $M_{\rm ZAMS} > 100\,M_{\odot}$, our PPISN rate would be negligible in any case. All of our models, except M9-M10, yield distributions that are broadly similar to the power-law component of the inferred distribution.  M9-M10 exhibit much steeper decrease at $M_{\rm BH,1}\gtrsim20\,M_{\odot}$. Models M5-M16 have a peak around $M_{\rm BH,1}\approx10\,M_{\odot}$, similarly to the inferred distribution. On the other hand,for  models M1-M4, this peak occurs at somewhat higher masses ($M_{\rm BH,1}\approx15$-$20\,M_{\odot}$), which is not supported by observations.

In Fig. \ref{fig:merger_efficiency}, we show the merging efficiency of each of our model variations (with standard stellar wind models) at each simulated metallicity. In Figure \ref{fig:HD_crossing_per_model_extensive}, we show the number of binaries of the stable channel in which any of the stars cross the Humphreys-Davidson limit, expressed as a fraction of all systems in the stable channel. In Figure \ref{fig:mass_lost_increased_lbv}, we show the estimated total mass lost due to LBV winds with a mass loss rate of $\dot{M_{\rm LBV}}=10^3M_{\odot}\rm{yr^{-1}}$ of those stars in the systems of the stable channel sources, which cross eventually the Humphreys-Davidson limit in the $\gamma = 2.5$, $\beta = 0.3$, $\zeta_{\rm ad,rad} = 4$, $T_{\rm eff}$IT model variation. 
In Fig. \ref{fig:mps}, we show density contours, which reflect the mass ratios and orbital separations of binaries at the onset of the second phase of mass transfer. These binaries were simulated with \textsc{SeBa} at a metallicity $Z = 0.0007$.  In Fig. \ref{fig:mps}, we limit the initial primary mass to $M_{\rm ZAMS,1} = 90\,M_{\odot}$. We show four model variations with different assumptions on $\gamma$ and $\beta$. At this stage, the initial primary star has already formed a black hole. If the first phase of mass transfer is Case B, then the mass of the black hole is $M_{\rm BH,1} = 35.44\, M_{\odot}$.  We distinguish binaries for which the first phase of mass transfer is Case A (shown by the green contour, hereafter `Case A binaries') and for which it is Case B or Case C (shown by the blue contour, hereafter `Case B binaries' and `Case C binaries', respectively). 
We also show the progenitors of GW sources as 2D histograms over the contours. We distinguish three types: CEE channel, stable channel with Case B binaries (`stable Case B', hereafter) and stable channel with Case A binaries (`stable Case A').


\bsp	
\label{lastpage}
\end{document}